\documentclass[12pt]{article}

\usepackage{fullpage}
\usepackage{amsmath,amssymb,amsthm,amscd,graphicx}
\usepackage{hyperref}
\usepackage{bbold}
\usepackage{amsfonts}
\usepackage{mathtools}
\usepackage{stmaryrd}
\SetSymbolFont{stmry}{bold}{U}{stmry}{m}{n}
\usepackage{caption}
\usepackage{subcaption}
\usepackage[style=alphabetic,backref,giveninits,maxbibnames=99,doi=true,isbn=false,url=true]{biblatex}
\usepackage[indent]{parskip}
\usepackage{geometry}
\usepackage{tikz}
\usetikzlibrary{calc,decorations.markings}
\usepackage{bm,tabularray}

\numberwithin{equation}{section}

\renewcommand\hat\widehat
\renewcommand\tilde\widetilde
\newcommand{\bea}{\begin{eqnarray}}
\newcommand{\eea}{\end{eqnarray}}
\newcommand{\be}{\begin{equation}}
\newcommand{\ee}{\end{equation}}

\newcommand{\AKV}{{\mathrm{AKV}}}

\newcommand\ii{\mathrm{i}} % imaginary unit
\newcommand\dif{\mathop{}\!\mathrm{d}} % differential
\newcommand\eu{\mathrm{e}} % euler
\newcommand\inst{\mathrm{inst}} % instantonic
\newcommand\reg{\mathrm{reg}} % instantonic
\newcommand\sing{\mathrm{sing}} % singular
\newcommand\QJK{\mathrm{QJK}} % Quantum Jeffrey-Kirwan
\newcommand\JK{\mathrm{JK}} % Jeffrey-Kirwan
\newcommand\e{\epsilon}

\newcommand\BR{\mathbb{R}}
\newcommand\BP{\mathbb{P}}
\newcommand\BC{\mathbb{C}}
\newcommand\BZ{\mathbb{Z}}

\newcommand\Q{\mathsf{Q}}
\newcommand\cF{\mathcal{F}}

\newcommand\cN{\mathcal{N}}

\newcommand\cH{\mathcal{H}} % hyperplane volume
\newcommand\bt{\bm{t}} % kahler class
\newcommand\bv{\bm{v}}
\newcommand\bgamma{\bm{\gamma}}
\newcommand\charge{q}
\newcommand\eGLSM{{$\epsilon$-GLSM}}
\newcommand\vol{{\rm{vol}}}
\newcommand\GW{{\rm{GW}}} 
\newcommand\semicl{{\rm{s.c.}}}

\DeclareMathOperator{\cD}{\mathcal{D}} % derivative wrt divisor
\DeclareMathOperator{\Ga}{\Gamma} % gamma function
\DeclareMathOperator{\Li}{Li} % polylog
\newtheorem{theorem}{Theorem}[section]

\newtheorem{example}[theorem]{Example}
\theoremstyle{definition}
\newtheorem{remark}[theorem]{Remark}

\newtheorem{definition}[theorem]{Definition}
\begin{document}
\hfill UUITP-18/23\\[25pt]

\begin{center} \Large
{\bf Symplectic cuts and open/closed strings I}
 \\[12mm] \normalsize
{\bf Luca Cassia${}^{a}$, Pietro Longhi${}^{b,c}$ and Maxim Zabzine${}^b$} \\
[8mm]
{\small\it ${}^a$Department of Mathematical Sciences, Durham University, \\
Lower Mountjoy, Stockton Road, Durham, DH1 3LE, United Kingdom}\\
{\small\it ${}^b$Department of Physics and Astronomy, Uppsala University,\\ Box 516, SE-75120 Uppsala, Sweden\\}
{\small\it ${}^c$Department of Mathematics, Uppsala University,\\ 
 Box 480, SE-75106 Uppsala, Sweden\\}

\end{center}
\vspace{7mm}
\begin{abstract}

This paper introduces a concrete relation between genus zero closed Gromov--Witten invariants of Calabi--Yau threefolds and genus zero open Gromov--Witten invariants of a Lagrangian $A$-brane in the same threefold.

Symplectic cutting is a natural operation that decomposes a symplectic manifold $(X,\omega)$ with a Hamiltonian $U(1)$ action into two pieces glued along an invariant divisor.
In this paper we study a quantum uplift of the cut construction defined in terms of equivariant gauged linear sigma models.
The nexus between closed and open Gromov--Witten invariants is a quantum Lebesgue measure associated to a choice of cut, that we introduce and study.
Integration of this measure recovers the equivariant quantum volume of the whole CY3, thereby encoding closed Gromov--Witten invariants.
Conversely, the monodromies of the quantum measure around cycles in K\"ahler moduli space
encode open Gromov--Witten invariants of a Lagrangian $A$-brane associated to the cut.

Both in the closed and the open string sector we find a remarkable interplay between worldsheet instantons and semiclassical volumes regularized by equivariance.
This leads to equivariant generating functions of GW invariants that extend smoothly across the entire moduli space, and which provide a unifying description of standard GW potentials.
The latter are recovered in the non-equivariant limit in each of the different phases of the geometry.

\end{abstract}

\eject

\tableofcontents

\section{Introduction}
In this work we introduce a new direct relation between genus zero closed and open topological string potentials.
We find an explicit and computable description of this relation by studying a stringy quantization of a classical geometric construction known as symplectic cutting.

A symplectic $d$-dimensional manifold $(M,\omega)$ with a Hamiltonian $U(1)$ action admits a well-known reduction to another manifold $(M_0,\omega_0) = (M,\omega)\sslash U(1)$ of dimension $d-2$, known as the symplectic quotient.
The symplectic cut of $M$ consists of the union of two parts $(\overline M_\pm,\omega_\pm)$ each of dimension $d$, glued along a codimension-two locus isomorphic to $(M_0,\omega_0)$. 
Symplectic cutting and the inverse operation of symplectic gluing are basic constructions of classical symplectic geometry and thanks to their functoriality property, it is possible to relate classical and quantum geometric data of $(M,\omega)$ to those of $(\overline M_\pm,\omega_\pm)$ and $(M_0,\omega_0)$.

\begin{figure}[!ht]
\centering
\includegraphics[width=0.35\textwidth]{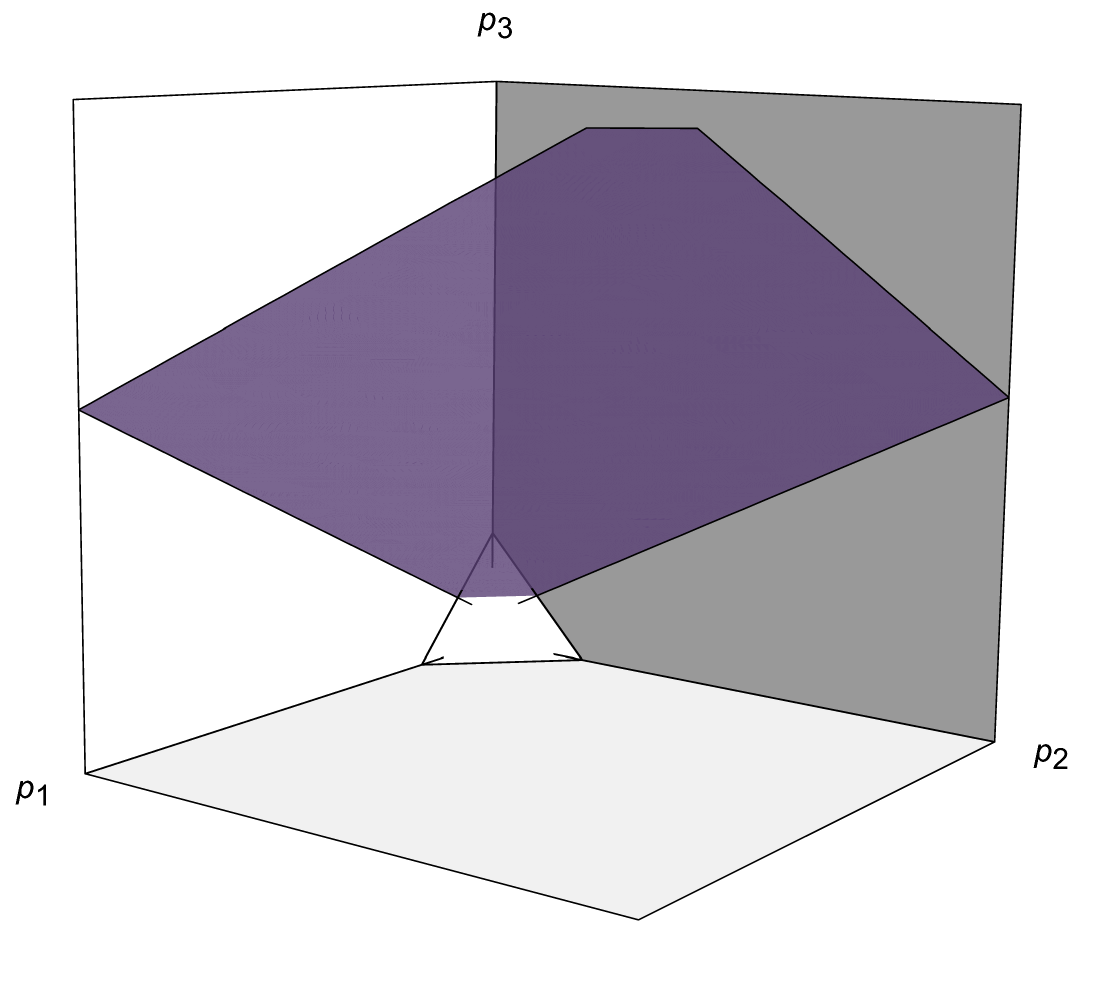}
\caption{Moment polytope of $O_{ \BP^2}(-3)$, with a hyperplane defining a symplectic cut.}
\label{fig:intro-ex}
\end{figure}

The symplectic cut construction has a very broad range of applications, and throughout the paper we focus on the setting of toric Calabi--Yau threefolds. 
Every toric CY3 $(X,\omega)$ admits a Hamiltonian $U(1)^3$ action, and the corresponding moment map defines a polytope with boundaries and corners where the action degenerates, see Figure~\ref{fig:intro-ex}.
The toric setting admits a natural class of symplectic cuts associated with hyperplanes of rational slope within the moment polytope of $(X,\omega)$.
As we explain in the main body of the paper, there is a close relationship between this class of symplectic cuts and a class of Lagrangian submanifolds known as ``toric Lagrangians'' studied extensively in the physics literature.

In this paper we initiate our study of \emph{quantum} symplectic cuts by considering equivariant
Gauged Linear Sigma Models (\eGLSM) with target $(X,\omega)$.%
\footnote{Here the equivariant parameters $\e_i$ are associated to a symplectic quotient description $(X,\omega) = \BC^{r+3}\sslash U(1)^r$ where they are related to rotations of the ambient space $\BC^{r+3}$. By ``quantum'' we refer to stringy instanton effects which promote classical cohomology to quantum cohomology. This should not be confused with quantization with respect to $g_s$.}
We focus on worldsheets with genus zero and one puncture (disks), and consider both $A$ and $B$-type boundary conditions.
While the study of $B$-branes in \eGLSM~is well developed, $A$-branes are much less understood.
One of the main results of this work is a proposal for including toric $A$-branes in the \eGLSM~framework. 
This proposal is motivated by the relation between symplectic cuts and toric Lagrangians in classical geometry, 
and is supported both by a physical interpretation and by explicit computations of the $A$-brane's superpotential.

\subsubsection*{Main results}

We now sketch the main results and the logical thread of the paper.

The partition function of an \eGLSM~with a space-filling $B$-brane computes the equivariant volume of $X$ corrected by genus-zero worldsheet instantons, schematically
\be
\label{eq:FD-split-intro}
 \cF^D(\bt,\e) = \cF_{\semicl}^D(\bt,\e) + \cF_\inst^D(\bt,\e) \,.
\ee
We will refer to $\cF^D$ as the ``quantum volume'' of $X$.%
Here the leading behaviour of the semiclassical part $\cF^D_{\semicl}\sim\vol_{\e}(X) + \dots$ in the equivariant parameters defines the regularized classical volume of $(X,\omega)$, and the instanton part collects contributions of order $\eu^{-\beta\cdot\bt}$ that vanish at large $\bt$, for $\beta$ the instanton charge.
The CY3 quantum volume $\cF^D$ is a solution of the equivariant Picard--Fuchs equations corresponding to a space-filling $B$-brane.\footnote{\label{foot:FD-GW}%
A disclaimer on terminology is in order. The quantum equivariant volume $\cF^D$ is defined as a 2d gauge theory partition function with $A$-type supersymmetry, and is known to obey equivariant Picard-Fuchs equations of the mirror \cite{Hori:2013ika}.
A connection to the genus-zero closed topological string free energy (i.e.\ the GW potential $\Phi^0_{\GW}$) arises through the Matone relation $\cF^D\sim \tilde{t}^a \partial_{\tilde{t}^a} \Phi^0_{\GW}  - 2 \Phi^0_{\GW}$, upon extracting the regular part in equivariant parameters~\cite{Cassia:2022lfj}.
However, here $\Phi^0_{\GW}(\tilde{t})$ can be regarded as the generating series of $A$-model (Gromov--Witten) invariants only when expressed as a function of the flat coordinates $\tilde{t}^a$ which are related to the K\"ahler moduli $t^a$ through the mirror map.
From a purely combinatorial viewpoint, this is the same relation that connects (stable) absolute quasi-maps invariants to equivariant Gromov--Witten invariants \cite{ciocan2010moduli, ciocan2014stable}, or equivalently Givental's $I$-function to the $J$-function \cite{givental1996equivariant}.
}
Other solutions 
correspond to quantum volumes of $B$-branes supported on toric divisors and their intersections, both compact and non-compact.

The first main result of this paper is the observation that equivariance ties together semiclassical and instanton contributions to quantum volumes in highly nontrivial ways, and that this leads to a unification of solutions of PF equations from different chambers of the K\"ahler moduli space into globally defined expressions.

A conventional way of expressing solutions to Picard--Fuchs equations in the physics literature is through power series in
$\eu^{-\beta\cdot\bt}$, which tends to obscure global analytic properties of the solutions.
By contrast, an important feature of the \eGLSM~framework is that it provides exact closed-form expressions for $\cF^D$, resummed to all orders in the worldsheet instanton parameters.
These provide control over the global analytic structure of $\cF^D(\bt,\e)$, which reveals remarkable phenomena that had been missed by previous studies based on series solutions to PF equations.\footnote{The global behaviour of $\cF^D$ is much better understood from a mathematical viewpoint,  thanks to work on the Crepant Transformation Conjecture. See Remark \ref{rmk:note-added}.}

To illustrate this point, let us consider the quantum volume of a space-filling brane in the resolved conifold (the computation can be found in Section~\ref{s:conifold})
\be\label{eq:conifold-FD-1-intro}
\begin{split}
 \cF^D (t, \e)  
 =&  \eu^{-\e_1 t} \Ga(\e_2-\e_1) \Ga(\e_1+\e_3) \Ga(\e_1+\e_4)
 \, {}_2F_1\left(\e_1+\e_3,\e_1+\e_4;\e_1-\e_2+1;\eu^{-t}\right) \\
 & + \eu^{-\e_2 t} \Ga(\e_1-\e_2) \Ga(\e_2+\e_3) \Ga(\e_2+\e_4)
 \, {}_2F_1\left(\e_2+\e_3,\e_2+\e_4;\e_2-\e_1+1;\eu^{-t}\right)\,.
\end{split}
\ee
This expression is suitable near the large volume point $t\to+\infty$, as ${}_2F_1$'s can be expanded as a power series in $\eu^{-t}$.
However, as we show in Section~\ref{s:conifold}, using analytic continuation it is possible to write an equivalent expression suitable for expansion near the conifold point $t\to 0$
\begin{multline}
\label{eq:conifold-FD-2-intro}
 \cF^D (t, \e) =
 \frac{\Ga(\e_1+\e_3)\Ga(\e_2+\e_3)\Ga(\e_1+\e_4)\Ga(\e_2+\e_4)}
 {\Ga(\e_1+\e_2+\e_3+\e_4)} \times\\
 \times \eu^{\e_3 t} \, {}_2F_1(\e_1+\e_3,\e_2+\e_3;\e_1+\e_2+\e_3+\e_4;1-\eu^t)~.
\end{multline}
The quantum volume $\cF^D$ and its splitting \eqref{eq:FD-split-intro} have remarkable analytic properties in $\bt$.
On the one hand, both $ \cF_{\semicl}^D(\bt,\e)$ and  $\cF_\inst^D(\bt,\e)$ feature jumps across different phases of $\bt$.
On the other hand, analytic continuation between \eqref{eq:conifold-FD-1-intro}-\eqref{eq:conifold-FD-2-intro} makes it manifest that the full expression for $\cF^D(\bt,\e) $ extends smoothly across all phases.
This involves highly nontrivial cancellations in the jumps, whose physical origin can be traced to worldsheet instantons of vanishing size.
It is worth noting that \emph{both} equivariance and worldsheet instanton effects play essential roles in the unification of quantum volumes into globally defined objects. 
While the former is a necessary regulator, classical equivariant volumes do not extend analytically across the moduli space. 
It is the resummation of instanton corrections to all orders that promotes classical equivariant volumes to global analytic functions.

To illustrate the other main results of the paper we switch to \eGLSM~with target manifolds defined by symplectic cuts of $X$.%
\footnote{This makes sense thanks to functoriality of the symplectic cutting and gluing operations.}
Let $X_0 = X\sslash U(1)$ be the quotient manifold along which the cut is performed, 
in the example of Figure~\ref{fig:intro-ex} this is a $U(1)^2$ fibration over the hyperplane.
The quantum volume, which is defined by an \eGLSM~with target space $X_0$, will be denoted $\cH^D(\bt,c,\e)$ and will play a central role throughout the paper.%
\footnote{Here $c$ is the $U(1)$ moment map parameter corresponding to the transverse position of the hyperplane.}
Exact all-instanton expressions for $\cH^D$ can be found in the main body of the paper for all examples that we consider.
There is a natural splitting
\be\label{eq:HD-splitting-intro}
	\cH^D(\bt,c,\e) = \cH_{\semicl}^D(\bt,c,\e) + \cH_\inst^D(\bt,c,\e) \,.
\ee
into the semiclassical volume $\cH_{\semicl}^D \sim\vol_\e(X_0)+\dots$, and instanton corrections $\cH_\inst^D$  of order $\eu^{-k \, c - \beta\cdot\bt}$. 
Similarly to the case of $\cF^D$, also the splitting \eqref{eq:HD-splitting-intro} of $\cH^D$ 
features remarkable analytic properties.
Both the classical equivariant volume $ \cH_{\semicl}^D(\bt,c,\e)$ and the instanton 
corrections $\cH_\inst^D(\bt,c,\e)$ feature jumps across different phases of $(\bt,c)$.
However the full expression for $\cH^D(\bt,c,\e) $ extends smoothly across all phases, thanks to delicate cancellations of the jumps from each piece.

The second main result of this paper is a direct relation between $\cH^D(\bt,c,\e)$ and closed topological strings on $X$. As remarked previously, Gromov--Witten invariants of $X$ are encoded by $\cF^D(\bt,\e)$ (through the mirror map, see footnote \ref{foot:FD-GW}).
The latter can be obtained directly from $\cH^D$ by integration along the real line
\be\label{eq:q-lebesgue-integral}
	\cF^D(\bt,\e) = \int_{-\infty}^{+\infty} \cH^D(\bt,c,\e)\, \dif c\,.
\ee
A natural interpretation of this formula comes from the following observation.
At large volume, $\cH^D$ reduces to the equivariant volume of $X_0$, 
and integration over $c$ corresponds to considering all possible transverse positions of the hyperplane (classically $X$ is foliated by copies of the hyperplane at different values of $c$).
This gives a computation of the equivariant (classical) volume of $X$ by Lebesgue integration.
Equation~\eqref{eq:q-lebesgue-integral} says that this relation continues to hold in the quantum theory with genus-zero worldsheet instantons taken into account. 
For this reason we refer to $\cH^D$ as the ``quantum Lebesgue measure''.\footnote{The alternative terminology ``quantum \emph{Cavalieri} measure'' may be equally, if not more, appropriate. Bonaventura Cavalieri (Milan 1598--Bologna 1647) introduced the idea of measure theory before Lebesgue, who later formalized it rigorously.
We thank Johannes Walcher for this erudite suggestion.}

The third main result we obtain is a direct connection between $\cH^D$ and the open string superpotential of a toric $A$-brane in $X$. 
As remarked earlier, there is a certain relation between symplectic cuts and toric Lagrangians, whose details are developed in the main body of the paper.
We define the equivariant superpotential $W(\bt,c,\e)$ of the toric $A$-brane through the relation
\be\label{eq:dcW-monodromy}
 \partial_c W(\bt,c,\e) = \frac{1}{2\pi\ii} \Big ( \cH^D(\bt,c+2\pi\ii,\e) - \cH^D(\bt,c,\e) \Big )\,,
\ee
where in the r.h.s.\ there appears the monodromy of the quantum Lebesgue measure under a shift of $c$ by $2\pi\ii$.
Note that in order to compute this it is crucial to have exact analytic expressions, which
is made possible by the \eGLSM~framework.
The open string potential for the toric brane is obtained by taking the regular terms in the $\e$ expansion of $W$ 
\be\label{eq:Weq-intro}
	W(\bt,c,\e) \propto W_{\sing}(\bt,c,\e) + W_\reg(\bt,c) +O(\e)\,.
\ee
In the main body of the paper we compute $W$ explicitly in several examples, and show that its regular part correctly reproduces earlier results derived from considerations of open-string mirror symmetry.
Once again there is a remarkable analytic structure underlying the splitting~\eqref{eq:Weq-intro}. 

We provide a physical interpretation of the monodromy relation \eqref{eq:dcW-monodromy} 
in terms of target space physics, by embedding topological strings into type IIA string theory on $X\times\BR^4$ with a D6 brane wrapping a domain wall stretched between toric Lagrangians at $c$ and $c+2\pi\ii$.
Finally, these results lead us to the following proposal for the computation of equivariant superpotentials of toric $A$-branes in toric Calabi--Yau threefolds
\be\label{eq:equiv-W-formula}
	W(\bt,c,\e) = \frac{1}{2\pi\ii} \int^{c+2\pi\ii}_{c} \cH^D (\bt,s,\e) \dif s  \,.
\ee
Our proposal is corroborated by physical reasoning and by exact computations in selected examples, concerning symplectic cuts of $\BC^3$, resolved conifold, and local $\BP^2$.
Note that this integral is very similar to the generating function of closed Gromov--Witten invariants \eqref{eq:q-lebesgue-integral}. Both feature the {same} integrand, but different contours.
It is indeed this specific choice of complex contour for the integral of the function $\cH^D$ which allows for the correspondence between symplectic cuts and disk counts on toric $A$-branes to work. In other words, we have a correspondence between:
\begin{center}
\begin{tblr}{ccc}
 {disk counts on framed toric Lagrangians \\ in a toric CY3 $X$}
 &$\leftrightarrow$&
 {monodromy of quantum volumes \\ of CY2 surfaces $X_0$}
\end{tblr}
\end{center}
Moreover, it appears that this choice of contour should correspond to a certain Seidel--Thomas twist for the 2-fold $X_0$ \cite{Brini:2013zsa}, however the precise details of the identification remain to be worked out.\footnote{We thank the anonymous referee for bringing this analogy to our attention.}

The analytic properties of the quantum Lebesgue measure $\cH^D$ play a central role in our work, ultimately establishing the link between open and closed string invariants. 
A remarkable feature of the analytic structure of $\cH^D$ is its universality: it is both invariant under a change of framing (of the toric brane), and under a change of topology of the brane across different chambers of the open string moduli space. 
We illustrate these points with exact computations of $\cH^D$ in examples below, where we further elaborate on these important structural properties.

The aim of this paper is to provide a self-contained and pedagogical illustration of the main ideas, with explicit computations for selected examples. To keep the paper readable we have postponed a more systematic treatment of several technical details, including the relation to mirror symmetry, to future works \cite{Cassia:2024txc}.

\subsubsection*{Relation to previous work}

Our work has several direct connections to previous literature on open and closed topological strings and to studies of Gromov--Witten theory in the context of symplectic cutting. Indeed many of the results that we obtain have been obtained in earlier work, but the connections that we establish among these fields, and our concrete approach to explicit results, are new.
Here we give a brief overview of how our work connects to previous literature.

The problem of computing genus-zero open Gromov--Witten invariants for toric $A$-branes was solved in the seminal papers \cite{Aganagic:2000gs, Aganagic:2001nx}. 
These works identify the moduli space of a toric $A$-brane with a circle fibration over the moduli space of the underlying toric Lagrangian. It is argued that worldsheet instantons regularize singularities in the moduli space and turn it into a smooth algebraic curve $\Sigma$.
The superpotential $W_\AKV$ was obtained from open periods of a certain 1-form on $\Sigma$, and shown to satisfy integrality properties in several examples. 

In this paper we consider a generalization of $W_{\AKV}$, the equivariant superpotential \eqref{eq:Weq-intro}.
As already stressed above, there is a stark contrast between the global behaviour of the equivariant superpotential and $W_{\AKV}$: while the former is encoded by $\cH^D$, which extends smoothly across the entire moduli space, the latter is only defined phase-by-phase and features jumps among different phases.
The relation between the two originates from the splitting \eqref{eq:Weq-intro} into singular and regular parts.
We show by explicit computation that the regular part of $W(\bt,c,\e)$ coincides exactly with $W_{\AKV}$ to all instanton orders, in several examples.
Agreement between $W_{\reg}$ and $W_{\AKV}$ is far from trivial however, since the definition \eqref{eq:equiv-W-formula} differs significantly from the Abel--Jacobi map proposed in \cite{Aganagic:2001nx, Aganagic:2000gs}.
A physical argument explaining why these compute the same quantity is sketched in Section~\ref{sec:physical-interpretation}.
While at the level of worldsheet instantons our results are simply in agreement with \cite{Aganagic:2001nx, Aganagic:2000gs},
the equivariant superpotential includes much more information. 
In particular, it is natural to expect that the leading singular terms  $W_{\sing}$ encode classical geometric data of the toric lagrangian.

A central role in our work is played by the quantum Lebesgue measure, which has not been studied previously.
A key property of $\cH^D$ is that it encodes both closed \eqref{eq:q-lebesgue-integral} and open \eqref{eq:equiv-W-formula} string potentials in a unified treatment.
Previous connections between closed and open strings have been obtained by means of a rather different nature.
In the context of toric Calabi--Yau threefolds both the open and the closed string potentials are encoded by the topological vertex formalism \cite{Aganagic:2003db} whose underpinnings lie on the relation between topological strings and Chern--Simons theory via large-$N$ geometric transitions \cite{Gopakumar:1998ki,Ooguri:1999bv}.
The topological vertex is a powerful framework that computes all-genus open string instantons in the presence of (stacked) toric $A$-branes in arbitrary positions.
In the absence of branes the vertex computes closed topological string instantons at all genera.
However, the relation between open and closed string invariants that we find appears to be of a starkly different nature.
In the context of the vertex, the closed and open string partition functions are related by a discrete sum over partitions, while in our case $W$ and $\cF^D$ are related by a change of contour in a continuous integral. 
It should be noted that $\cF^D$ does not coincide with the closed string free energy, but is related to it in a nontrivial way (see footnote \ref{foot:FD-GW}).
It seems reasonable to expect that a relation between our framework and topological vertex may be better understood in terms of the mirror $B$-model. The open-closed string relation in that context was elucidated in \cite{Bouchard:2007ys}, and we plan to return to this in \cite{Cassia:2024txc}.

Another approach to the relation between open and closed strings in genus zero was developed by \cite{Mayr:2001xk, Lerche:2001cw}. This approach relates the toric $A$-brane open string superpotential to the closed topological strings on a Calabi--Yau fourfold. While the geometric setup appears to be rather different from ours, there are certain analogies with our framework. In particular one may regard the fourfold as an extension of the CY3, where the additional direction corresponds to the position $c$ of the hyperplane associated to the toric Lagrangian. 
A difference with our work is that the domain of applicability of the fourfold setup appears to have certain limitations due to technical issues.

From the GLSM perspective our work advances a concrete proposal for the inclusion of $A$-branes in the partition function,  namely formula \eqref{eq:equiv-W-formula}.
The inclusion of $B$-branes is well understood following the thorough first-principles study of \cite{Hori:2013ika}. 
However much less is known about $A$-branes.
A proposal due to \cite{Govindarajan:2000ef, Govindarajan:2001zk} shows that the open string superpotential $W_{\AKV}$ can be obtained by formulating a GLSM with target space an auxiliary ``boundary'' toric variety. Open string superpotentials for toric branes in toric Calabi--Yau threefolds (both in genus zero and higher) were shown to arise from a localization computation on the moduli space of maps with Lagrangian boundaries, see  \cite{Katz:2001vm, Graber:2001dw} for seminal works and \cite{Brini:2010sw, Brini:2011ij, Brini:2013zsa, Brini:2014fea} for later developments.

While the relation between topological strings and symplectic cuts has not been considered in the physics literature, several studies of Gromov--Witten theory for symplectic cuts have appeared in the mathematics literature. 
In \cite{Li:1998hba, ionel1998gromov, 2000math.....10217I, ionel2003relative, 2014arXiv1404.1898T} a relation between Gromov--Witten invariants on a manifold and its cut has been proposed.
More recently \cite{2010arXiv1003.4325F} established a relation between open Gromov--Witten invariants for Lagrangian submanifolds, and the symplectic cut of its ambient space. While the class of geometries considered in these papers does not exactly match with the ones that we study, many of the ideas involved are closely related. 
Similar relations between local GW invariants and log GW invariants (counts of maps with maximal tangency to a certain divisor) have been discussed in recent works \cite{2017arXiv171205210V, Choi:2018eam, Bousseau:2019bwv} and a correspondence to open Gromov--Witten invariants for Lagrangian submanifolds was proposed in \cite{Bousseau:2020fus}. 
Finally, the open/closed relation involving CY4 of \cite{Mayr:2001xk,Lerche:2001cw} has been recently examined and reformulated from a rigorous mathematical viewpoint in \cite{Liu:2021eeb, Liu:2022swc, Yu:2023yey}.

\bigskip

There certainly appears to be much common ground with all these works, we believe that elaborating on these connections in greater detail would be very interesting.

\subsubsection*{Organization of the paper}

Section~\ref{sec:geometric-review} contains a review of basic constructions in symplectic geometry, including symplectic cuts and toric Lagrangians. In this section we also describe a relation between cuts in toric CY3 and toric Lagragians that is central to the rest of our work. The section also includes a brief review of equivariant GLSMs, which is the main computational tool in the paper.

Later sections contain a detailed exposition of the main ideas with explicit computations of quantum volumes and superpotentials in three different examples of increasing complexity: $\BC^3$ (Section~\ref{sec:C3}), resolved conifold (Section~\ref{s:conifold}) and local $\BP^2$ (Section~\ref{s:P2}).

Section~\ref{sec:physical-interpretation} provides a string theoretic interpretation of the computations performed through GLSM techniques in previous sections. Here we explain how our formula for the $A$-brane equivariant superpotential relates to the original computation of toric brane potentials via mirror symmetry.

Section~\ref{sec:summary} summarizes the salient technical aspects of the main results and suggestions for future work.

\subsubsection*{Acknowledgements}

We are grateful to Tobias Ekholm, Marcos Mari\~no, Nikita Nekrasov, Mauricio Romo and Johannes Walcher for illuminating discussions, and the anonymous referee for careful reading and helpful suggestions.
The work of L.\ C.\ is supported in part by STFC grant ST/T000708/1.
The work of P.\ L.\ is supported by the Knut and Alice Wallenberg Foundation grant KAW2020.0307.

\subsubsection*{Data Availability}

Data sharing not applicable to this article as no datasets were generated or analysed during the current study.

\section{Symplectic cuts and Lagrangians}\label{sec:geometric-review}

Two of the most basic constructions in symplectic geometry are \emph{cutting} and \emph{gluing}.
Given the ubiquity of symplectic manifolds in physics, and especially in string theory,  
it is somewhat surprising that neither of these operations has found many physical applications.
In this section we review the construction of symplectic cutting originally introduced in \cite{lerman1995symplectic}.\footnote{For the inverse construction, known as symplectic gluing, see \cite{gompf1995new}.}
For completeness we include basic background on symplectic geometry, referring to \cite{da2008lectures} for a more extensive exposition.

\subsection{Geometric background}

Let $(M,\omega)$ be a symplectic manifold endowed with a $G$ action $\psi:G\to{\rm Diff}(M)$.
The action is said to be a symplectomorphism if it preserves the symplectic structure $\psi(g)^\ast\omega = \omega$ for all $g\in G$.
Furthermore, the action is said to be Hamiltonian if there exists a moment map $\mu:M\to\mathfrak{g}^\ast$ such that 
given any generator $\xi\in\mathfrak{g}$, the function $\langle\mu,\xi\rangle$ is a Hamiltonian for $\xi$.
This means that if $v(\xi)\in\Ga(TM)$ is a vector field associated to the action of $\xi$,
then it is related to the gradient of the Hamiltonian by contraction with $\omega$,
\be
 \dif \langle\mu,\xi\rangle = \iota_{v(\xi)}\,\omega\,.
\ee
We will be mainly concerned with actions by abelian groups $G\cong U(1)^r$ where the
moment map $\mu:M\to\BR^{r}$ is real-valued.

A value $\bt\in\BR^r$ for the moment map is said to be regular if $U(1)^r$ acts
freely on the preimage $\mu^{-1}(\bt)\subset M$.
The symplectic (Marsden--Weinstein) quotient $M\sslash U(1)^r$ is defined as the
symplectic manifold obtained by taking the ordinary quotient by the $U(1)^r$-action
on the level set $\mu^{-1}(\bt)$, for a regular value $\bt$ \cite{marsden1974reduction}.

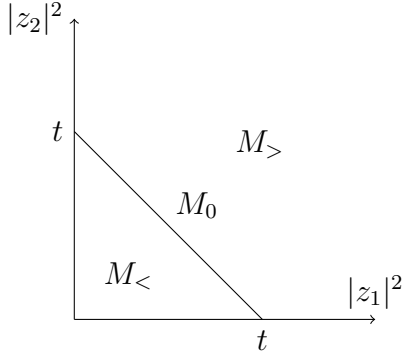
\begin{figure}[!ht]
\begin{center}
\begin{tikzpicture}
\draw[->] (0,0) -- (4,0) node[anchor=south]{$|z_1|^2$};
\draw[->] (0,0) -- (0,4) node[anchor=east]{$|z_2|^2$};
\draw[-] (2.5,0) node[anchor=north]{$t$} -- (0,2.5) node[anchor=east]{$t$};
\draw (1.2,1.2) node[anchor=south west]{$M_0$};
\draw (2,2) node[anchor=south west]{$M_>$};
\draw (0.25,0.25) node[anchor=south west]{$M_<$};
\end{tikzpicture}
\caption{The quotient $\BC^2\sslash U(1)$ with moment map $\mu(z)=|z_1|^2+|z_2|^2$. $M_0\cong S^3$ is the locus where $\mu(z)=t>0$, which in this case, is described by a $U(1)^2$ fibration over an interval such that on each of the endpoints one of the circles shrinks to a point.}
\label{fig:quotient-example}
\end{center}
\end{figure}

\begin{example}[Symplectic quotient]
\label{ex:quotient}
Let $M=\BC^2$ with the standard symplectic structure 
$$
 \omega = -\ii \, (\dif z_1\wedge\dif \bar z_1+\dif z_2\wedge\dif \bar z_2)
 = \dif \theta_1\wedge\dif |z_1|^2+\dif \theta_2\wedge\dif |z_2|^2\,. 
$$
The $U(1)$ action $\psi(\alpha):(z_1,z_2)\mapsto(\eu^{\ii\alpha}z_1,\eu^{\ii\alpha}z_2)$
induces the vector field $v=\partial_{\theta_1}+\partial_{\theta_2}$.
The corresponding moment map is
$$
 \mu(z) = |z_1|^2+|z_2|^2\,,
$$
which satisfies $\dif\mu=\iota_{v}\omega$.
The level set $\mu^{-1}(t)$ for $t>0$ is an $S^3$ of radius $\sqrt{t}$.
The $U(1)$ action corresponds to rotations of the Hopf fiber,
therefore the quotient gives the Hopf base $S^2\simeq\BP^1$
$$
 \BC^2\sslash U(1) = \mu^{-1}(t) / U(1) = \BP^1\,.
$$
Here $(\BP^1,\omega_{\BP^1})$ is a symplectic manifold whose volume is $t$,
as inherited by the ambient $\omega$ according to Marsden--Weinstein's theorem.
This construction is sketched in Figure~\ref{fig:quotient-example}.

\end{example}

Before proceeding, it is worthwhile to stress a few general properties of $U(1)$
symplectic quotients as illustrated by the previous example.
The first property is that symplectic quotients produce manifolds of real
dimension two lower than the dimension of $M$.
The second observation is that regular level sets $\mu^{-1}(t)$ are
real-codimension one submanifolds of $M$, therefore they subdivide $M$ into the
disjoint union of three manifolds
\be\label{eq:M-mu-decomposition}
 M = M_< \cup M_{0} \cup M_>
\ee
where $M_0 = \mu^{-1}(t)$ has real-codimension $1$, while $M_{\lessgtr}$ are
preimages of $\{s\in\BR,\,s\lessgtr{t}\}$ and have the same dimension as $M$.
For instance, in Example~\ref{ex:quotient} $M_0\simeq S^3$,
$M_<\simeq B^4$ and $M_>\simeq S^3\times\BR$.

The third and last property concerns the case of toric manifolds,
\be
\label{eq:X-sslash-A}
 X=\BC^N\sslash U(1)^r
\ee
where $M=\BC^N$ and the torus action of $U(1)^r$ is induced by the
natural action of $U(1)^N$ on $\BC^N$, via a choice of an embedding
of $U(1)^r\subset U(1)^N$, for $r<N$.
Using that $\mathrm{Hom}(U(1)^{r},U(1)^{N})\cong\BZ^{r\times N}$,
such an embedding is equivalently specified by a choice of a matrix of integer
charges $Q_i^a$ and the associated moment map $\mu_Q:\BC^{N}\to\BR^r$ is given by
\be
\label{eq:X-def-s-quot}
 \mu_Q^a(z) = \sum_{i=1}^{N} Q^a_i |z_i|^2 \,,\qquad(a=1,\dots,r)\,.
\ee
with $z_i$ local complex coordinates on $\BC^{N}$.
The level sets of $\mu_Q$ are parametrized by K\"ahler moduli $\bt=(t^1,\dots,t^r)\in\BR^r$
and the geometry of $X$ depends on the choice of $\bt$ through the
moment map condition
\be
 \mu_Q(z) = \bt\,.
\ee
In general, the K\"ahler moduli space is divided into open connected subsets of
regular values of $\bt$, known as \emph{chambers} or \emph{phases}\footnote{In some of the chambers the value of $\bt$ might be such that the $U(1)^r$-action on the preimage is not free but it has finite stabilizers. When this is the case, the corresponding quotient is better described as an orbifold or a toric Deligne--Mumford stack.} and manifolds obtained from different values of $\bt$ are complex isomorphic iff they belong to the same chamber.\footnote{Otherwise, different chambers are related by birational equivalence, see \cite{1994alg.geom..7007M, Li:1998hba, COATES20181002, mclean2020birational} for discussions of how quantum cohomology behaves under transitions.}

In this case, the ambient space $\BC^N$ can be seen as a $U(1)^N$-fibration over a convex hull $\BR_{\geq0}^N$.
The level set $\mu_Q^{-1}(\bt)$ corresponds to the restriction of this fibration
to a real-codimension $r$ submanifold in the base cut-out by the moment map equation.
This real submanifold of $\BR_{\geq0}^N$ is called the \emph{moment polytope}\footnote{
The moment polytope is an actual polytope only when $X$ is compact, however,
by abuse of notation, we use the name moment polytope also in the non-compact case.} of $X$.
The quotient by the $U(1)^r$-action on $M_0$ then reduces the $U(1)^N$-fibration to
an $U(1)^{N-r}$-fibration over the moment polytope.

In Example~\ref{ex:quotient} the convex hull is the positive quadrant of $\BR^2$
parametrized by $(|z_1|^2,|z_2|^2)$, and the level set $M_0$ is a $U(1)^2$
fibration over a segment.
The map $U(1)\to U(1)^{2}$ is the diagonal embedding given by the matrix $Q=(1,1)$
and the corresponding $U(1)$-quotient identifies $(\theta_1,\theta_2)\sim(\theta_1+\alpha,\theta_2+\alpha)$
leaving the local coordinate $\theta_1-\theta_2$ as the angle variable on $\BP^1$.

\subsection{Symplectic cutting}

Let $(M,\omega)$ be a symplectic manifold with a $U(1)$ action and associated moment map $\mu$.
We consider the symplectic manifold $(Z,\omega_Z)$ where
\be
 Z = M\times\BC\,,\qquad \omega_Z = \omega -\ii\dif w\wedge\dif \bar w\,.
\ee
with $w$ the coordinate on the $\BC$ factor.
The $U(1)$ action on $M$ can be lifted to $Z$ by 
\be\label{eq:Psi-plus-action}
 \Psi_+(\alpha): \ (m,w)\mapsto (\psi(\alpha)\cdot m, \eu^{-\ii\alpha} w)\,.
\ee
with corresponding moment map
\be
 \nu_+(m,w) = \mu(m) - |w|^2\,.
\ee
The symplectic quotient is then
\be\label{eq:s-cut-plus}
 \overline{M}_+ = Z\sslash U(1)
 = \nu_+^{-1}(c) / U(1)
 = \{\mu(m) - |w|^2 = {c}\} / U(1)\,.
\ee
This is a manifold with the same dimension as $M$, but generically with different topology. 
To understand the structure of $\overline{M}_+$ consider the foliation of $M$ by level sets of $\mu$.
\begin{itemize}
\item
Given a fixed ${c}$, if $m\in M_<$ then $\mu(m)-{c}<0$ and there are no solutions to $\nu_+={c}$.
Therefore the region $M_<\subset M$ does not contribute to \eqref{eq:s-cut-plus}, i.e.\ $\nu_+^{-1}({c})\cap(M_<\times\BC)=\emptyset$.
\item
Points $m\in M_>$ satisfy $\mu(m)-{c}>0$. Correspondingly, we have $\nu_+^{-1}({c})\cap(M_>\times\BC)=M_>\times S^1$, where the circle direction corresponds to the phase of $w$.
The $U(1)$-action in \eqref{eq:Psi-plus-action} acts diagonally on $M_>\times S^1$ and the quotient is isomorphic to $M_>$ itself.
This description also makes it manifest that we can think of $M_>$ as a fibration over $\BR_{\geq0}$ (parametrized by $|w|^2>0$) with fibers $\mu^{-1}({c}+|w|^2)$.
\item
Finally, points $m\in M_0$ satisfy $\mu(m)-{c}=0$ and we have that $\nu_+^{-1}({c})\cap(M_0\times\BC)=M_0\times\{0\}$.
After the quotient by $U(1)$ this gives precisely $M\sslash U(1)$.
\end{itemize}

The space $\nu_+^{-1}(c)$ is therefore a circle fibration over $M_>\cup M_0$ with $S^1$ shrinking over $M_0$.
Topologically, the $U(1)$ quotient just takes away the circle over $M_>$, but acts non-trivially on the ``boundary'' component $M_0$.
As a result $\overline{M}_+$ can be described as a fibration over $\BR_{\geq0}$ with
fiber isomorphic to $\mu^{-1}(c+s)$ over a generic point $s>0$
and fiber $M_0/U(1)$ over the origin $s=0$.

A similar construction applies to a different $U(1)$ action on $Z$,
\be
 \Psi_-(\alpha): \ (m,w)\mapsto (\psi(\alpha)\cdot m, \eu^{\ii\alpha} w)\,.
\ee 
with corresponding moment map
\be
 \nu_-(m,w) = \mu(m) + |w|^2\,.
\ee
The symplectic quotient is then
\be
 \overline{M}_-
 = Z\sslash U(1)
 = \nu_-^{-1}(c) / U(1)
 = \{\mu(m) + |w|^2 = c\} / U(1)\,.
\ee
The manifold $\overline{M}_-$ has the same dimension as $M$ and it can be
described as a fibration over $\BR_{\leq0}$ with fiber isomorphic to $\mu^{-1}(c+s)$
over a generic point $s<0$ and fiber $M_0/U(1)$ over the origin $s=0$.

Observe that both manifolds $\overline{M}_\pm$ contain $M_0$ as a real-codimension two submanifold,%
\footnote{In the case when $M$ is a toric variety, the manifolds $\overline{M}_\pm$ are also toric and $M_0$ is a toric divisor of both.}
and in fact we are led to the following natural definition.
\begin{definition}[Symplectic cut]
The \textit{symplectic cut} of $M$ with respect to the moment map $\mu$
is the union
\be
 \overline{M}_-\cup_{M_0}\overline{M}_+\,,
\ee
which can be seen as a fibration of $\mu^{-1}(c+s)$ over $\BR$ with degenerate fiber $\mu^{-1}(c)/U(1)$ over the origin.
\end{definition}

The construction of $\overline M_\pm$ may appear somewhat exotic at first,
but it is in fact quite familiar in the context of toric geometry,
as the next example shows.

\begin{example}[Symplectic cut of $\BC^2$]\label{ex:s-cut}
If $M=\BC^2$ with the standard symplectic structure and moment map $\mu=|z_1|^2+|z_2|^2$, the manifolds $\overline M_-$
and $\overline M_+$ can be identified with $\BP^2$ and $O_{\BP^1}(-1)$, respectively.
Let $c>0$, then
\be
\begin{split}
 \overline M_- &= \{|z_1|^2 + |z_2|^2 + |w|^2 = {c}\} / U(1)_- = \BP^2 \\
 \overline M_+ &= \{|z_1|^2 + |z_2|^2 - |w|^2 = {c}\} / U(1)_+ = O_{\BP^1}(-1) \\
\end{split}
\ee
where subscripts of $U(1)_\pm$ denote the actions $(z_1,z_2,w)\mapsto (\eu^{\ii\alpha}z_1,\eu^{\ii\alpha}z_2,\eu^{\mp\ii\alpha}w)$.

\begin{figure}[!ht]
 \centering
 \begin{subfigure}[b]{0.3\textwidth}
  \centering
  \begin{tikzpicture}
   \draw[fill=lightgray,thick,-] (0,0) -- (2,0) -- (0,2) -- (0,0);
  \end{tikzpicture}
  \caption{}
  \label{fig:Mbar-toric}
 \end{subfigure}
 \hspace*{.1\textwidth}
 \begin{subfigure}[b]{0.3\textwidth}
  \centering
  \begin{tikzpicture}
   \fill[lightgray] (3,0) -- (2,0) -- (0,2) -- (0,3) -- (3,3);
   \draw[fill=lightgray,thick,-] (3,0) -- (2,0) -- (0,2) -- (0,3);
  \end{tikzpicture}
  \caption{}
  \label{fig:Mbar+toric}
 \end{subfigure}
 \caption{Moment polytopes of $\overline M_-$ and $\overline M_+$.}
 \label{fig:convex-hulls-example}
\end{figure}
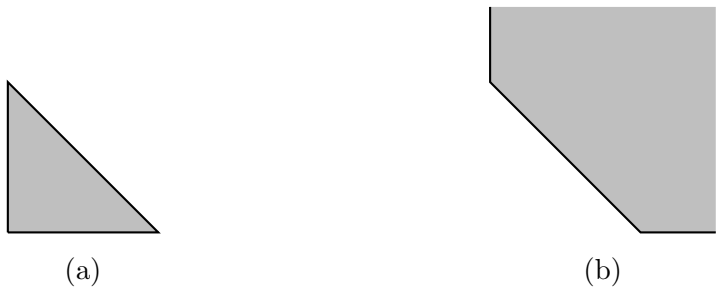

As toric manifolds, $\overline{M}_{\pm}$ are described by $U(1)^2$ fibrations
over convex hulls shown in Figure~\ref{fig:convex-hulls-example}. 
The \emph{gluing locus} is a $\BP^1$ which is codimension-two in both $\overline{M}_{\pm}$.
In the case of $\overline{M}_-$ this is the divisor corresponding to the line at infinity,
while in $\overline{M}_+$ it is the exceptional divisor corresponding to the zero-section of the tautological line bundle.
This construction explicitly shows that $\BP^2\cong\BC^2\cup\BP^1$ and that $O_{\BP^1}(-1)$ is a blow-up of $\BC^2$ at the origin. 
The parameter ${c}$ controls the volume of $\BP^1$.

It is helpful to consider the two pieces of the symplectic cut $\overline{M}_\pm$ together. 
We introduce the coordinate $s\in\BR$, and identify $s=|w|^2$ for $\overline{M}_+$ while $s=-|w|^2$ for $\overline{M}_-$.
The coordinate $|w|^2$ takes values in $\BR_{\geq 0}$ for $\overline{M}_+$
while it takes values in $[0,{c}]$ for $\overline{M}_-$.\footnote{Observe that $\mu^{-1}(c+s)=\emptyset$ if $s<-c$.}
Therefore the two manifolds $\overline M_\pm$ map to the half-line
\be
 s \in [-c,+\infty) \,,
\ee
where the part over $-c\leq s\leq 0$ describes $\overline M_-$, while $s\geq0$ describes $\overline M_+$. 
The union of these spaces is the symplectic cut of $\BC^2$, glued along the quotient locus $\BP^1$,
\be\label{eq:s-cut-decomposition}
 \overline{M}_{-}\cup_{\BP^1}\overline{M}_{+}\,. 
\ee
Their moment polytopes are shown in Figure~\ref{fig:C2-cut}.
It is suggestive to compare this with the decomposition $M_< \cup {M_0} \cup M_>$ of $\BC^2$ in Figure~\ref{fig:quotient-example}. Topologically one may be viewed as a $U(1)$ quotient of the other, where the $U(1)$ only acts on $M_0$.

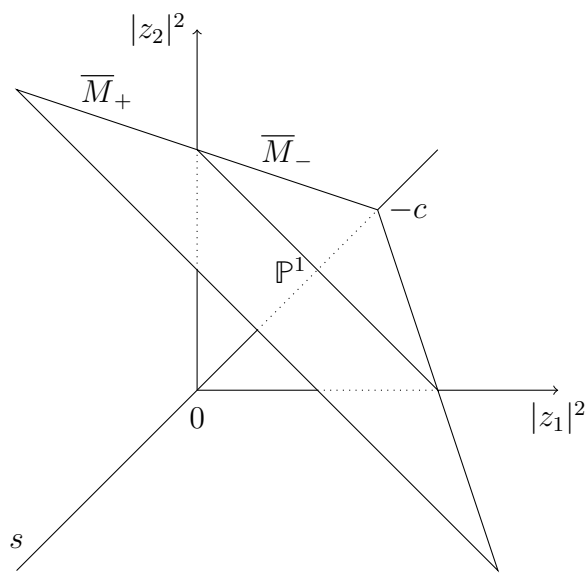
\begin{figure}[ht!]
\begin{center}
\usetikzlibrary{math}
\tikzmath{\xscale=0.8;\yscale=0.8;}
  \begin{tikzpicture}
   \draw (0*\xscale,-0.1*\yscale) node[anchor=north]{$0$};
   \draw (3*\xscale,3*\yscale) node[anchor=west]{$-c$};
   \draw[-] (0,2*\yscale) -- (0,0);
   \draw[dotted,-] (0,2*\yscale) -- (0,4*\yscale);
   \draw[->] (0,4*\yscale) -- (0,6*\yscale) node[anchor=east]{$|z_2|^2$};
   \draw[-] (1*\xscale,1*\yscale) -- (-3*\xscale,-3*\yscale);
   \draw (-3*\xscale,-2.8*\yscale) node[anchor=south]{$s$};
   \draw[dotted,-] (1*\xscale,1*\yscale) -- (3*\xscale,3*\yscale);
   \draw[-] (3*\xscale,3*\yscale) -- (4*\xscale,4*\yscale);
   \draw[-] (0,0) -- (2*\xscale,0);
   \draw[dotted,-] (2*\xscale,0) -- (4*\xscale,0);
   \draw[->] (4*\xscale,0) -- (6*\xscale,0) node[anchor=north]{$|z_1|^2$};
   \draw[-] (4*\xscale,0) -- (5*\xscale,-3*\yscale) -- (-3*\xscale,5*\yscale) -- (0,4*\yscale) node [midway,above] {$\overline{M}_+$} -- (4*\xscale,0) node [midway,left] {$\BP^1$} -- (3*\xscale,3*\yscale) -- (0,4*\yscale) node [midway,above] {$\overline{M}_-$};
  \end{tikzpicture}
\caption{Symplectic cut of $\BC^2$ along $\BP^1$.}
\label{fig:C2-cut}
\end{center}
\end{figure}
\end{example}

\subsection{Cuts of toric Calabi--Yau threefolds}

A toric Calabi--Yau threefold $X$ admits a description as a symplectic quotient 
\be
 X = \BC^{r+3} \sslash U(1)^r
\ee
such that the matrix of charges describing the $U(1)^r$-action satisfies
\be\label{eq:CY-cond}
 \sum_{i=1}^{r+3} Q^a_i = 0\,,\qquad(a=1,\dots,r)\,,
\ee
If this condition is satisfied, the canonical bundle of $X$ is trivial and therefore the quotient is a Calabi--Yau manifold.
However, it is well known that the metric induced by the ambient space $\BC^{r+3}$ is not Ricci-flat. 
When considering string theory on $X$, the gauged linear sigma model (GLSM)
on $\BC^{r+3}$ with standard K\"ahler form is expected to flow in the
infrared to a CFT with a metric that is Ricci-flat at large radius \cite{Witten:1993yc}.
Since the RG flow only affects D-terms and not the superpotential,
properties of topological strings and mirror symmetry can be studied using the GLSM.

We will now describe symplectic cuts of toric threefolds. 
The main takeaway of the upcoming discussion is that the cut is defined by the
choice of an affine hyperplane with rational slope inside the ambient space $\BC^{r+3}$.

In order to describe a symplectic cut of $X$ let us start from the ambient space $\BC^{r+3}$.
This space admits a natural Hamiltonian action by the torus $U(1)^{r+3}$ given by
\be
 (\alpha_1,\dots,\alpha_{r+3})\cdot (z_1,\dots, z_{r+3})
 = (\eu^{\ii\alpha_1}z_1,\dots,\eu^{\ii\alpha_{r+3}}z_{r+3})
\ee
This toric action gives a presentation of $\BC^{r+3}$ as a $U(1)^{r+3}$ fibration
over the moment polytope $B=\BR_{\geq 0}^{r+3}$.
Fix a choice of subgroup $U_\charge(1)\subset U(1)^{r+3}$, where the label $\charge=(\charge_1,\dots,\charge_{r+3})$
can be identified with a vector of integer charges via the isomorphism $\mathrm{Hom}(U(1),U(1)^{r+3})\cong\BZ^{r+3}$.
The moment map for the induced $U_\charge(1)$ action is given by
\be
 \mu_\charge(z) = \sum_{i=1}^{r+3} \charge_i\, |z_i|^2 \in\mathrm{Lie}\left(U_\charge(1)\right)\cong\BR\,.
\ee
Let $c\in\BR$ be a regular value and consider the level set
\be\label{eq:toric-hyperplane-ambient}
 M_0(c) := \{z\in\BC^{r+3}\,|\, \mu_\charge(z) = c\}\,.
\ee
For generic $(\charge,c)$, $M_0(c)$ is a fibration of $U(1)^{r+3}$ over the intersection of
the moment polytope $B$ with an affine hyperplane $h\subset\BR^{r+3}$
\be
\begin{array}{ccc}
 U(1)^{r+3} &\to& M_0(c) \\
 && \downarrow \\
 && h\cap B
\end{array}
\ee
where $h$ is given by the equation $\charge_1 |z_1|^2+\dots+\charge_{r+3}|z_{r+3}|^2=c$.
We similarly define the two half-spaces on either side of $M_0(c)$ as
\be
 M_{\lessgtr}(c) := \{z\in\BC^{r+3}\,|\, \mu_\charge(z) \lessgtr c\}\,.
\ee
In the following the dependence of $M_0$ on $c$ will be implicit.

Since all toric group actions commute with each other, it follows that $M_0, M_{\lessgtr}$ descend to a hyperplane and half-spaces in $X$, respectively
\be
 M_0, M_{\lessgtr}\subset \BC^{r+3} 
 \quad\rightarrow\quad
 X_0, X_{\lessgtr} \subset X\,,
\ee
after the quotient by $U(1)^r$. Moreover, the $U_\charge(1)$-action descends to a Hamiltonian action on $X$
(assuming $\charge$ is generic) \cite{delzant1988hamiltoniens, lerman1994symplectic, lerman1995symplectic},
therefore $\charge$ defines a symplectic cut of $X$
\be\label{eq:CY3-cut-decomposition}
	X\quad\rightarrow\quad
	\overline X_- \cup_{X_0} \overline X_+
\ee
where the cut is along the symplectic quotient of $X$ by the $U_\charge(1)$-action
\be\label{eq:X0}
 X_0 = X \sslash U_\charge(1) = \mu_Q^{-1}(\bt)\cap\mu_\charge^{-1}(c)/U(1)^r\times U_\charge(1)\,.
\ee

\subsection{Toric Lagrangian \texorpdfstring{$A$}{A}-branes}\label{sec:Abrane-cuts}

Special Lagrangian submanifolds of toric Calabi--Yau threefolds admit a universal construction as torus fibrations over linear subspaces with rational slope \cite{harvey1982calibrated}.
We will be mainly interested in the case of ``toric Lagrangians'', whose properties under Mirror Symmetry are discussed in \cite{Aganagic:2000gs, Aganagic:2001nx}. After reviewing basic properties of toric Lagrangians, we will show how they are naturally related to symplectic cuts.

A toric Lagrangian $L\subset X$ can be described as a $U(1)^2$ fibration over an affine line $\ell$ in the moment polytope of $X$.
We now describe the construction of each piece in turn by starting with Lagrangian submanifolds in the ambient space $\BC^{r+3}$.
 
The line $\ell$ is defined by considering two affine hyperplanes $h^1, h^2$ in the moment polytope $B$ of $\BC^{r+3}$
\be\label{eq:toric-hyperpl}
 h^b : \quad \sum_{i=1}^{r+3} \charge^b_i |z_i|^2 = c^b \,,\qquad (b=1,2)
\ee
and taking their intersection with the pre-image of the moment map in \eqref{eq:X-def-s-quot}
\be\label{eq:muXinv}
 \mu_Q^{-1}({\bt}):\quad \sum_{i=1}^{r+3} Q^{a}_i |z_i|^2 = t^{a} \,,\qquad ({a}=1,\dots, r)\,.
\ee
Provided $\charge^b_i$ are chosen suitably, together \eqref{eq:toric-hyperpl} and \eqref{eq:muXinv} specify $r+2$ linearly independent conditions in $\BR_{\geq 0}^{r+3}$, thus defining an affine half-line. 
If we denote by $\bv=(v^1,v^2,\dots,v^{r+3})$ its slope, an explicit parametrization for this line is
\be\label{eq:half-line-ambient}
 \BR_{\geq0} \bv + \bv_0,
\ee
where $\bv_0$ is a point on the boundary of $\BR_{\geq 0}^{r+3}$
(at least one of its entries vanishes) and $\sum_i \charge^b_i v^i =\sum_i Q^a_i v^i =0$ for all $a$ and $b$.
This descends to an affine half-line $\ell$ in the moment polytope of $X$.

The $U(1)^2$ fibers over $\ell$ are constructed as follows.
In $\BC^{r+3}$ one considers the sub-torus $T_{\bv} \subset U(1)^{r+3}$
of rank $r+2$ defined by the condition
\be\label{eq:T2+r}
 T_{\bv} : \quad \sum_i v^i \,\theta_i = 0\,,
\ee
where $v^i\in\BZ$ are the entries of the slope vector $\bv$. 
In other words, if we identify $\bv$ with a choice of $U_{\bv}(1)$ subgroup of $U(1)^{r+3}$,
then we can define $T_{\bv}$ as the cokernel of the inclusion and we have the short exact sequence of abelian groups
\be
 0\to U_{\bv}(1) \to U(1)^{r+3} \to T_{\bv} \to 0\,.
\ee
The manifold $L$ obtained by fibering $T_{\bv}$ over the half-line \eqref{eq:half-line-ambient}
is Lagrangian, i.e. the pullback of $\omega$ vanishes. 
In order for this to be \emph{special} Lagrangian, the pullback of the holomorphic
3-form on $X$ to $L$ must be a real multiple of the volume form. 
This requires that
\be\label{eq:special-condition}
 \sum_i \charge_i^b = 0\,, \qquad (b=1,2)\,.
\ee
Condition \eqref{eq:special-condition} restricts the possible orientations of
$\bv$ within the moment polytope. In fact since the Calabi--Yau condition
also requires that $\sum_i Q^a_i=0$ for $a=1,\dots,r$, it follows that the
slope vector must be $\bv = (1,\dots,1)$. 

We have now constructed a special Lagrangian submanifold of $\BC^{r+3}$.
This descends to a special Lagrangian $L$ in the toric Calabi--Yau $X$.
The half-line \eqref{eq:half-line-ambient} descends to a half-line $\ell$ in
the moment polytope of $X$, while the torus $T_{\bv}$ descends to a $U(1)^2$ fibered
over $\ell$ after the symplectic quotient by $U(1)^r$.
With these conditions satisfied, $L$ is special Lagrangian and can be used to
model $A$-branes at weak string coupling.

The topology of $L$ depends on the choice of $c^b$. It is easy to see that for generic $c^b$,
$L$ will have a boundary at finite distance, i.e.\ at the locus where
$L$ intersects one the walls of the moment polytope.
It is then necessary to restrict to certain choices of these parameters to ensure that boundaries are not present.
For example, if $X=\BC^3$ then $r=0$ and $\ell$ can be chosen to be $\{|z_1|^2-|z_3|^2=c^1\} \cap \{|z_2|^2-|z_3|^2=c^2\}$.
The vector $\bv = (1,1,1)$ defines a 2-torus $\theta_1+\theta_2+\theta_3=0$,
locally parameterized by $(\theta_1,\theta_2,-\theta_1-\theta_2)$.
If both $c^1,c^2$ are positive then $\ell$ is an affine line that intersects
the locus $|z_3|^2=0$ at $|z_b|^2=c^b$ for $b=1,2$. At this point the $U(1)^3$-fiber
of $\BC^3$ degenerates to $U(1)^2$ parametrized by $(\theta_1,\theta_2)$.
Therefore $L$ has topology $U(1)^2\times\BR_{\geq 0}$.
If instead we choose, say $c^1>0$ with $c^2=0$ then the half-line $\ell$ ends
on the intersection $\{|z_2|^2=0\}\cap\{|z_3|^2=0\}$ where $\theta_2$ and $\theta_3$ circles shrink.
This implies that $U(1)^2$ reduces to $U(1)$ there, and $L$ has now topology $\BR^2\times S^1$.
Tuning $c^2=0$ shrinks away the finite-distance boundary of $L$. 

Generally speaking, the requirement that $L$ has topology $\BR^2\times S^1$
implies that the affine line $\ell$ ends on the intersection of (projections of)
two toric divisors in the moment polytope of~$X$.
In turn, this implies that one of the two hyperplanes $h^b$ \emph{can} be chosen
to be homogeneous, i.e. one can always choose $c^1=c$ and $c^2=0$ after a
suitable linear combination of the charge vectors $\charge^b_i$.
The hyperplane equations \eqref{eq:toric-hyperpl} then become
\be\label{eq:toric-brane-hyperplanes}
 h^1:\ \sum_{i=1}^{r+3} \charge^1_i |z_i|^2 = c\,,
 \qquad 
 h^2:\ \sum_{i=1}^{r+3} \charge^2_i |z_i|^2 = 0\,.
\ee
Even after fixing the choice of vanishing $c^2$, there is a residual \emph{framing ambiguity} in the description of toric Lagrangians \cite{Aganagic:2001nx}. Indeed the definition of the first hyperplane can always be shifted by the second charge vector
\be
 \charge^1_i \sim \charge^1_i + f \cdot \charge^2_i \qquad (f\in\BZ)
\ee
and for any $f$ these define the same Lagrangian classically.
The framing ambiguity is however resolved by quantum effects.

The connection between \emph{framed} toric lagrangians and symplectic cuts of
toric Calabi--Yau threefolds is now easily stated.
Both are defined by a choice of an affine hyperplane 
\be\label{eq:Abrane-cut-planes}
 h\equiv h^1
\ee
in the moment polytope $B$ of the ambient space $\BC^{r+3}$. 
However, the correspondence is not one-to-one for at least two reasons:
\begin{itemize}
\item
While symplectic cuts can be defined for generic choices of $h$, the affine
hyperplane $h^1$ associated to $L$ must obey the condition \eqref{eq:special-condition}. 
\item
Toric Lagrangians also involve in their definition a second hyperplane $h^2$, which is homogeneous.
However, this will not play a role in the connection to symplectic cuts for
reasons that will become clear later and further elaborated in \cite{Cassia:2024txc}.
The main point is that a symplectic cut can describe toric Lagrangians in more
than one phase, e.g. for $c^1>0, c^2=0$ and for $c^1=0, c^2<0$. 
\end{itemize}

\subsection{Equivariance and abelian GLSM}\label{sec:equiv-GLSM}

So far we have reviewed basic constructions of classical symplectic geometry.
We close this section by introducing two further constructions that will be
central in our story, one of them is equivariance and the other is the
gauged linear sigma model.
The presentation and conventions follow closely \cite{Cassia:2022lfj}.

Recall from Section~\ref{ex:quotient} that toric Calabi--Yau threefolds admit
a description as abelian symplectic quotients $X=\BC^N\sslash U(1)^r$ with $N=r+3$,
where the $U(1)^r$ action is specified by a charge matrix $Q_i^a$ encoding the
moment map in \eqref{eq:X-def-s-quot}.

To define the equivariant volume of $X$ we introduce $N$ equivariant parameters
$\e_i$ associated to the natural $U(1)^N$ action on $\BC^N$ by rotations of
the coordinates $z_i$. We define $p_i=|z_i|^2$ to be the so called \emph{momenta},
i.e.\ coordinates on the moment polytope of the ambient space.
The equivariant volume is then 
\be\label{volume-original}
 \cF(\bt, \e) = \frac{1}{(2\pi \ii)^N} \int_{\BC^N}
 \prod_{i=1}^N \dif z_i\wedge\dif\overline{z}_i~
 \eu^{-\sum_i \e_i|z_i|^2} \prod_{a=1}^r \delta\left(\mu^a_Q(z) - t^a\right)~,
\ee

If $X$ is compact it is possible to take $\e=0$ unambiguously, and in this
limit $\cF(\bt,\e)$ recovers the standard symplectic volume of $X$
(with symplectic structure induced by that of $\BC^N$).
However, the Calabi--Yau condition \eqref{eq:CY-cond} implies that $X$ is
always non-compact in our case.
This implies on the one hand that we always need to require that $\Re(\e_i)>0$
for the convergence of the integral.
On the other hand, non-compactness also means that the geometric interpretation
of the equivariant volume is more subtle \cite{Nekrasov:2021ked}.
Equivariance regularizes the infinite classical volume of $X$,
this phenomenon will play a crucial role in many of our constructions later.

The formula \eqref{volume-original} can equivalently be written as the contour integral
\be\label{eq:equivVolContour}
 \cF(\bt,\e) = \oint_{\JK} \prod_{a=1}^r \frac{\dif\phi_a}{2\pi\ii}
 \frac{\eu^{\sum_a \phi_a t^a}}{\prod_{i=1}^N (\e_i + \sum_a\phi_a Q_i^a)}\,.
\ee
To obtain this expression one trades the $\delta$-functions for their integral
representation and performs integration over coordinates $z_i$. 
In particular this means that the integration of $\phi_a$ is taken along $\ii\BR$.
To fully define the integral, one must then choose a closure for these contours,
and this is done according to the Jeffrey--Kirwan (JK) prescription \cite{Goldin:2003dis} related
to the choice of chamber for $\bt$.

The threefold $X$ has $N$ toric divisors defined by $z_i=0$ for $i\in\{1,\dots,N\}$.
To each divisor one can associate a differential operator defined as follows
\be
 \cD_i = \e_i + \sum_{a=1}^r Q_i^a \frac{\partial}{\partial t^a}~. 
\ee
such that $\cD_i\cF$ computes the equivariant volume of that divisor.
The equivariant volume $\cF(\bt,\e)$ is annihilated by specific products of
$\cD_i$'s whenever the corresponding divisors do not intersect.
These constraints on $\cF$ encode the equivariant cohomology relations of $X$.
In this sense, more generally $\cF(\bt,\e)$ can be regarded as a generating
function of equivariant integrals of all relevant characteristic classes of $X$.

The equivariant volume admits a natural quantum deformation, where classical
geometry is replaced by a 2d gauged linear sigma model from a disk $D$ into $X$
\cite{Witten:1993yc, Hori:2013ika, Honda:2013uca, Sugishita:2013jca}.
The symplectic quotient \eqref{eq:X-sslash-A} is replaced by a two-dimensional
$(2,2)$ theory of $N$ chiral fields coupled to $r$ gauge fields with charges $Q_i^a$.
Equivariance can be introduced by identifying $\e_i$ as mass parameters of the chiral fields. 
Localization allows to express the partition function as a contour integral over
the Coulomb branch, where the integrand includes contributions from 1-loop
determinants of the matter fields
\be\label{disk-general-contour}
 \cF^D(\bt,\e,\lambda) = \lambda^{-N} \oint_{\QJK}
 \prod_{a=1}^r \frac{\dif\phi_a}{2\pi\ii} 
 \,\eu^{\sum_a \phi_a t^a} \prod_{i=1}^N
 \Ga\Big( \frac{\e_i+\sum_a\phi_a Q_i^a}{\lambda} \Big)\,.
\ee

The GLSM partition function \eqref{disk-general-contour} can be regarded as a
direct generalization of the equivariant volume  contour integral \eqref{eq:equivVolContour},
where the rational functions are replaced by $\Ga$ functions. 
Indeed the two are related by the limit 
\be
 \cF(\bt,\e) = \lim_{\lambda^{-1}\to 0} \cF^D(\bt,\e,\lambda)\,.
\ee

To make the match precise, one should also specify how the ``quantum'' contour
QJK is taken according to the choice of chamber for $\bt$ \cite{Bonelli:2013mma}.
The relevant prescription for selected examples will be explained below, while
the more general definition of QJK contours can be found in \cite[Definition~3.1]{Cassia:2022lfj}.

We refer to $\cF^D(\bt,\e,\lambda)$ as the \emph{disk partition function},
or \emph{quantum volume} of $X$.
An important remark is that this should \emph{not} be confused with the
topological string free energy, but should instead be regarded as a
regularization of the period associated to a space-filling brane on~$X$.
The quantum volume is however related to the free energy, since by mirror symmetry
it encodes the genus zero closed Gromov--Witten invariants \cite{Cassia:2022lfj}.
Since it is a (non-compact) period, $\cF^D(\bt,\e,\lambda)$ satisfies equivariant
Picard--Fuchs (PF) equations, i.e.\ it is annihilated by the PF operator
\be\label{eq:equiv-PF}
 \prod_{\{i|\sum_a \gamma_a Q^a_i>0\}} \left(\lambda^{-1} \cD_i\right)_{\sum_a \gamma_a Q^a_i}
 - \eu^{-\lambda \sum_a \gamma_a t^a} 
 \prod_{\{i|\sum_a \gamma_a Q^a_i\leq 0\}} \left(\lambda^{-1} \cD_i\right)_{-\sum_a \gamma_a Q^a_i}
\ee
where $(z)_n=\prod_{k=0}^{n-1}(z+k)$ is the Pochhammer symbol\footnote{Observe that, for $n$ a positive integer, the Pochhammer symbol $(z)_n$ is polynomial in $z$, therefore it make sense to substitute $z$ with a differential operator such as $\lambda^{-1}\cD_i$.} (see \eqref{eq:Pochhammer}) and $\bgamma=(\gamma^1,\dots,\gamma^r)\in\BZ^r$ is an integer vector subject to
certain chamber-dependent positivity constraints.
Indeed, there is one such operator for each $\bgamma$, but only finitely many of
them are independent.
We refer to \cite{Cassia:2022lfj} for more details on the general structure of
equivariant PF operators, while their explicit form will be given explicitly
for the examples that we consider below.
These PDEs encode the equivariant quantum cohomology and reduce to the standard
Picard--Fuchs equations in the limit $\e_i\to 0$.
Equivariant PF equations encode all equivariant periods of $X$,
including the overall volume $\cF^D$, as solutions with prescribed semiclassical asymptotics.
In this sense, the set of equivariant PF equations can be taken as the defining
property of the quantum geometry described by the GLSM.
In particular, the GLSM partition function is the unique solution with asymptotics
\be\label{eq:GLSM-asymptotics}
 \int_X \eu^{\omega - H_\e} \, \hat\Ga(TX)
\ee
where $\omega$ is the symplectic form of $X$, $H_\e$ is the Hamiltonian
for a one-parameter subgroup of the torus $T$ acting on $\BC^N$,
and $\hat\Ga(TX)$ is the insertion of an equivariant Gamma class of the tangent \cite{Hori:2013ika, Nekrasov:2021ked, Cassia:2022lfj}.
Other interesting equivariant quantities can be studied by inserting different
equivariant classes under the integral \eqref{disk-general-contour}.

Moreover, it is known that the disk function $\cF^D$ itself can be regarded as a classical integral of the form \eqref{eq:GLSM-asymptotics} together with the insertion of an
appropriate equivariant cohomology class. This class is the equivariant Givental $I$-function of $X$ \cite{givental1996equivariant} which allows to take into account all contributions coming from the instanton corrections. In fact, it was shown in \cite[Proposition 7.2]{Cassia:2022lfj} that $\cF^D$ admits the alternative integral expression
\be\label{eq:cFD-quiv-loc}
 \cF^D = \int_X \eu^{\omega - H_\e} \, \hat\Ga(TX) \, I_X 
 = \int_{X^T} \frac{\iota^\ast(\eu^{\omega - H_\e} \,
 \hat\Ga(TX) \, I_X )}{e_T(N_{X^T/X})}
\ee
where the second equality follows from equivariant localization to the
$T$-fixed locus $X^T\xhookrightarrow{\iota}X$.
In particular, we observe that the localization formula \eqref{eq:cFD-quiv-loc} is reminiscent of the definition of the ``equivariant $H$-function'' in \cite[Sec.~4.4.2]{Iritani:2009ani}.

It is interesting to observe that
using the representation of the $\Ga$ function
\be
 \Ga (x) = \int^\infty_0 y^{x-1} \eu^{-y} \dif y~,~~~~\Re(x) > 0
\ee
the equivariant quantum volume \eqref{disk-general-contour} can be rewritten as follows
\be\label{disk-general-y}
 \cF^D(\bt, \e, \lambda) = \lambda^{-N} \int^{\ii\infty}_{-\ii \infty} \prod_{a=1}^r \frac{\dif\phi_a}{2\pi\ii}
 ~\eu^{\sum_a \phi_a t^a} \int_0^\infty \prod_{i=1}^N \dif y_i ~y_i^{\lambda^{-1}(\e_i + \sum_a \phi_a Q_i^a)-1} \eu^{-\sum_i y_i} 
\ee
where the contour for the $\phi$'s is deformed appropriately.
This manipulation parallels the Hori--Vafa derivation of mirror symmetry for
toric Calabi--Yau threefolds \cite{Hori:2000kt}, but it has the advantage of
being on a firm mathematical footing thanks to equivariance,
since a proper analytic definition of the integral requires $\Re(\e_i)>0$.
Another advantage of this representation over the original QJK contour
prescription \eqref{disk-general-contour}, is that in \eqref{disk-general-y}
it becomes manifest that the quantum volume does not depend on the choice of chamber.
The smooth transition across different phases of the geometry of $X$ is a familiar
feature of mirror symmetry, and can be understood physically as a consequence
of quantum effects that smooth out classical singularities \cite{Ooguri:1996me}.

In this paper we will not develop a general theory of this type of integrals,
leaving further details about the general formalism to the upcoming work \cite{Cassia:2024txc}.

\begin{remark}[Conventions]
In the remainder of this paper we will set the deformation parameter
$\lambda=1$ to lighten formulae, since restoring this parameter is straightforward.
\end{remark}

\begin{remark}[Note added] \label{rmk:note-added}
Because the system of PF equations solved by the periods associated to $X$ is of hypergeometric type, one can use the theory of Mellin--Barnes integrals to write their general solutions as analytic functions defined over the extended K\"ahler moduli space, as explained for instance in \cite{Horja:2000hyp}.
Power series solutions associated to specific choices of chambers/phases can then be recovered by deforming the contour of integration to a closed path encircling some of the poles of the Mellin--Barnes integrand in a way which explicitly depends on the value of the K\"ahler moduli $t^a$. These contour integral representations, turn out to coincide with the Coulomb branch integrals in \eqref{disk-general-contour} as we will show in the examples below.
Similar techniques have been employed to study the analytic behavior under toric crepant transformations of the $I$-function itself, which has been shown to be a global section of the equivariant PF system associated to $X$.
The study of the relation between analytic continuation and wall
crossings in Gromov–-Witten theory has received much attention until recently and
it culminated in the proof of the \emph{Crepant Transformation Conjecture} for toric orbifolds \cite{COATES20181002} (see also \cite{Coates:2008wal,Coates:2009com,Coates:2009wal,Iritani:2009ani,Brini:2013zsa,Brini:2014fea}).
We thank the anonymous referee for bringing our attention to these points.
\end{remark}

\section{\texorpdfstring{$\BC^3$}{C3}}\label{sec:C3}

Having reviewed the classical notion of symplectic cuts in the previous section,
we now turn to discussing the role of cuts in topological string theory.
We start from the simplest example, namely the flat $\BC^3$.
While this CY3 is trivial as a toric quotient, it still admits interesting symplectic cuts.
Another feature of this example is that the key relations between open and closed
string partition functions encoded by cuts take an especially simple form.

\subsection{Symplectic cut and equivariant volumes}

Consider $\BC^3$, viewed as a $U(1)^3$ fibration over $\BR^3_{\geq 0}$ with coordinates $(p_1, p_2, p_3)$ and $p_i=|z_i|^2$.
There is a natural $U(1)^3$ action that rotates fibers and which degenerates at the boundaries of $\BR^3_{\geq 0}$.
Introducing three equivariant parameters for the $U(1)^3$ action on $\BC^3$ denoted $(\e_1, \e_2, \e_3)$, 
we can compute the equivariant volume of $\BC^3$ by a straightforward application of \eqref{volume-original} where we assume $\e_i>0$
\be\label{eq:C3-eVol}
 \cF (\e) = \frac{1}{\e_1 \e_2 \e_3}~.
\ee

\begin{figure}[!ht]
\centering
\begin{subfigure}[b]{0.39\textwidth}
\centering
\includegraphics[width=\textwidth]{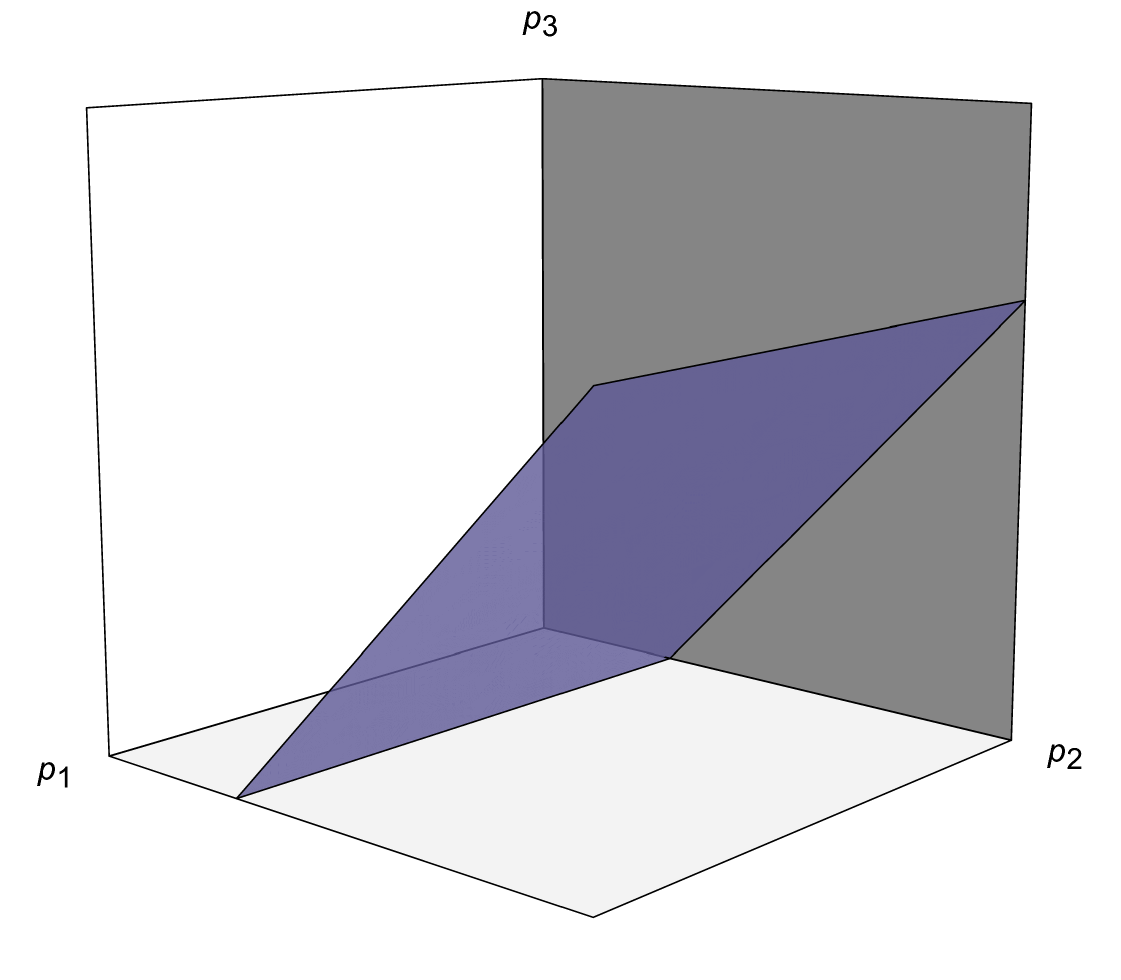}
\caption{Phase 1: $c>0$}
\end{subfigure}
\hfill
\begin{subfigure}[b]{0.39\textwidth}
\centering
\includegraphics[width=\textwidth]{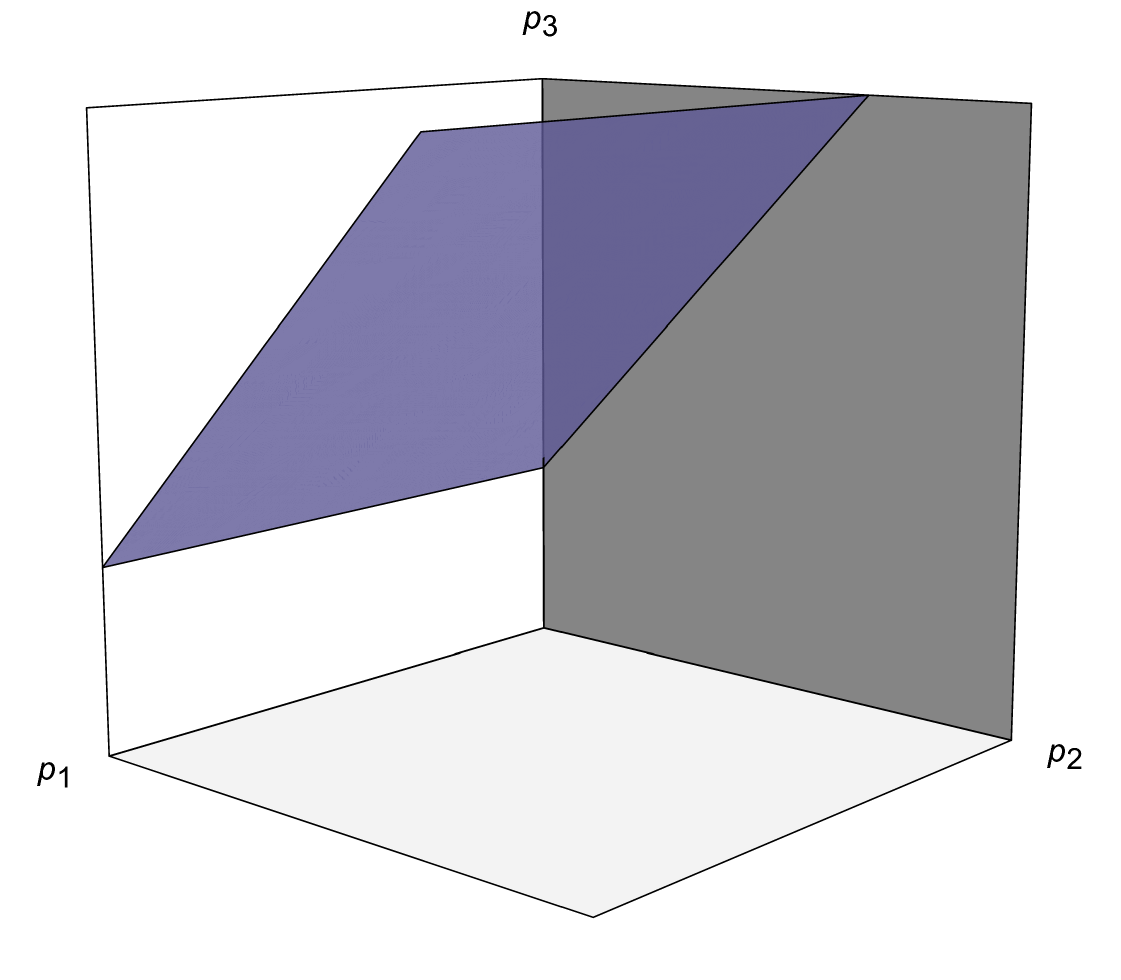}
\caption{Phase 2: $c<0$}
\end{subfigure}
\caption{Hyperplane for a symplectic cut of $\BC^3$ in two different phases.}
\label{fig:C3cutphases}
\end{figure}

Consider the symplectic cut of $\BC^3$ defined by the hyperplane 
\be\label{eq:C3-momentmap}
 p_2 - p_3=c~\,.
\ee
This is the analogue of \eqref{eq:toric-hyperplane-ambient}, and it divides $\BC^3$ into two disconnected pieces. 
The corresponding charge matrix is 
\be\label{C3-plane-Q}
 \charge = (0,1,-1)
\ee
The projection of this hyperplane to the base $\BR^3_{\geq 0}$ is shown in Figure~\ref{fig:C3cutphases}.

Taking the symplectic quotient $\BC^3\sslash U(1)$ defined by the moment map \eqref{eq:C3-momentmap}
defines a Calabi--Yau twofold with K\"ahler parameter $c$. 
We denote its equivariant volume by $\cH(c,\e)$, which can be computed by means of \eqref{eq:equivVolContour}
\be
 \cH (c,\e) = \frac{1}{\e_1} \oint_{\JK} \frac{\dif\phi}{2\pi\ii}
 \frac{\eu^{c\phi}}{(\e_2+\phi)(\e_3-\phi)}\,.
\ee
The overall factor $\e_1^{-1}$ originates from integration on the free direction with coordinate $z_1$. 
The choice of contour depends on the sign of $c$ through the Jeffrey--Kirwan prescription. 
If the value of $c$ is positive then we have to take the first pole and so the equivariant volume is
\be\label{eq:C3eVol1}
 \cH_{(1)} (c, \e) = \frac{\eu^{-\e_2 c}}{\e_1 (\e_3+\e_2)}~,~~~~(c>0)
\ee
and if $c$ is negative then the second pole is relevant  
\be\label{eq:C3eVol2}
 \cH_{(2)} (c,\e) = \frac{\eu^{\e_3 c}}{\e_1 (\e_3+\e_2)}~,~~~~(c<0)~. 
\ee
The equivariant volume of the twofold therefore depends on the sign of $c$, at least at the classical level.

The twofold defines a cut of $\BC^3$ along a locus whose transverse coordinate is $c$, the K\"ahler parameter. 
One may expect that integrating over all values of $c$ would return the equivariant volume of $\BC^3$. 
This is indeed the case
\be\label{eq:classical-volume-C3}
 \int_{-\infty}^{\infty} \cH (c,\e) \dif c =  
 \int_{-\infty}^0 \frac{\eu^{\e_3 c}}{\e_1 (\e_3+\e_2)} \dif c
 + \int_0^\infty \frac{\eu^{-\e_2 c}}{\e_1 (\e_3+\e_2)} \dif c
 = \frac{1}{\e_1 \e_2 \e_3} = \cF(\e)~. 
\ee
Notice that it is crucial to perform the computation equivariantly, as well as to use different expressions for the CY2 volume in different phases of $c$, since $\cH(\e,c)$ is continuous but not smooth at $c=0$.
In the spirit of Lebesgue integration, one may regard $\cH$ as a sort of measure for computing the volume of $\BC^3$ by slicing it along copies of the 2-fold.

\subsection{Quantum cut and quantum volumes}

We now examine how the simple geometric considerations of the previous section generalize at the quantum level.
To this end we consider a GLSM from a disk to $\BC^3$ with a space-filling brane wrapping the whole CY3.
The partition function, as defined by \eqref{disk-general-contour}, evaluates to
\be
 \cF^D (\e) = \Ga (\e_1) \Ga (\e_2) \Ga (\e_3)\,.
\ee
This may be regarded as the quantum generalization of the equivariant volume
\eqref{eq:C3-eVol}, which is recovered in the limit $\lambda\to \infty$ after restoring the dependence on $\lambda$ as in \eqref{disk-general-contour}.
The trivial nature of this function is related to the absence of Gromov--Witten invariants for $\BC^3$. 

In a similar spirit we may compute the quantum volume of the twofold associated with the symplectic cut \eqref{eq:C3-momentmap}.
We can define its volume by means of the GLSM construction discussed in Section~\ref{sec:equiv-GLSM}. Indeed since the twofold is described by a symplectic quotient of $\BC^3$ we can compute its quantum volume by wrapping a space-filling brane on it.
The equivariant periods of this system are annihilated by the equivariant Picard--Fuchs operators \eqref{eq:equiv-PF} associated to the charge matrix \eqref{C3-plane-Q}
\be
 \cD_2 - \eu^{-c} \cD_3\,,
\ee
with $\cD_2 = \e_2 + \partial_c$ and $\cD_3 = \e_3 - \partial_c$.
The unique solution with semiclassical asymptotics \eqref{eq:GLSM-asymptotics}
is the contour integral \eqref{disk-general-contour}
\be
\label{eq:C3HD}
 \cH^D = \Ga(\e_1) \oint_{\QJK} \frac{\dif\phi}{2\pi\ii}~\eu^{c\phi} \Ga(\e_2+\phi) \Ga(\e_3-\phi)\,,
\ee
where the overall factor $\Ga(\e_1)$ corresponds to the free $\BC$ direction with coordinate $z_1$.

As in the classical case, there is a choice of contour involved in the definition of $\cH^D$, which is determined by $c$. 
The GLSM for the twofold has two phases determined by the sign of $c$.
Here the ``quantum'' prescription (QJK) consists of taking the integral along the imaginary axis in $\phi$ and closing with a semicircle at infinity either to the left or to the right,
as in Figure~\ref{fig:image}.
\begin{figure}[ht]
 \centering
\begin{subfigure}{0.45\textwidth}
 \centering
	\begin{tikzpicture}[x=0.5cm,y=0.5cm]
    \def\pole{\small$\bullet$}
    \def\epsx{1}
    \def\epsy{1}
    \def\eepsx{-2.5}
    \def\eepsy{-2}
    \def\dlam{0.75}
    \def\smallradius{0.33}
	\begin{scope}
	\clip (0,0) circle (5);
	\end{scope}
	\draw [->](-6.5,0)--(6.5,0) node[above right] {$\text{Re}(\phi)$};
	\draw [->](0,-6.5)--(0,6.5) node[above] {$\text{Im}(\phi)$};
	\draw [->,very thick,black](0,-6)--(0,\eepsy-\smallradius*\dlam)
    --(\eepsx,\eepsy-\smallradius*\dlam) arc(-90:-270:\smallradius*\dlam)
    --(0,\eepsy+\smallradius*\dlam)--(0,\epsy-\smallradius*\dlam)
    --(\epsx,\epsy-\smallradius*\dlam) arc(-90:90:\smallradius*\dlam)
    --(0,\epsy+\smallradius*\dlam) --(0,6) arc(90:270:6);
    \filldraw[black] (\epsx,\epsy) circle (1pt) node[above] {$\hspace{4pt}-\e_2$};
    \filldraw[black] (\epsx-\dlam,\epsy) circle (1pt) node[right] {};
    \filldraw[black] (\epsx-2*\dlam,\epsy) circle (1pt) node[right] {};
    \filldraw[black] (\epsx-3*\dlam,\epsy) circle (1pt) node[right] {};
    \filldraw[black] (\epsx-4*\dlam,\epsy) circle (1pt) node[right] {};
    \filldraw[black] (\epsx-5*\dlam,\epsy) circle (1pt) node[right] {};
    \filldraw[black] (\epsx-6*\dlam,\epsy) circle (1pt) node[right] {};
	\node at (\epsx-7.25*\dlam,\epsy) {\dots};
    \filldraw[black] (\eepsx,\eepsy) circle (1pt) node[below] {$\e_3\hspace{16pt}$};
    \filldraw[black] (\eepsx+\dlam,\eepsy) circle (1pt) node[right] {};
    \filldraw[black] (\eepsx+2*\dlam,\eepsy) circle (1pt) node[right] {};
    \filldraw[black] (\eepsx+3*\dlam,\eepsy) circle (1pt) node[right] {};
    \filldraw[black] (\eepsx+4*\dlam,\eepsy) circle (1pt) node[right] {};
    \filldraw[black] (\eepsx+5*\dlam,\eepsy) circle (1pt) node[right] {};
    \filldraw[black] (\eepsx+6*\dlam,\eepsy) circle (1pt) node[right] {};
    \filldraw[black] (\eepsx+7*\dlam,\eepsy) circle (1pt) node[right] {};
    \filldraw[black] (\eepsx+8*\dlam,\eepsy) circle (1pt) node[right] {};
	\node at (\eepsx+9.5*\dlam,\eepsy) {$\cdots$};
	\end{tikzpicture}
\caption{Integration contour in the phase $c>0$.}
\label{fig:subim1}
\end{subfigure}
\hspace{0.05\textwidth}
\begin{subfigure}{0.45\textwidth}
	\begin{tikzpicture}[x=0.5cm,y=0.5cm]
    \def\pole{\small$\bullet$}
    \def\epsx{1}
    \def\epsy{1}
    \def\eepsx{-2.5}
    \def\eepsy{-2}
    \def\dlam{0.75}
    \def\smallradius{0.33}
	\begin{scope}
	\clip (0,0) circle (5);
	\end{scope}
	\draw [->](-6.5,0)--(6.5,0) node[above right] {$\text{Re}(\phi)$};
	\draw [->](0,-6.5)--(0,6.5) node[above] {$\text{Im}(\phi)$};
	\draw [->,very thick,black](0,-6)--(0,\eepsy-\smallradius*\dlam)
    --(\eepsx,\eepsy-\smallradius*\dlam) arc(-90:-270:\smallradius*\dlam)
    --(0,\eepsy+\smallradius*\dlam)--(0,\epsy-\smallradius*\dlam)
    --(\epsx,\epsy-\smallradius*\dlam) arc(-90:90:\smallradius*\dlam)
    --(0,\epsy+\smallradius*\dlam) --(0,6) arc(90:-90:6);
    \filldraw[black] (\epsx,\epsy) circle (1pt) node[above] {$\hspace{4pt}-\e_2$};
    \filldraw[black] (\epsx-\dlam,\epsy) circle (1pt) node[right] {};
    \filldraw[black] (\epsx-2*\dlam,\epsy) circle (1pt) node[right] {};
    \filldraw[black] (\epsx-3*\dlam,\epsy) circle (1pt) node[right] {};
    \filldraw[black] (\epsx-4*\dlam,\epsy) circle (1pt) node[right] {};
    \filldraw[black] (\epsx-5*\dlam,\epsy) circle (1pt) node[right] {};
    \filldraw[black] (\epsx-6*\dlam,\epsy) circle (1pt) node[right] {};
    \filldraw[black] (\epsx-7*\dlam,\epsy) circle (1pt) node[right] {};
	\node at (\epsx-8.25*\dlam,\epsy) {\dots};
    \filldraw[black] (\eepsx,\eepsy) circle (1pt) node[below] {$\e_3\hspace{16pt}$};
    \filldraw[black] (\eepsx+\dlam,\eepsy) circle (1pt) node[right] {};
    \filldraw[black] (\eepsx+2*\dlam,\eepsy) circle (1pt) node[right] {};
    \filldraw[black] (\eepsx+3*\dlam,\eepsy) circle (1pt) node[right] {};
    \filldraw[black] (\eepsx+4*\dlam,\eepsy) circle (1pt) node[right] {};
    \filldraw[black] (\eepsx+5*\dlam,\eepsy) circle (1pt) node[right] {};
    \filldraw[black] (\eepsx+6*\dlam,\eepsy) circle (1pt) node[right] {};
    \filldraw[black] (\eepsx+7*\dlam,\eepsy) circle (1pt) node[right] {};
	\node at (\eepsx+8.5*\dlam,\eepsy) {$\cdots$};
	\end{tikzpicture}
\caption{Integration contour in the phase $c<0$.}
\label{fig:subim2}
\end{subfigure}
\caption{Choices of integration contour in the $\phi$-plane. The poles of the integrand in \eqref{eq:C3HD} are located at $\phi=-\e_2-n$ for $n\in\mathbb{Z}_{\geq0}$ and $\phi=\e_3+n$ for $n\in\mathbb{Z}_{\geq0}$ where it is assumed that $\e_2$ and $\e_3$ are in general position such that no pole of the first set coincides with any pole of the second set.
The part of the contour along the imaginary axis is curved so that the poles of $\Ga(\e_2+\phi)$ lie on the left and the poles of $\Ga(\e_3-\phi)$ lie on the right. The contour is then closed with a semicircle of infinite radius either to the left or to the right, according to the phase.}
\label{fig:image}
\end{figure}

In phase 1, we have $c>0$ and the relevant poles are located at $\phi=-\e_2-n$ with $n=0,1,\dots$ 
giving a series in~$\eu^{-c}$
\be\label{eq:C3Hd1}
\begin{split}
	\cH^D_{(1)} (c,\e) 
	& = \eu^{-\e_2 c} \Ga(\e_1) \Ga(\e_3+\e_2) \sum_{n=0}^\infty (-\eu^{-c})^n \frac{(\e_2+\e_3)_n}{n!} \\
	& = \Ga(\e_1) \Ga(\e_2+\e_3) \frac{\eu^{-\e_2c}}{(1+\eu^{-c})^{\e_2+\e_3}}
\end{split}\qquad 
\begin{split}
	(c>0)
\end{split}
\ee
In phase 2, we have $c<0$ and the relevant poles are $\phi=\e_3+n$, $n=0,1,\dots$ 
giving a series in~$\eu^{c}$
\be\label{eq:C3Hd2}
\begin{split}
 	\cH^D_{(2)} (c, \e) 
	& = \eu^{\e_3 c} \Ga(\e_1) \Ga(\e_3+\e_2) \sum_{n=0}^\infty 
 (-\eu^{c})^n \frac{(\e_2+\e_3)_n}{n!} \\
 & = \Ga(\e_1) \Ga(\e_3+\e_2)
 \frac{\eu^{\e_3 c}}{(1+\eu^{c})^{\e_2+\e_3}}
\end{split}
\qquad
\begin{split}
	(c<0)
\end{split}
\ee

Remarkably, the expressions computed in the two different phases coincide exactly
\be\label{H-forC3}
 \cH^D(c,\e) = \cH^D_{(1)} (c,\e) =\cH^D_{(2)} (c,\e)\,.
\ee
This is in contrast with the two different results for the classical equivariant volume obtained in \eqref{eq:C3eVol1} and \eqref{eq:C3eVol2}, which are recovered in the classical limit ($\lambda\to\infty$).
In the quantum symplectic cut we can define a \emph{single} function $\cH^D$ valid for all phases of the GLSM, corresponding to the quantum volume of the associated twofold.

The unification of different phases is due to quantum effects, as disk instantons with vanishing size weighted by $\eu^{\pm c}$ become dominant near $c=0$ and smooth out the transition. 
This type of phenomenon is well known in string theory \cite{Ooguri:1996me}, and more specifically in studies of mirror symmetry.
Indeed the agreement of quantum volumes in different phases can be demystified by observing that the general definition \eqref{disk-general-contour} adopted earlier can be reformulated as in \eqref{disk-general-y} which gives in this case
\be
\label{eq:gamma-integral-C3-cut}
 \cH^D(c,\e) = \Ga(\e_1) \int_{-\ii\infty}^{\ii\infty} \frac{\dif\phi}{2\pi\ii} \eu^{\phi c} \int_0^\infty \dif y_1 \int_0^\infty \dif y_2
 ~ y_1^{\e_2+\phi-1} y_2^{\e_3-\phi-1} \eu^{-y_1-y_2}~.
\ee
This integral can be performed explicitly to give \eqref{H-forC3}, and it is manifest that the answer will be the same for both signs of $c$. 
The fact that this integral presentation manifestly explains the smooth transition from $c<0$ to $c>0$ is reminiscent of the mirror description of GLSMs, we will return to this point in Section~\ref{sec:physical-interpretation}.

Finally, we comment on the second key property of symplectic cuts that we encountered in the classical setting.
In \eqref{eq:classical-volume-C3} we have seen that integrating the equivariant volume of the twofold over $c$ gives the equivariant volume of $\BC^3$.
This statement generalizes to the full quantum theory as follows
\be
 \int^\infty_{-\infty} \cH^D (c,\e) \dif c = \Ga (\e_1) \Ga (\e_2) \Ga (\e_3) = \cF^D(\e)~,
\ee
as can be verified directly by using the integral representation of the beta function \eqref{int-beta-function}.

In the quantum theory $\cH^D$ again plays the role of a ``Lebesgue measure'', we henceforth refer to $\cH^D$ as the \emph{quantum Lebesgue measure}.
An interesting application of this measure is to compute the volume of the two components of the symplectic cut decomposition
\eqref{eq:CY3-cut-decomposition} which in this case reads
\be
 \BC^3 \quad\rightarrow\quad
 \overline X_- \cup_{X_0} \overline X_+\,,
\ee
with volumes defined by
\be\label{eq:C3-half-quantum-volumes}
 \cF^D_< (c,\e) = \int^c_{-\infty} \cH^D (s,\e) \dif s \,,\qquad
 \cF^D_> (c,\e) = \int_c^{+\infty} \cH^D (s,\e) \dif s \,,
\ee
obeying the sum rule
\be\label{eq:C3-volumes-sum}
 \cF^D (\e) = \cF^D_< (c,\e) + \cF^D_> (c,\e) \,.
\ee
The half-volumes can be evaluated explicitly with the aid of \eqref{HG-int-1}
\be\label{C3-F+hyper}
 \cF^D_< (c,\e)
 = \frac{\Ga(\e_1)\Ga(\e_2+\e_3)}{\e_3} \eu^{\e_3 c} \,  {}_2 F_1 (\e_2+\e_3,\e_3;\e_3+1;-\eu^c)~,
\ee
and noting the relation
\be
 \cF^D_> (c,\e_1, \e_2, \e_3) = \cF^D_< (-c, \e_1, \e_3, \e_2)~.
\ee

\subsection{Holomorphic disk potential from quantum Lebesgue measure}

We have shown by direct computation that the quantum Lebesgue measure encodes
the quantum volume of the Calabi--Yau threefold. We will now show that $\cH^D$
also captures information about the open string superpotential in the presence
of an $A$-brane supported on the toric Lagrangian associated to the cut.

First we recall how the genus zero part of the toric brane superpotential is
computed by open-string mirror symmetry, following \cite{Aganagic:2000gs, Aganagic:2001nx}.
According to the definition of toric Lagrangian reviewed in Section~\eqref{sec:Abrane-cuts},
we consider a brane defined by the affine line \eqref{eq:toric-brane-hyperplanes} 
\be\label{eq:C3toricbranedata}
 p_2 - p_3 = c \,,\qquad p_1 - p_2 = 0 \,.
\ee
Choosing the phase $c<0$ gives a Lagrangian $L$ that is a $U(1)^2$ fibration over an affine
half-line $\ell$ ending at $p_1=p_2=0$, $p_3=-c$. Since $b_1(L)=1$ its
moduli space is one-dimensional, and corresponds to displacements of the line
parameterized by $c$, along the $p_3$-axis.
Other phases of the Lagrangian moduli space are reached by going through the
origin of $\BC^3$ where $L$ becomes singular.

In the quantum theory this picture gets modified. An $A$-brane on $L$ is described
at weak string coupling by an abelian local system, whose holonomy complexifies $c$
into the open string modulus $x=\eu^{c}$ valued in $\BC^\ast$.
The moduli space of the $A$ brane in the absence of disk instantons would
therefore be the union of three cones, with a shared tip at $c=0$.
Disk instantons smooth out the singularity, and give a moduli space that is
topologically a sphere with three punctures.
This can be described as the algebraic curve
\be\label{eq:C3-curve}
 \Sigma: \ 1 + x + y = 0\quad\subset \BC^\ast\times\BC^\ast\,,
\ee
where a choice of framing for the toric brane has been implicitly made.

The geometry of $\Sigma$ encodes contributions of open string instantons that
smooth out the singularity of the bare classical moduli space.
The disk potential can therefore be recovered from $\Sigma$, by the following formula
\be\label{eq:C3superpotential}
 W_{\AKV}(x) = -\int^x\log (-y(x)) \, \dif\log x = \Li_2(-x)\,.
\ee

Next we explain how to recover \eqref{eq:C3superpotential} from the quantum Lebesgue measure.
As we explained in Section~\ref{sec:Abrane-cuts}, to a framed toric Lagrangian $L$
we can naturally associate a symplectic cut. The key step in the identification
is the choice of affine hyperplane in the description of $L$, and to adopt it
as the defining plane for the cut, see \eqref{eq:Abrane-cut-planes}. 
We claim that the genus zero part of the open-string superpotential is related
to the monodromy of the quantum Lebesgue measure as follows
\be\label{C3-defW}
 \partial_c W(c,\e) = \frac{1}{2\pi\ii} \Big ( \cH^D(c+2\pi\ii,\e) - \cH^D(c,\e) \Big )\,,
\ee
or by the equivalent integrated expression in terms of the quantum half-volume
\be\label{C3-defW-1}
	W(c,\e) = \frac{1}{2\pi\ii} \Big( \cF^D_<(c+2\pi\ii,\e) - \cF^D_<(c,\e) \Big)~.
\ee
More precisely $W(c,\e)$ is an equivariant generalization of the open string
superpotential. The relation between the two will be illustrated shortly with
explicit computations.

In these definitions we have promoted $c$ from a real-valued variable to a
complex-valued one, and introduced analytic continuation in the complex $c$-plane.
Indeed, an important feature of both $\cH^D$ and $\cF^D_>$ is that they are
multivalued as functions of $\eu^c$, and this ensures that the monodromy along
a path from $\eu^{c}$ to $\eu^{c+2\pi\ii}$ is actually non-trivial.
It is straightforward to see from \eqref{eq:C3Hd1}-\eqref{eq:C3Hd2} that $\cH^D$
has branching at three points located at
\be\label{eq:C3-branching}
 \eu^c \in\{ 0,-1,\infty\}\,,
\ee
and that these three monodromies are all different. 
Notice how the three branch points coincide with positions of the punctures
of $\Sigma$ in \eqref{eq:C3-curve}.

For illustration we consider the region $c<0$. In this region the appropriate
expansion is in terms of $\eu^c$ as given in \eqref{eq:C3Hd2}, and its monodromy is simply
\begin{multline}\label{eq:DeltaHD-C3}
	\cH^D(c+2\pi\ii,\e)-\cH^D(c,\e)
	 =
	(\eu^{2\pi\ii\e_3} - 1)
	\Ga(\e_1)\Ga(\e_2+\e_3) \frac{\eu^{\e_3 c}}{(1+\eu^c)^{\e_2+\e_3}}
	\\
	 =  
	2\pi\ii\, \frac{\e_3}{\e_1}
	\left[
	\frac{1 + \pi \ii \e_3}{\e_2+\e_3}
	-
	\gamma \frac{\e_1+\e_2+\e_3}{\e_2+\e_3}
	+
	\frac{\e_3}{\e_2+\e_3} c
	- \log(1+\eu^c)
	+O(\e_i)
	\right]
\end{multline}
According to \eqref{C3-defW}, to obtain $W(c,\e)$ we should take the primitive in $c$, which is easily done using \eqref{C3-F+hyper}
\be
\begin{split}
 	W(c, \e) 
 	&=
 	\frac{\eu^{2\pi\ii\e_3}-1}{2\pi\ii} \, \frac{\Ga(\e_1)\Ga(\e_2+\e_3)}{\e_3} \eu^{\e_3 c}
    \,  {}_2 F_1 (\e_2+\e_3,\e_3;\e_3+1;-\eu^c)
	\\
	&= \frac{\e_3}{\e_1} \left [
	\frac{1+ \pi \ii \e_3}{\e_2+\e_3}  c + \gamma\frac{\e_1+\e_2+\e_3}{\e_2+\e_3} c
	+
	\frac{\e_3}{\e_2+\e_3} \frac{c^2}{2} + \Li_2(-\eu^c)
     + O(\e) \right ]~.
\end{split}
\ee

We finally come to the relation between $W(c,\e)$ and the genus zero open string superpotential.
The latter can be extracted from the former by taking the \emph{instanton part}
of the \emph{regular} term, i.e.\ the terms of degree $O(\e^0)$ where we
discard polynomials in $c$ (we will refer to these as \emph{semiclassical} terms).
In this case we find
\be
 W^{\inst}_{\reg} (c) = \Li_2 (-\eu^c)~,
\ee
up to an overall normalization factor $\e_3/\e_1$. This coincides exactly with
the open string superpotential \eqref{eq:C3superpotential}.

\paragraph{Remarks:}
\begin{enumerate}
\item
We have illustrated how relations \eqref{C3-defW}-\eqref{C3-defW-1} encode $W_{\AKV}$ for a specific choice of phase  of the Lagrangian $A$-brane ($c<0$), and in a specific choice of framing. However the relation holds in greater generality. 
As observed in \eqref{H-forC3} the function $\cH^D$ extends to all phases, however its monodromy will be different at $c>0$ and $c<0$ due to the pole at $\eu^{c}=-1$, recall \eqref{eq:C3-branching}. 
Computing the monodromy in the new phase will give $W_\AKV$ for a toric brane on leg $p_2$ (see Figure~\ref{fig:C3cutphases}). 
There is also monodromy around $\eu^c=-1$, which is related to the superpotential for a brane along leg $p_1$.

\item
It is not difficult to show that $\cH^D$ is invariant under a change of framing.
More precisely under a change of framing for the brane \eqref{eq:C3toricbranedata}, one considers a new GLSM where \eqref{C3-plane-Q} is replaced by $(f,1-f,-1)$ for $f\in \BZ$. The K\"ahler modulus $c$ also changes, according to the identification $x=\eu^c$ and to the fact that under framing we have the change of variables $x\to xy^f$ \cite{Aganagic:2001nx}.
The net result is that $\cH^D$ is \emph{framing invariant}. Nevertheless, appropriate shifts \eqref{C3-defW} recover the framing-dependent superpotential $W_{\AKV}$ for any value of $f\in\BZ$. We refer to \cite{Cassia:2024txc} for more details.

\item
It is an important question how to fix the normalization factor from first principles. For the moment we leave this to future work, but in what follows we will see that fixing the overall normalization for the toric brane in $\BC^3$ fixes the overall normalization for other examples. 
Also observe that imposing the Calabi--Yau condition $\e_1+\e_2+\e_3=0$ reduces the regular semiclassical term to $-\pi\ii c-c^2/2$.
In general, extracting the regular semiclassical terms may be ambigous, and require additional geometric input \cite{Cassia:2022lfj}. 
As it turns out, geometries without compact toric divisors like $\BC^3$ are especially pathological when it comes to extracting instanton and semiclassical regular contributions.
We will see in Section~\ref{s:P2} that in geometries with a compact divisor some of these ambiguities are absent if we simply impose the Calabi--Yau condition on $\e$'s, and that the remaining ones can be handled by comparing with the normalization prescription adopted in the case of $\BC^3$.

\end{enumerate}

In conclusion, the genus zero $A$-brane superpotential can be computed in the GLSM by combining  \eqref{C3-defW-1} with \eqref{eq:C3-half-quantum-volumes}, resulting into the following simple formula in terms of the quantum Lebesgue measure
\be
 W(c,\e) = \frac{1}{2\pi\ii} \int^{c+2\pi\ii}_{c} \cH^D (s,\e) \dif s  \,.
\ee
In exponentiated coordinates $x=\eu^c$, this corresponds to the holonomy of $\cH^D$ around a small loop around a puncture of $\Sigma$, corresponding to the phase in which $W$ is being evaluated.
The transport of branes around such loops appeared in \cite{Aganagic:2001nx} in the definition of open string flat coordinates. We will elaborate more on this point and its physical interpretation in Section~\ref{sec:physical-interpretation}.

\section{Resolved conifold}\label{s:conifold}

In this section we consider the resolved conifold $O(-1)\oplus O(-1) \to \BP^1$,
which is the symplectic quotient $\BC^4\sslash U(1)$ defined by the charge matrix 
\be
 Q = (1,1,-1,-1)\,.
\ee
Denoting by $z_i$ coordinates on $\BC^4$ and defining momenta $p_i := |z_i|^2$, the moment map constraint reads
\be
 p_1 + p_2 - p_3 - p_4 = t\,.
\ee
The classical conifold geometry has two phases, corresponding to the sign of the K\"ahler parameter $t$.
Physics is essentially the same in both of these and we shall focus on the phase $t>0$.

\subsection{Symplectic cut and equivariant volumes}

The equivariant volume for the resolved conifold is defined by \eqref{eq:equivVolContour}, which gives 
\be\label{eq:conifold-contour-integral}
\begin{aligned}
 \cF(t,\e)= & \oint_{(1),(2)} \frac{\dif\phi}{2\pi \ii} ~ \frac{\eu^{\phi t}}
 {(\e_1+\phi)(\e_2+\phi)(\e_3-\phi)(\e_4-\phi)} \\
 = & \frac{\eu^{-\e_1 t}}{(\e_2-\e_1) (\e_3+\e_1) (\e_4+\e_1)}
 + \frac{\eu^{-\e_2 t}}{(\e_1-\e_2) (\e_3+\e_2) (\e_4+\e_2)} 
\end{aligned}
\ee
where $\oint_{(1),(2)}$ denotes that for $t>0$ the JK prescription instructs us to take the poles at $-\e_1$ and $-\e_2$. 
In the limit $t\to\infty, \e_2\to 0$ with $\e_2\cdot t\to 0$, this recovers the equivariant volume of $\BC^3$ obtained in \eqref{eq:C3-eVol} up to relabeling of $\e_i$

\begin{figure}[!ht]
\centering
\begin{subfigure}[b]{0.32\textwidth}
\centering
\includegraphics[width=\textwidth]{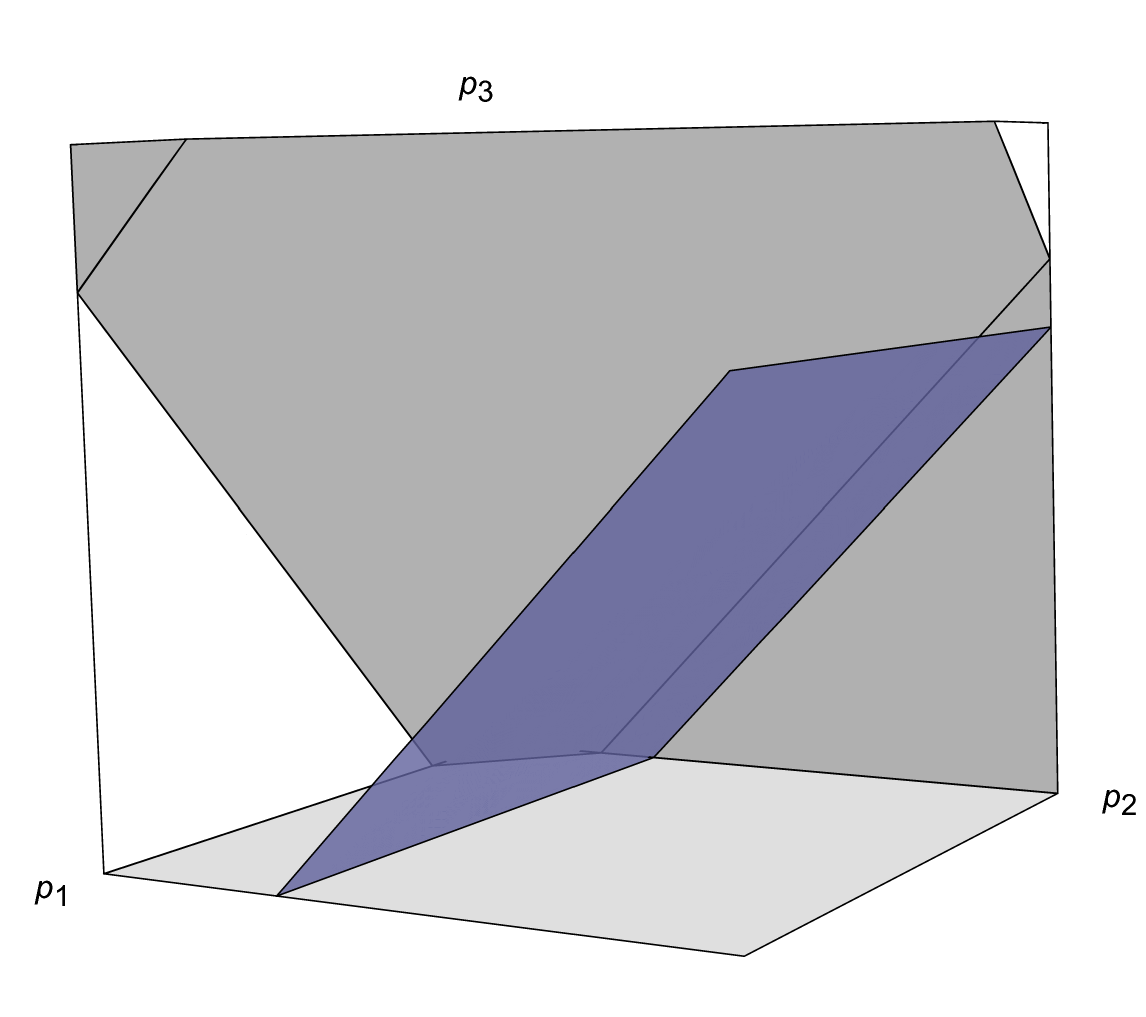}
\caption{Phase 1: $c>t>0$}
\end{subfigure}
\hfill
\begin{subfigure}[b]{0.32\textwidth}
\centering
\includegraphics[width=\textwidth]{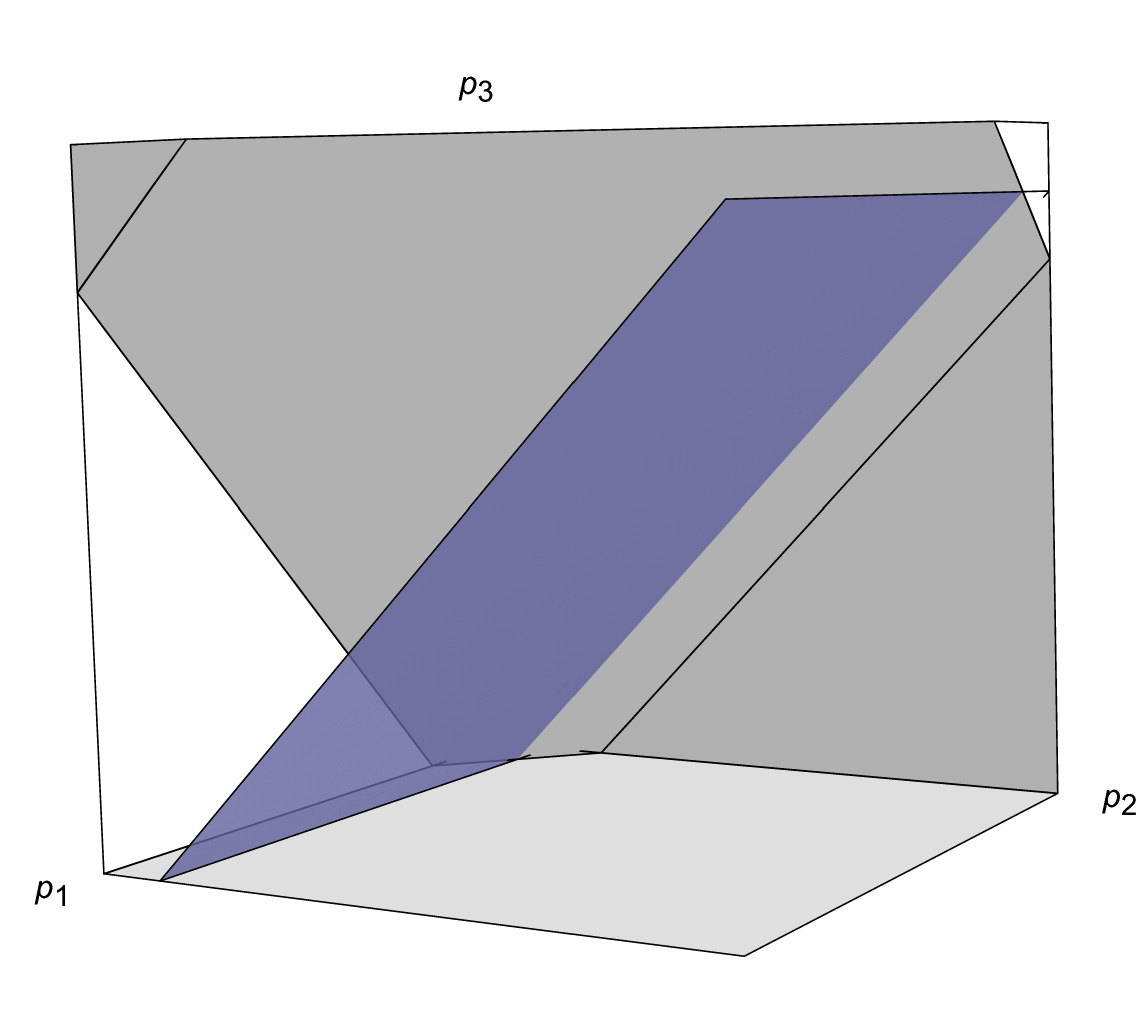}
\caption{Phase 2: $t>c>0$}
\end{subfigure}
\hfill
\begin{subfigure}[b]{0.32\textwidth}
\centering
\includegraphics[width=\textwidth]{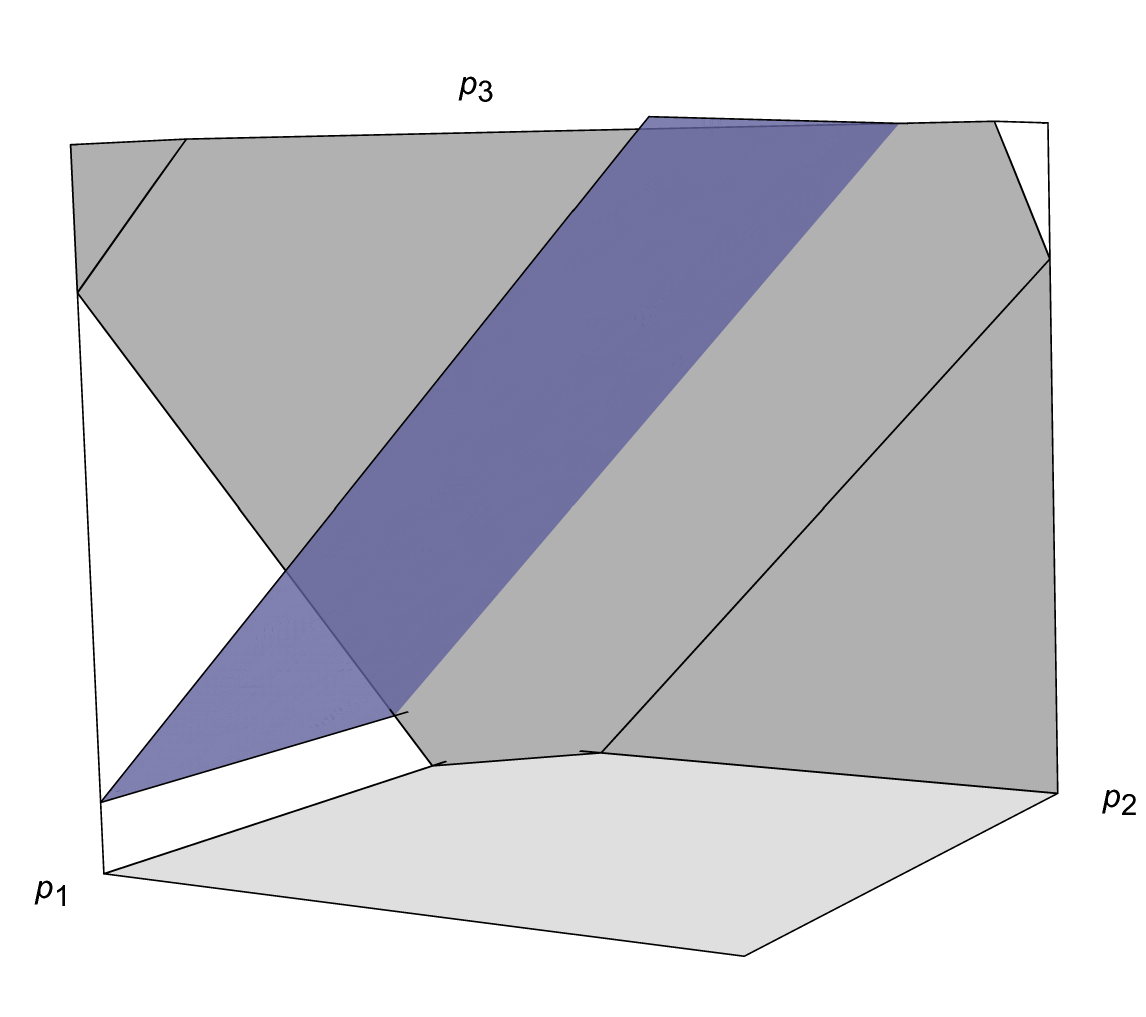}
\caption{Phase 3: $t>0>c$}
\end{subfigure}
\caption{Hyperplane for a symplectic cut of the resolved conifold in three different phases.}
\label{fig:conifoldcutphases}
\end{figure}

Consider the symplectic cut defined by the hyperplane with charges
\be
 q=(0,1,-1,0)\,.
\ee
The moment map constraints become
\be\label{conifold-plane-2}
\begin{aligned}
 & p_1 + p_2 - p_3 - p_4 = t\\
 & p_2-p_3=c     
\end{aligned}
\ee
corresponding to the ``augmented'' charge matrix
\be\label{conifold-plane-1}
 \left (\begin{array}{cccc}
 1 & 1 & -1 & -1\\
 0 & 1 & -1 & 0
\end{array} \right )
\ee
The projection of this hyperplane to the moment polytope is shown in Figure~\ref{fig:conifoldcutphases}.

Taking the symplectic quotient $\BC^4\sslash U(1)^2$ defined by the moment map \eqref{conifold-plane-2}
defines a Calabi--Yau twofold.
Although this twofold is non-compact, equivariance regularizes its volume and retains dependence on both K\"ahler moduli $(t,c)$. This can be computed by applying the definition \eqref{eq:equivVolContour} as follows
\be\label{eq:conifold-sym-quot}
 \cH(t,c,\e) = \oint_{\JK} \frac{\dif\phi_1}{2\pi \ii} \frac{\dif \phi_2}{2\pi \ii} 
 ~ \frac{\eu^{\phi_1 t} \eu^{\phi_2 c}}
 {(\e_1+\phi_1)(\e_2+\phi_1+\phi_2)(\e_3-\phi_1-\phi_2)(\e_4-\phi_1)}\,.
\ee
In this case the Jeffrey--Kirwan prescription for the contour includes three different phases
\bea
&& \cH_{(1)} (t,c,\e) = \frac{\eu^{ \e_4 t}\eu^{-(\e_2+\e_4)c}}{(\e_1+\e_4)(\e_3+\e_2)}~,~~~~c>t>0\\
&& \cH_{(2)} (t,c,\e) = \frac{\eu^{-\e_1 t}\eu^{(\e_1-\e_2)c}}{(\e_3+\e_2)(\e_4+\e_1)}~,~~~~t>c>0\\
&& \cH_{(3)} (t,c,\e) = \frac{\eu^{-\e_1 t}\eu^{(\e_3+\e_1)c}}{(\e_3+\e_2)(\e_4+\e_1)}~,~~~~t>0>c 
\eea
where in the first case poles $(2,4)$ are taken, in the second case the poles are $(1,2)$
and in third case the poles $(1,3)$. 

Since the twofold defines a cut along a locus with transverse coordinate $c$, one may again expect
that integrating over $c$ would return the equivariant volume of the resolved conifold. 
Indeed this works out as expected
\be\label{eq:conifold-volume-addition}
 \int_{-\infty}^\infty \dif c~ \cH (t,c,\e)
 = \int_{-\infty}^0 \dif c~ \cH_{(3)}
 + \int_{0}^t \dif c~ \cH_{(2)}
 + \int_t^{\infty} \dif c~ \cH_{(1)} = \cF(t,\e)~.
\ee

From the viewpoint of the symplectic quotient this can actually be expected. Indeed in \eqref{eq:conifold-sym-quot} we have essentially added an extra moment map to \eqref{eq:conifold-contour-integral}, and then in \eqref{eq:conifold-volume-addition} we have removed it by integrating over $c$.\footnote{Since $\phi_i\in \ii \BR$ in the finite-distance part of JK contours, the integral over $c\in\BR$ gives a delta function.}

\subsection{Quantum cut and quantum volumes}

The quantum volume of the resolved conifold is computed by a GLSM from a disk to $O(-1)\oplus O(-1)\to \BP^1$ with a space-filling brane. 
The defining property of $\cF^D(t, \e)$ is that it is annihilated by the equivariant PF operator
\be\label{PF-conifold}
 \cD_1 \cD_2 - \eu^{-t} \cD_3 \cD_4 \equiv (\e_1 + \partial_t) (\e_2 + \partial_t)
 - \eu^{-t} (\e_3 - \partial_t) (\e_4 - \partial_t) 
\ee
and it has prescribed semiclassical asymptotics \eqref{eq:GLSM-asymptotics}. 

The general solution is \eqref{disk-general-contour}, and it can be computed explicitly in different phases.
At large volume it is expanded as a series in $\eu^{-t}$ 
\be\label{eq:conifold-q-volume-1}
\begin{split}
 \cF^D (t, \e)  
 =&  \eu^{-\e_1 t} \Ga(\e_2-\e_1) \Ga(\e_1+\e_3) \Ga(\e_1+\e_4)
 \, {}_2F_1\left(\e_1+\e_3,\e_1+\e_4;\e_1-\e_2+1;\eu^{-t}\right) \\
 & + \eu^{-\e_2 t} \Ga(\e_1-\e_2) \Ga(\e_2+\e_3) \Ga(\e_2+\e_4)
 \, {}_2F_1\left(\e_2+\e_3,\e_2+\e_4;\e_2-\e_1+1;\eu^{-t}\right)\,,
\end{split}
\ee
for $t>0$ or as a series in $\eu^{t}$ 
\be\label{eq:conifold-q-volume-2}
\begin{split}
 \cF^D (t, \e)
 =&  \eu^{\e_3 t} \Ga(\e_1+\e_3) \Ga(\e_2+\e_3) \Ga(\e_4-\e_3)
 \, {}_2F_1\left(\e_1+\e_3,\e_2+\e_3;\e_3-\e_4+1;\eu^t\right) \\
 & + \eu^{\e_4 t} \Ga(\e_1+\e_4) \Ga(\e_2+\e_4) \Ga(\e_3-\e_4)
 \, {}_2F_1\left(\e_1+\e_4,\e_2+\e_4;\e_4-\e_3+1;\eu^t\right)\,. \\
\end{split}
\ee
for $t<0$.

Near the conifold point instead it can be given as a series in $1-\eu^{t}$
\begin{multline}
\label{eq:conifold-q-volume-3}
 \cF^D (t, \e) =
 \frac{\Ga(\e_1+\e_3)\Ga(\e_2+\e_3)\Ga(\e_1+\e_4)\Ga(\e_2+\e_4)}
 {\Ga(\e_1+\e_2+\e_3+\e_4)} \times \\
 \times \eu^{\e_3 t} \, {}_2F_1(\e_1+\e_3,\e_2+\e_3;\e_1+\e_2+\e_3+\e_4;1-\eu^t)~.
\end{multline}
All three expressions are equivalent under analytic continuation. See Appendix~\ref{a:conifold} for derivations of these results.

In a similar spirit we may compute the quantum volume of the twofold associated with the symplectic cut \eqref{conifold-plane-2}.
The equivariant periods of this system are annihilated by the equivariant Picard--Fuchs operators \eqref{eq:equiv-PF} associated to the charge matrix \eqref{conifold-plane-1} 
\be\label{PF-plane-conifold}
\begin{aligned}
 & \cD_2 - \eu^{-c} \cD_3 \equiv (\e_2 +\partial_t + \partial_c)-\eu^{-c} (\e_3 -\partial_t - \partial_c)~, \\
 & \cD_1 - \eu^{-(t-c)} \cD_4 \equiv (\e_1 + \partial_t) -\eu^{-(t-c)} (\e_4 -\partial_t)~.
\end{aligned}
\ee
One can observe that the system \eqref{PF-plane-conifold} recovers the threefold PF system \eqref{PF-conifold} for $c$-independent solutions. 
Alternatively, any solution of \eqref{PF-plane-conifold} integrated over $c$ will give rise to a solution of \eqref{PF-conifold}. 
The system \eqref{PF-plane-conifold} resembles a sort of separation of variables
for the PDE associated to the operator in \eqref{PF-conifold}.

The quantum volume for the hyperplane \eqref{conifold-plane-2} that defines the symplectic cut 
can be computed by considering a GLSM with a space-filling brane defined by charges \eqref{conifold-plane-1}
\be
 \cH^D(t,c,\e) = \oint_{\QJK} \frac{\dif \phi_1}{2\pi\ii} \frac{\dif \phi_2}{2\pi \ii} 
 \eu^{t\phi_1} \eu^{c\phi_2} \Ga(\e_1 + \phi_1) \Ga (\e_2 + \phi_1 + \phi_2)
 \Ga(\e_3 - \phi_1 - \phi_2) \Ga(\e_4- \phi_1)
\ee
The choice of contour is determined by the Jeffrey--Kirwan prescription, and it depends on the values of $c$ and $t$.
This integral corresponds to the unique solution to \eqref{PF-plane-conifold} with asymptotics \eqref{eq:GLSM-asymptotics}.

In phase 1 we have $c>t>0$ and $\cH^D_{(1)}$ is computed via $-\oint_{(2,4)}$,
meaning that we take the poles $\e_2 + \phi_1 + \phi_2 = - n$ and $\e_4 - \phi_1 = - m$
for $n=0,1,\dots$ and $m=0,1,\dots$ and the minus in front of the integral arises from the orientation 
of the contour. The sum over poles converges to a closed form expression
\be
\begin{aligned}
 \cH^D_{(1)}
 &= \sum_{n,m=0}^\infty \eu^{(\e_4+n)t} \eu^{-(\e_2+\e_4+m+n)c} \frac{(-1)^{n+m}}{n!m!} \Ga(\e_1+\e_4+n) \Ga(\e_3+\e_2+m) \\
 &= \Ga(\e_1+\e_4) \Ga(\e_3+\e_2) \frac{\eu^{\e_4(t-c)} \eu^{-\e_2c}}{(1+\eu^{t-c})^{\e_1+\e_4}(1+\eu^{-c})^{\e_3+\e_2}}~.
\end{aligned}
\ee

In phase 2 we have $t>c>0$, and we compute via $\oint_{(1,2)}$ taking the poles $\e_1 + \phi_1=-n$, $\e_2 + \phi_1+\phi_2=-m$ for $n=0,1,..$ and $m=0,1,\dots$
\be
\begin{aligned}
 \cH^D_{(2)}
 &= \sum_{n,m=0}^\infty \eu^{-(\e_1+n)t} \eu^{(\e_1 -\e_2+n-m)c} \frac{(-1)^{n+m}}{n!m!} \Ga(\e_3+\e_2+m) \Ga(\e_4+\e_1+n) \\
 &= \Ga(\e_3+\e_2) \Ga(\e_4+\e_1)
 \frac{\eu^{\e_1(-t+c)} \eu^{-\e_2c}}{(1+\eu^{-t+c})^{\e_1+\e_4} (1+\eu^{-c})^{\e_3+\e_2}}~.
\end{aligned}
\ee

In phase 3 we have 
$t>0>c$ and the JK prescription in this case is to use the contour $-\oint_{(1,3)}$,
which encircles the poles $\e_1 + \phi_1 =-n$ and $\e_3- \phi_1 - \phi_4 = - m$ for $n=0,1,\dots$ and $m=0,1,\dots$, and where the overall sign arises from the orientation of the contour
\be\label{eq;conifoldHd3}
\begin{aligned}
 \cH^D_{(3)}
 &= \sum_{n,m=0}^\infty \eu^{-(\e_1+n)t} \eu^{(\e_3+\e_1+m+n)c} \frac{(-1)^{n+m}}{n!m!} \Ga(\e_2+\e_3+m) \Ga(\e_4+\e_1+n) \\
 &= \Ga(\e_2+\e_3) \Ga(\e_4+\e_1)
 \frac{\eu^{\e_1(-t+c)} \eu^{\e_3 c}}{(1+\eu^{-t+c})^{\e_4+\e_1} (1+\eu^c)^{\e_2+\e_3}}~.
\end{aligned}
\ee

It is easy to check that expressions for $\cH^D$ in the three different phases agree exactly
\be\label{con-HD}
 \cH^D (t,c,\e) = \cH^D_{(1)}=\cH^D_{(2)}=\cH^D_{(3)} \,.
\ee
As in the case of $\BC^3$, this unification of phases can be understood as the consequence of disk instantons that smooth out singularities in the classical moduli space of the symplectic cut hyperplane.

The smoothness of $\cH^D$ is rather surprising from the viewpoint of the computation by summation over poles, since the three phases correspond to three different contours.
However this property becomes manifest if we observe that the general definition \eqref{disk-general-contour} adopted earlier can be reformulated as in \eqref{disk-general-y} 
\begin{multline}
 \cH^D (t,c,\e) = \int_{-\ii \infty}^{\ii \infty} \frac{\dif \phi_1 \dif \phi_2}{(2\pi \ii)^2} \eu^{\phi_1 t + 
 \phi_2 c} \int_0^\infty \dif y_1 \dif y_2 \dif y_3 \dif y_4 \times \\
 \times y_1^{\e_1 +\phi_1 -1} y_2^{\e_2+\phi_1+\phi_2 -1} 
 y_3^{\e_3 -\phi_1 - \phi_2 -1} y_4^{\e_4 - \phi_1-1} \eu^{-y_1-y_2-y_3-y_4}~.
\end{multline}

Once again $\cH^D$ plays the role of a quantum Lebesgue measure: integrating over $c$ recovers the quantum volume of the whole Calabi--Yau threefold
\be
	\int_{-\infty}^{\infty} \cH^D(t,c,\e)\,\dif c =  \cF^D(t,\e)\,,
\ee
with $\cF^D$ as given in \eqref{eq:conifold-q-volume-1}-\eqref{eq:conifold-q-volume-3}. 
The proof of this identity is straightforward by using the integral representation \eqref{int-HG-2} of the hypergeometric series ${}_2F_1$.

Using $\cH^D$ we can compute the volumes of the two components of the symplectic cut decomposition of the resolved conifold
\be
 O(-1)\oplus O(-1)\to \BP^1 \quad\rightarrow\quad
 \overline X_- \cup_{X_0} \overline X_+\,.
\ee
Introducing half-volumes as in \eqref{eq:C3-half-quantum-volumes}-\eqref{eq:C3-volumes-sum}, we obtain
\be\label{con-int-FD+}
\begin{split}
 \cF^D_< (t,c,\e) 
 & = \Ga(\e_2+\e_3) \Ga(\e_4+\e_1) \eu^{-\e_1 t}
 \int^c_{-\infty} \frac{\eu^{(\e_1 +\e_3) s}}
 {(1+\eu^{-t+s})^{\e_4+\e_1} (1+\eu^s)^{\e_2+\e_3}} \dif s \\
 & = \Ga(\e_2+\e_3) \Ga(\e_4+\e_1) \eu^{\e_1(-t+c)+\e_3 c}
 \int^1_{0} \frac{w^{\e_1 +\e_3-1}}{(1+\eu^{-t+c} w)^{\e_4+\e_1} (1+ \eu^c w)^{\e_2+\e_3}} \dif w \\
 & = \frac{\Ga(\e_2+\e_3) \Ga(\e_4+\e_1)}{\e_1+\e_3}
 \eu^{\e_1(-t+c)+\e_3 c} \times \\
 &\quad\quad\quad\times
 F_1 (\e_1+\e_3,\e_2+\e_3,\e_1+\e_4;\e_1+\e_3+1,-\eu^c,-\eu^{-t+c})~.
\end{split}
\ee
where $F_1$ is the Appell hypergeometric function of \eqref{int-AppelF1-function}.
The two half-volumes obey the relation
\be
 \cF^D_> (t, c, \e_1, \e_2, \e_3, \e_4)
 = \cF^D_< (t, t-c, \e_2, \e_1, \e_4, \e_3)\,.
\ee

\subsection{Holomorphic disk potential from quantum Lebesgue measure}

We now discuss how the quantum Lebesgue measure computes the open string disk potential for a framed toric $A$-brane in the resolved conifold.
We consider a Lagrangian described by the affine line
\be
 p_2 - p_3 = c\,,\qquad p_4 - p_2 = 0\,.
\ee
Choosing the phase $c<0$ gives a Lagrangian $L$ that is a $U(1)^2$ fibration over an affine half-line $\ell$ ending at $p_4=p_2=0$, $p_3=-c$ and $p_1=t-c$.
The genus-zero quantum corrected moduli space of an $A$-brane wrapped on $L$ is described by the algebraic curve
\be\label{eq:conifold-curve}
	\Sigma: \ 1 + x + y + \Q x y = 0\quad\subset \BC^\ast\times\BC^\ast\,,
\ee
where $x=\eu^c$ with $c\in \BC$ complexified by the holonomy of the abelian local system on $L$, $\Q=\eu^{-t}$ with $t\in \BC$ complexified by the flux of the $B$-field on $\BP^1$, and a choice of framing has been implicitly made.
Note that topologically $\Sigma$ is a four-punctured sphere.
The disk potential is given by the formula
\be\label{eq:conifoldsuperpotential}
	W_{\AKV}(x) = -\int^x\log (-y(x)) \, \dif\log x 
	= \Li_2(-x) - \Li_2(-\Q x)\,.
\ee

To recover this result from the quantum Lebesgue measure we define the equivariant superpotential $W(t,c,\e)$ in terms of the monodromy 
\be\label{Conifold-defW}
	\partial_c W(t,c,\e) = \frac{1}{2\pi\ii} \Big ( \cH^D(t,c+2\pi\ii,\e) - \cH^D(t,c,\e) \Big )\,.
\ee
Note that $\cH^D(t,c,\e)$ is multivalued as a function of $\eu^c$, with branching at the four points
\be\label{eq:conifold-punctures}
 \eu^c \in\{ 0,-1,-\eu^t,\infty\}\,,
\ee
corresponding to the four punctures of $\Sigma$.
The computation of the monodromy in \eqref{Conifold-defW} therefore depends on the choice of phase for $c$.
To recover the toric brane potential \eqref{eq:conifoldsuperpotential} we should work in phase 3 ($t>0>c$), where we may use expression \eqref{eq;conifoldHd3} for $\cH^D$
\begin{multline}
	\cH^D(c+2\pi\ii,\e)-\cH^D(c,\e)
	=
	(\eu^{2\pi\ii (\e_1+\e_3)} - 1) 
	\frac{\Ga(\e_2+\e_3)}{(1+\eu^c)^{\e_2+\e_3}} \eu^{\e_3 c}
	\cdot
	\frac{\Ga(\e_1+\e_4)}{(1+\eu^{c-t})^{\e_1+\e_4}} \eu^{\e_1 (c-t)}
	\\
	=
	\frac{2\pi\ii(\e_1+\e_3)}{(\e_2+\e_3)(\e_1+\e_4)}
	\Big[
	1 - \gamma (\e_1+\e_2+\e_3+\e_4 )
	+(c+\pi\ii)(\e_1+\e_3)
	-t\e_1 \\
	-(\e_2+\e_3)\log(1+\eu^c)
	-(\e_1+\e_4)\log(1+\eu^{c-t})
	+O(\e^2)
	\Big]~,
\end{multline}
where in the last step we took a Laurent expansion at $\e_i=0$. 
The equivariant superpotential is obtained by taking the primitive in $c$
\be
\begin{split}
	W(t,c,\e) & = 
	\frac{1}{2\pi\ii} \left(\cF^D_< (t,c+2\pi \ii,\e)-\cF^D_< (t,c,\e) \right)
	\\
	& = \frac{1}{2\pi \ii} (\eu^{2\pi \ii (\e_1+\e_3)} - 1)
	\frac{\Ga(\e_2+\e_3) \Ga(\e_4+\e_1)}{\e_1+\e_3} \eu^{-\e_1 t} \eu^{(\e_1 + \e_3)c}\\
	&\times 
	F_1 (\e_1+\e_3,\e_2+\e_3,\e_1+\e_4;\e_1+\e_3+1,-\eu^c,-\eu^{-t+c})~,
\end{split}
\ee
where there monodromy of the $F_1$ function is trivial in this phase.
Imposing the Calabi--Yau condition $\e_1+\e_2 + \e_3 + \e_4 =0$ and expanding around $\e_i=0$ we find
\be
 W(t,c,\e) =
 \frac{\e_1+\e_3}{\e_1+\e_4}
 \left[
 -\frac{c^2}{2} \frac{\e_1+\e_3}{\e_1+\e_4}
 + c t \frac{\e_1}{\e_1+\e_4}
 - \pi\ii c\frac{\e_1+\e_3}{\e_1+\e_4}
 + \Li_2( -\eu^c) - \Li_2 (- \eu^{c-t}) + O(\e) \right ]~.
\ee
The instanton part of the regular terms is
\be
 W^{\inst}_{\reg} (t,c) =  \Li_2(-\eu^c) - \Li_2 (-\eu^{c-t}) ~,
\ee
in perfect agreement with \eqref{eq:conifoldsuperpotential} up to the overall normalization factor.
To fix this we can simply adopt the same normalization as for $\BC^3$ (by comparing them in the appropriate limit $t\rightarrow \infty$, as remarked below \eqref{eq:conifold-contour-integral}).

\paragraph{Remarks:}
As in the case of $\BC^3$ it should be stressed that, while we have illustrated the relation between $\cH^D$ and $W_\AKV$ is a specific choice of phase $(c<0)$ and framing, the story holds much more generally. Since $\cH_D$ is defined for all values of $c$, it is straightforward to analytically continue to other phases, and consider its monodromies around any of the punctures \eqref{eq:conifold-punctures}. This will give the superpotential $W_\AKV$ for any of the corresponding branes.
Moreover $\cH^D$ has once again the property of being \emph{framing invariant}, while recovering the open string superpotential for any choice of framing.

\section{Local \texorpdfstring{$\BP^2$}{P2}}\label{s:P2}

In this section we consider the toric Calabi--Yau geometry
$O(-3)\to\BP^2$ also known as local $\BP^2$ or the canonical line bundle of $\BP^2$.
This space can be defined as the symplectic quotient $\BC^4\sslash U(1)$ with charge matrix 
\be
 Q = (1 , 1 , 1 , -3)
\ee
The corresponding moment map constraint is
\be
\label{eq:moment-map-localP2}
 p_1 + p_2 + p_3 - 3p_4 = t\,.
\ee
and we shall work at $t>0$ from now on.

\subsection{Symplectic cut and equivariant volumes}

The equivariant volume of local $\BP^2$, as defined by \eqref{eq:equivVolContour}, is given 
by
\begin{multline}
\label{locP2-eqvol}
	\cF (t, \e) 
	=  \oint_{(1),(2),(3)} \frac{\dif\phi}{2\pi \ii} ~ 
	 \frac{\eu^{\phi t}}{(\e_1+\phi)(\e_2+\phi)(\e_3+\phi)(\e_4-3\phi)}
	\\
	= 
	\frac{\eu^{-\e_1 t}}{(\e_2-\e_1)(\e_3-\e_1)(\e_4+3\e_1)}
	+\frac{\eu^{-\e_2 t}}{(\e_1-\e_2)(\e_3-\e_2)(\e_4+3\e_2)}
	+\frac{\eu^{-\e_3 t}}{(\e_1-\e_3)(\e_2-\e_3)(\e_4+3 \e_3)}
\end{multline}
where the poles at $-\e_1$, $-\e_2$ and $-\e_3$ have been taken as a consequence of the JK prescription for the phase $t>0$. 

\begin{figure}[!ht]
\centering
\begin{subfigure}[b]{0.32\textwidth}
\centering
\includegraphics[width=\textwidth]{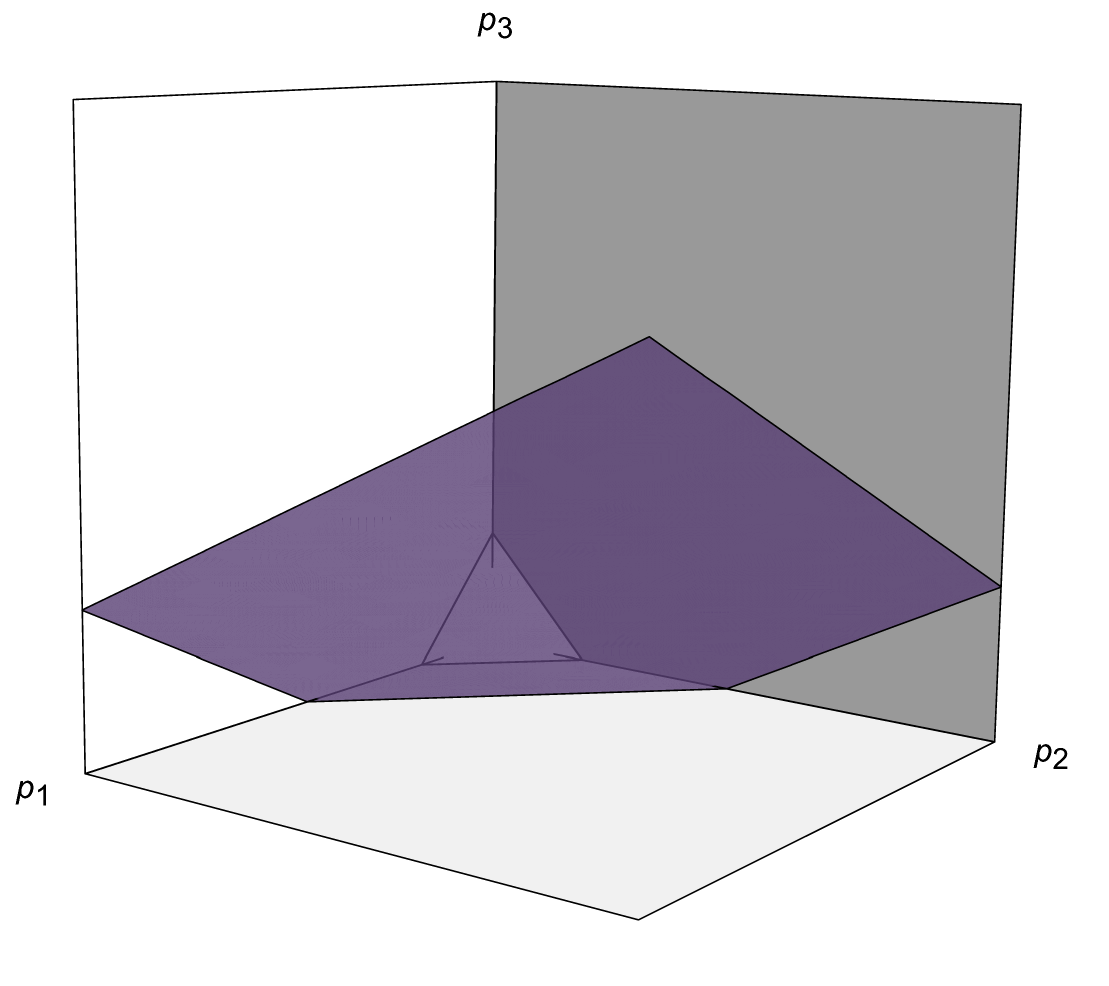}
\caption{Phase 1: $c,t>0$}
\end{subfigure}
\hfill
\begin{subfigure}[b]{0.32\textwidth}
\centering
\includegraphics[width=\textwidth]{localP2phase2.pdf}
\caption{Phase 2: $0>c>-t$}
\end{subfigure}
\hfill
\begin{subfigure}[b]{0.32\textwidth}
\centering
\includegraphics[width=\textwidth]{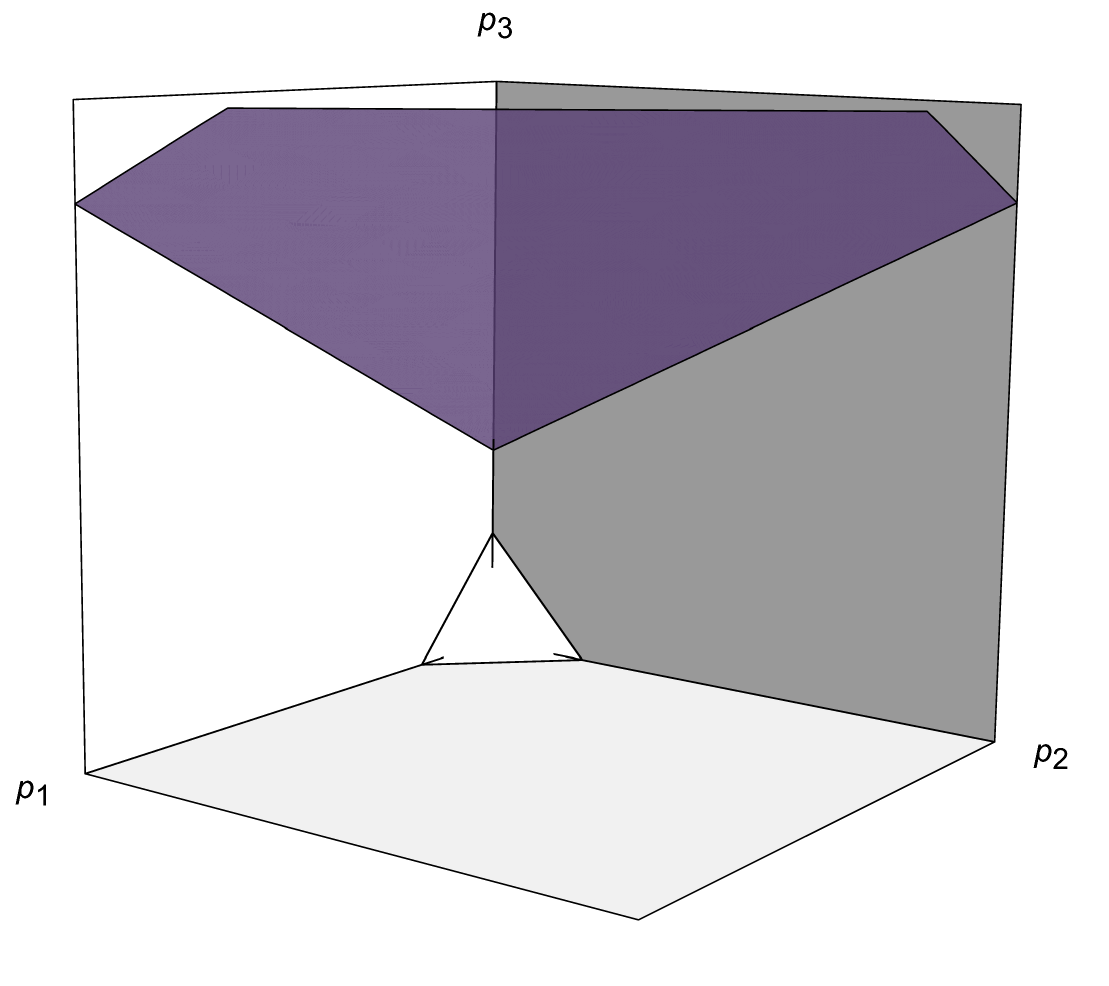}
\caption{Phase 3: $c<-t<0$}
\end{subfigure}
\caption{Hyperplane for a symplectic cut of local $\BP^2$ in three different phases.}
\label{fig:P2cutphases}
\end{figure}

We consider a symplectic cut of local $\BP^2$ defined by the hyperplane of charge
\be
 q = (0 , 0 , -1 , 1)\,.
\ee
The moment map constraints become
\be\label{locP2-plane-2}
\begin{aligned}
 & p_1 + p_2 + p_3 - 3 p_4 = t\\
 & -p_3+p_4=c     
\end{aligned}
\ee
corresponding to the “augmented” charge matrix
\be\label{locP2-plane-1}
\left (\begin{array}{cccc}
 1 & 1 & 1 & -3\\
 0 & 0 & -1 & 1
\end{array} \right )\,.
\ee
The projection of this hyperplane to the moment polytope is shown in Figure~\ref{fig:P2cutphases}.

The symplectic quotient $\BC^4\sslash U(1)^2$ defined by the moment map constraints
in \eqref{locP2-plane-2} defines a Calabi--Yau twofold whose equivariant volume is
\be\label{eq:localP2-sym-quot}
 \cH(t,c,\e) = \oint_{\JK} \frac{\dif\phi_1}{2\pi \ii} \frac{\dif \phi_2}{2\pi \ii} 
 ~ \frac{\eu^{\phi_1 t} \eu^{\phi_2 c}}
 {(\e_1+\phi_1)(\e_2+\phi_1)(\e_3+\phi_1-\phi_2)(\e_4-3\phi_1+\phi_2)}\,.
\ee
The Jeffrey--Kirwan prescription for the contour admits three different choices of phase
\bea
 && \cH_{(1)} = \frac{\eu^{-\e_2(t+3c)-\e_4 c}}{(\e_1-\e_2)(2\e_2+\e_3+\e_4)}
 + \frac{\eu^{-\e_1(t+3c)-\e_4 c}}{(\e_2-\e_1)(2\e_1+\e_3+\e_4)}~,~~~~~c,t>0 \\
 && \cH_{(2)} = \frac{\eu^{-\e_2(t+c)+\e_3 c}}{(\e_1-\e_2)(2\e_2+\e_3+\e_4)}
 + \frac{\eu^{-\e_1(t+c)+\e_3 c}}{(\e_2-\e_1)(2\e_1+\e_3+\e_4)}~,~~~~~ -t < c <0 \\
 && \cH_{(3)} = \frac12\frac{\eu^{\frac12\e_3(t+3c)+\frac12\e_4(t+c)}}{(\e_1+\frac12(\e_3+\e_4))(\e_2+\frac12(\e_3+\e_4))}~,~~~~~~~~~~~~~~~~~~~~~~~~~ c < - t < 0
\eea
where in $\cH_{(1)}$ the two sets of poles that contribute are $(1,4)$ and $(2,4)$,
while for $\cH_{(2)}$ we take poles $(1,3)$ and $(2,3)$,
and for $\cH_{(3)}$ only the poles $(3,4)$ contribute. 

Integrating over all values of $c$ returns the equivariant volume of local $\BP^2$ 
\be\label{eq:localP2-volume-addition}
 \int_{-\infty}^\infty \dif c~ \cH
 = \int_{-\infty}^{-t} \dif c~ \cH_{(3)}
 + \int_{-t}^0 \dif c~ \cH_{(2)}
 + \int_0^{\infty} \dif c~ \cH_{(1)} = \cF(t,\e)  \,.
\ee
As in previous examples, this identity can be expected on general grounds, since the integration over $c$ removes the second
moment map condition in \eqref{locP2-plane-2}.

\subsection{Quantum cut and quantum volumes}

The quantum volume of local $\BP^2$ is computed by a GLSM from a disk to the total space of the bundle $O(-3)\to \BP^2$, together with a space-filling brane. 
By definition $\cF^D(t,\e)$ is the solution of the equivariant PF equation
\be\label{locP2-PF}
 \left[\cD_1 \cD_2 \cD_3 - \eu^{-t} (2 + \cD_4) (1 + \cD_4) \cD_4\right] \cF^D(t,\e) =0
\ee
with the prescribed semiclassical asymptotics \eqref{eq:GLSM-asymptotics},
where we have
\be
 \cD_1 = \e_1 + \partial_t \,, \quad
 \cD_2 = \e_2 + \partial_t \,, \quad
 \cD_3 = \e_3 + \partial_t \,, \quad
 \cD_4 = \e_4 - 3 \partial_t \,.
\ee
The general solution is given by \eqref{disk-general-contour} and can be computed explicitly, see Appendix~\ref{a:locP2}.
At large volume it can be expanded as a series in $\eu^{-t}$ 
\be\label{eq:localP2-q-volume-1}
\begin{split}
 \cF^D (t, \e)  
 =&\,
 \eu^{-\e_1 t} \Ga(\e_2-\e_1) \Ga(\e_3-\e_1) \Ga(3\e_1+\e_4) \times \\
 & \quad \times \, {}_3F_2\left(\e_1+\frac{\e_4}{3},\e_1+\frac{\e_4}{3}+\frac13,\e_1+\frac{\e_4}{3}+\frac23;\e_1-\e_2+1,\e_1-\e_3+1;-27\eu^{-t}\right) \\
 & + \eu^{-\e_2 t} \Ga(\e_1-\e_2) \Ga(\e_3-\e_2) \Ga(3\e_2+\e_4) \times \\
 & \quad \times \, {}_3F_2\left(\e_2+\frac{\e_4}{3},\e_2+\frac{\e_4}{3}+\frac13,\e_2+\frac{\e_4}{3}+\frac23;\e_2-\e_1+1,\e_2-\e_3+1;-27\eu^{-t}\right) \\
 & + \eu^{-\e_3 t} \Ga(\e_1-\e_3) \Ga(\e_2-\e_3) \Ga(3\e_3+\e_4) \times \\
 & \quad \times\, {}_3F_2\left(\e_3+\frac{\e_4}{3},\e_3+\frac{\e_4}{3}+\frac13,\e_3+\frac{\e_4}{3}+\frac23;\e_3-\e_1+1,\e_3-\e_2+1;-27\eu^{-t}\right)~.
\end{split}
\ee

The quantum volume for the hyperplane \eqref{locP2-plane-2} that defines the symplectic cut 
can be computed by considering a GLSM with a space-filling brane defined by charges \eqref{locP2-plane-1}.
The equivariant PF operators of the GLSM are
\be\label{locP2-plane-PF}
 \cD_1 \cD_2 - \eu^{-t-3c} (1+\cD_3) \cD_3 \,, \quad\quad\quad
 \cD_4 - \eu^{-c} \cD_3 \,,
\ee
with
\be
 \cD_1 = \e_1 + \partial_t \,, \quad
 \cD_2 = \e_2 + \partial_t \,, \quad
 \cD_3 = \e_3 + \partial_t - \partial_c \,, \quad
 \cD_4 = \e_4 - 3 \partial_t + \partial_c \,.
\ee
Similar to the conifold case, the two PF systems \eqref{locP2-PF} and \eqref{locP2-plane-PF} are related to each 
other. 
For example, taking a solution of the system \eqref{locP2-plane-PF} and integrating over $c$
(provided that this integral is well-defined) gives a solution of the system \eqref{locP2-PF}. 

The quantum volume of the twofold is the unique solution to \eqref{locP2-plane-PF} with asymptotics \eqref{eq:GLSM-asymptotics}. 
The general solution can be expressed by means of the GLSM partition function with a space-filling brane on the twofold \eqref{disk-general-contour}, which in this cases specializes to
\be\label{locP2-HD-def}
 \cH^D (t,c, \e) = \oint_{\QJK} \frac{\dif\phi_1\dif\phi_2}{(2\pi \ii)^2}~\eu^{t\phi_1+c\phi_2}
 \Ga(\e_1+\phi_1) \Ga(\e_2+\phi_1) \Ga(\e_3+\phi_1-\phi_2)
 \Ga(\e_4-3\phi_1+\phi_2)~,
\ee
where the contour depends on the values of $t$ and $c$ through
the Jeffrey--Kirwan prescription.

In phase 1 we have $c,t>0$ and the poles that contribute to \eqref{locP2-HD-def} are the two sets $(1,4)$, $(2,4)$.
This means that we take towers $\e_1+\phi_1=-n$, $\e_4 - 3\phi_1 + \phi_2=-m$ ($n,m =0,1,2,\dots$)
plus the towers $\e_2 +\phi_1 = -n$, $\e_4 - 3\phi_1 + \phi_2=-m$ ($n,m =0,1,2,\dots$). 
The result is
\be
\begin{aligned}
 \cH^D_{(1)} =&  
 \sum_{n=0}^\infty \sum_{m=0}^\infty
 \frac{(-1)^m}{n!m!}
 \frac{\Ga(\e_2-\e_1)}{(\e_1-\e_2+1)_n}
 \Ga(m+2n+2\e_1+\e_3+\e_4)
 \,\eu^{-(t+3c)(n+\e_1)-c(m+\e_4)} \\
 & +  \sum_{n=0}^\infty \sum_{m=0}^\infty
 \frac{(-1)^m}{n!m!}
 \frac{\Ga(\e_1-\e_2)}{(\e_2-\e_1+1)_n}
 \Ga(m+2n+2\e_2+\e_3+\e_4)
 \,\eu^{-(t+3c)(n+\e_2)-c(m+\e_4)}\,. 
 \end{aligned}
\ee
This expression can be resummed into the following closed-form 
\be\label{eq:P2Hd1}
\begin{aligned}
 \cH^D_{(1)}
 =&
 \Ga(\e_2-\e_1) \Ga(2\e_1+\e_3+\e_4)
 \frac{\eu^{-\e_1(t+3c)-\e_4 c} }{ (1+\eu^{-c})^{2\e_1+\e_3+\e_4} } \times \\
 & \quad\quad\times \, {}_2F_1\left(\e_1+\frac{\e_3}{2}+\frac{\e_4}{2},\e_1+\frac{\e_3}{2}+\frac{\e_4}{2}+\frac12;\e_1-\e_2+1;\frac{4\eu^{-t-3c}}{(1+\eu^{-c})^2}\right) \\
 & + \Ga(\e_1-\e_2) \Ga(2\e_2+\e_3+\e_4)
 \frac{ \eu^{-\e_2(t+3c)-\e_4 c} }{ (1+\eu^{-c})^{2\e_2+\e_3+\e_4} } \times \\
 & \quad\quad\times \, {}_2F_1\left(\e_2+\frac{\e_3}{2}+\frac{\e_4}{2},\e_2+\frac{\e_3}{2}+\frac{\e_4}{2}+\frac12;\e_2-\e_1+1;\frac{4\eu^{-t-3c}}{(1+\eu^{-c})^2}\right)~\,.
\end{aligned}
\ee
In phase 2 we have $-t < c <0 $ and in this case the relevant sets of poles are the two towers labeled by $(1,3)$, $(2,3)$
\be\label{eq:P2Hd2}
\begin{aligned}
 \cH^D_{(2)} 
 =&
 \sum_{n=0}^\infty\sum_{m=0}^\infty\frac{(-1)^m}{n!m!}\frac{\Ga(\e_2-\e_1)}{(\e_1-\e_2+1)_n}\Ga(m+2n+2\e_1+\e_3+\e_4)
 \,\eu^{-(t+c)(n+\e_1)+c(m+\e_3)} \\
 &+
 \sum_{n=0}^\infty\sum_{m=0}^\infty\frac{(-1)^m}{n!m!}\frac{\Ga(\e_1-\e_2)}{(\e_2-\e_1+1)_n}\Ga(m+2n+2\e_2+\e_3+\e_4)
 \,\eu^{-(t+c)(n+\e_2)+c(m+\e_3)} \\
\end{aligned}
\ee
where an overall sign has been taken into account due to the orientation of the QJK contour. 
Again, these series admit resummations in terms of hypergeometric functions
\be
\begin{aligned}
 \cH^D_{(2)} =&
 \Ga(\e_2-\e_1) \Ga(2\e_1+\e_3+\e_4)
 \frac{\eu^{-\e_1(t+c)+\e_3 c}}{ (1+\eu^c)^{2\e_1+\e_3+\e_4}} \times\\
 &\quad\quad\times \, {}_2F_1\left(\e_1+\frac{\e_3}{2}+\frac{\e_4}{2},\e_1+\frac{\e_3}{2}+\frac{\e_4}{2}+\frac12;\e_1-\e_2+1;\frac{4\eu^{-t-c}}{(1+\eu^c)^2}\right) \\
 & + \Ga(\e_1-\e_2) \Ga(2\e_2+\e_3+\e_4)
 \frac{\eu^{-\e_2(t+c)+\e_3 c}}{ (1+\eu^c)^{2\e_2+\e_3+\e_4}} \times\\
 &\quad\quad\times \, {}_2F_1\left(\e_2+\frac{\e_3}{2}+\frac{\e_4}{2},\e_2+\frac{\e_3}{2}+\frac{\e_4}{2}+\frac12;\e_2-\e_1+1;\frac{4\eu^{-t-c}}{(1+\eu^c)^2}\right) \\
\end{aligned}
\ee
In phase 3 we have $c < - t<0$ and the contour picks up the single tower of poles $(3,4)$
\begin{multline}\label{eq:P2Hd3}
 \cH^D_{(3)} 
 = \frac12 \sum_{n=0}^{\infty} \sum_{m=0}^{\infty} \frac{(-1)^{m+n}}{m! n!}
 \Ga\left(\e_1+\frac12(m+n+\e_3+\e_4)\right)
 \Ga\left(\e_2+\frac12(m+n+\e_3+\e_4)\right) \times \\
 \times \eu^{\frac12c(m+3n+3\e_3+\e_4)+\frac12 t(m+n+\e_3+\e_4)}
\end{multline}
where again an overall sign comes from the orientation of the contour. 
This can be resummed to the closed form expression
\be\label{eq;P2Hd3-resum}
\begin{aligned}
 \cH^D_{(3)} 
 = &
 \frac12 \eu^{\frac12\e_3(t+3c)+\frac12\e_4(t+c)}
 \Big\{
 \Ga\left(\e_1+\frac{\e_3}{2}+\frac{\e_4}{2}\right)
 \Ga\left(\e_2+\frac{\e_3}{2}+\frac{\e_4}{2}\right) \times \\
 &\quad\quad\times\, {}_2F_1\left(\e_1+\frac{\e_3}{2}+\frac{\e_4}{2},
 \e_2+\frac{\e_3}{2}+\frac{\e_4}{2};
 \frac12;\frac14 \eu^{t+c} (1+\eu^c)^2\right) \\
 & - \eu^{\frac12(t+c)} (1+\eu^c)
 \Ga\left(\e_1+\frac{\e_3}{2}+\frac{\e_4}{2}+1\right)
 \Ga\left(\e_2+\frac{\e_3}{2}+\frac{\e_4}{2}+1\right) \times \\
 &\quad\quad\times\, {}_2F_1\left(\e_1+\frac{\e_3}{2}+\frac{\e_4}{2}+1,
 \e_2+\frac{\e_3}{2}+\frac{\e_4}{2}+1;\frac32;\frac14 \eu^{t+c} (1+\eu^c)^2\right)
 \Big\}
\end{aligned}
\ee

The equivalence of the three expressions under analytic continuation follows from identities among hypergeometric series \eqref{eq:rule1}, hence
\be
 \cH^D_{(1)} = \cH^D_{(2)} = \cH^D_{(3)} =: \cH^D (t,c,\e)~.
\ee
Worth of mention is a fourth expression for $\cH^D$ that holds near the ``conifold'' point $t=c=0$ 
\begin{multline}\label{diskP2-full}
 \cH^D (t,c,\e) =
 2^{2\e_1+\e_3+\e_4+1} \Ga(\e_1+\e_2+\e_3+\e_4)
 B(2\e_1+\e_3+\e_4,2\e_2+\e_3+\e_4)
 \frac{\eu^{-\e_1(t+3c)-\e_4 c}}{(1+\eu^{-c})^{2\e_1+\e_3+\e_4}}
 \times \\
 \times\,
 {}_2F_1\left(\e_1+\frac{\e_3}{2}+\frac{\e_4}{2},\e_1+\frac{\e_3}{2}+\frac{\e_4}{2}+1;
 \e_1+\e_2+\e_3+\e_4+\frac12;1-\frac{4\eu^{-t-3c}}{(1+\eu^{-c})^2}\right)~,
\end{multline}
which is related to \eqref{eq:P2Hd1} via the hypergeometric identity \eqref{eq:rule2}.

As in previous examples, the unification of phases is a consequence of disk instantons that smooth out singularities in the classical moduli space of the symplectic cut hyperplane, and it is best understood in terms of the integral presentation \eqref{disk-general-y}.

Once again $\cH^D$ plays the role of a quantum Lebesgue measure: integrating over $c$ recovers the quantum volume of the whole Calabi--Yau threefold. In addition, we can use $\cH^D$ to compute the volumes of the two components of the symplectic cut decomposition $\cF^D_\lessgtr(t,c,\e)$. 
The computations are straightforward, we leave details to the interested reader.

\subsection{Holomorphic disk potential from quantum Lebesgue measure}

To discuss the open string disk potential for a framed toric $A$-brane in local $\BP^2$ we consider the Lagrangian modeled on the affine line
\be
 p_4 - p_3 = c\,,\qquad p_1 - p_2 = 0\,.
\ee
Choosing the phase $c<-t<0$ gives a Lagrangian $L$ that is a $U(1)^2$ fibration over an affine half-line $\ell$ ending at $p_1 = p_2 = 0$, $p_3 = - \frac12(t+3c)$ and $p_4 = - \frac12(t+c)$.
The genus-zero quantum corrected moduli space of an $A$-brane wrapped on $L$ is described by the algebraic curve
\be
	\Sigma: \ 1+y+x + \Q x^3 y^{-1} = 0 \quad\subset \BC^\ast\times\BC^\ast\,.
\ee
This is a genus-one curve with three punctures, and the brane that we study corresponds to the puncture at $x=\infty$.
For the purpose of computing the superpotential we need to change coordinates.
First we rescale the $y$-variable by substituting $y\to - y \sqrt{-\Q x^3}$ to ensure that $|y|\to 1$ as $x\to\infty$. 
Afterwards we redefine $x\to 1/x$ for convenience, so that the expansion is now again at $x=0$. 
These steps give the new curve
\be\label{eq:localP2-curve}
 1 + x + \sqrt{-\frac{\Q}{x}} \,y^{-1} - \sqrt{-\frac{\Q}{x}} y = 0\,,
\ee
from which it is straightforward to compute the superpotential
\be\label{eq:P2superpotential}
 W_{\AKV}(x) = -\int^x\log (-y(x)) \, \dif\log x 
\ee
by solving for $y\equiv y(x)$,
\be\label{eq:logy-localP2}
 y(x) =
 - \frac12(-x\Q^{-1})^{1/2}(1+x)
 - \sqrt{1-\tfrac14 x\Q^{-1}(1+x)^2}\,.
\ee
We can now identify $x=\eu^c$ and $\Q=\eu^{-t}$ with $c,t\in \BC$ to compare with the GLSM predictions.
Since the superpotential \eqref{eq:P2superpotential} is rather involved in this case, it will be more convenient to reproduce its logarithmic derivative
$\partial W_{\AKV}/\partial\log x=-\log(-y)$ given by \eqref{eq:logy-localP2}.

To recover \eqref{eq:logy-localP2} from the quantum Lebesgue measure we consider the relation between the equivariant superpotential $W(t,c,\e)$ and the monodromy 
\be\label{localP2-defW}
 \partial_c W(t,c,\e) =
 \frac{1}{2\pi\ii} \Big ( \cH^D(t,c+2\pi\ii,\e) - \cH^D(t,c,\e) \Big )\,.
\ee
As in previous examples, note that $\cH^D(t,c,\e)$ is multivalued as a function of $\eu^c$ and has branching at punctures of $\Sigma$.
This needs to be taken into account when computing the monodromy \eqref{localP2-defW}, as it will end up depending on the phase of $c$.

For the toric brane we are studying, the relevant phase is the third one 
($c < - t<0$). From the expression for $\cH^D$ obtained in \eqref{eq:P2Hd3} we can immediately compute the monodromy\footnote{An alternative derivation is to use the resummed expression \eqref{eq;P2Hd3-resum} and note that the hypergeometric function is analytic inside the unit disk.}
\be
\begin{split}
 & \cH^D(c+2\pi\ii)-\cH^D(c) \\
 & =
 \frac12 \sum_{n=0}^{\infty} \sum_{m=0}^{\infty} \frac{(-1)^{m+n}}{m! n!}
 \Ga\left(\e_1+\frac12(m+n+\e_3+\e_4)\right)
 \Ga\left(\e_2+\frac12(m+n+\e_3+\e_4)\right) \\
 & \times \left( \eu^{\pi \ii (m+3n+3\e_3+\e_4)} - 1\right)
 \eu^{\frac12c(m+3n+3\e_3+\e_4)+\frac12 t(m+n+\e_3+\e_4)} \\
 & = 2\pi\ii
 \Big\{
 \frac{3\e_3+\e_4}{(2\e_1+\e_3+\e_4)(2\e_2+\e_3+\e_4)} \\
 & +
 \frac{3\e_3+\e_4}{(2\e_1+\e_3+\e_4)(2\e_2+\e_3+\e_4)}
 \left(
 \frac{c+\pi\ii}{2} (3\e_3+\e_4)
 +\frac{t}{2} (\e_3+\e_4)
 - \gamma (\e_1+\e_2+\e_3+\e_4)
 \right) \\
 & 
 -\log\left[
 \frac12(-\eu^{t+c})^{\frac12}(1+\eu^c)
 + \sqrt{1-\tfrac14\eu^{t+c}(1+\eu^c)^2}
 \right] + O(\e)
 \Big\}
\end{split}
\ee
where we took a Laurent expansion in $\e$'s around zero. 
The instanton part of the regular terms is therefore
\be
 \partial_c W^{\inst}_{\reg} (t,c) = 
 -\log\left[
 \frac12(-\eu^{t+c})^{1/2}(1+\eu^c)+\sqrt{1-\tfrac14\eu^{t+c}(1+\eu^c)^2}
 \right] ~,
\ee
in perfect agreement with \eqref{eq:P2superpotential}-\eqref{eq:logy-localP2}.

\paragraph{Remarks:}
As in previous examples, we can compute $W_\AKV$ from $\cH^D$ in other phases, as well as other choice of framing.
The quantum Lebesgue measure $\cH^D$ is framing-invariant, but its monodromy (computed in appropriate coordinates) recovers the framing-dependent superpotential $W_{\AKV}$ in any choice of framing.
Going to other phases we find that we can again obtain $W_{\AKV}$.
Fixing the overall normalization is an important but delicate matter, whose systematic discussion we leave to future work. 

\section{Monodromy as domain wall tension}\label{sec:physical-interpretation}

We have presented concrete evidence that the monodromy of the quantum Lebesgue measure computes the open topological string genus-zero superpotential for toric $A$-branes in several different examples. The key relation is 
\be\label{eq:dW-deltaHD}
	\partial_c W(\bt, c,\e) = \frac{1}{2\pi\ii} \Big ( \cH^D(\bt, c+2\pi\ii,\e) - \cH^D(\bt, c,\e) \Big )\,,
\ee
or its integrated version 
\be
	W(\bt, c,\e) = \frac{1}{2\pi\ii} \Big( \cF^D_<(\bt, c+2\pi\ii,\e) - \cF^D_<(\bt, c,\e) \Big)~\,.
\ee
where  $W(\bt, c,\e)$ is an equivariant generalization of the toric brane superpotential from which the latter can be recovered.

In this section we sketch the physical meaning of these formulae by drawing a parallel between our approach and previous works on open string mirror symmetry.
We stress from the outset that, while the overall physical interpretation generally agrees with previous proposals, the details of its realization will differ significantly.
As already remarked in the introduction, a key feature of the equivariant superpotential $W(\bt, c,\e)$ is that it comes from $\cH^D(\bt, c,\e)$, which extends smoothly over the whole K\"ahler moduli space. 
By contrast, open string superpotentials considered in previous literature are always defined within a single phase of the moduli space, and typically feature jumps from one phase to another.
In these regards equivariance acts as a regulator of certain divergencies in the open string potential, and $W(\bt, c,\e)$ provides a unifying description of previous results obtained in different phases.

\subsection{Superpotentials from open string mirror symmetry}

We start with a review of the computation of open-string superpotentials for Lagrangian $A$-branes, as originally studied in the works \cite{Aganagic:2000gs, Aganagic:2001nx} in the absence of equivariance.
In these papers it was argued that the superpotential counting holomorphic disks with boundary on $L$ arises by integrating a certain 1-form $\lambda_{\AKV}$ along a 1-chain on the mirror curve $\Sigma$
\be\label{eq:AKV-proposal}
 W_{\AKV}(\bt,c) = \int^{c} \lambda_{\AKV}\,.
\ee
In this picture $\Sigma$ is the quantum moduli space of the toric $A$-brane supported on a special Lagrangian $L$, after corrections by genus-zero curves. 
Formula \eqref{eq:AKV-proposal} is natural from the viewpoint of mirror symmetry adapted to open strings, as it relates curve counts to 1-chain integrals.

The first piece of information that goes into the definition \eqref{eq:AKV-proposal} is the choice of 1-chain. 
Its definition depends on the choice of phase for $L$ in the semiclassical picture. As reviewed in Section~\ref{sec:Abrane-cuts}, there is classically a finite number of distinct phases for $L$, characterized by the choice of an edge of the toric diagram where the affine half-line $\ell$ ends.
In classical geometry, the moduli space of the $A$-brane is a circle fibration over the toric diagram with singularities at toric vertices.
The circle direction corresponds in classical geometry to the holonomy modulus of a $U(1)$ local system on $L$, which has $b_1=1$.

In the quantum theory, singularities are smoothed out by zero-area disks, and the moduli space is described by $\Sigma$. 
When the phase of $L$ corresponds to an {external} line of the toric diagram, the 1-chain appearing in the definition \eqref{eq:AKV-proposal} is a contour that begins at the corresponding puncture on $\Sigma$ and terminates at a point $(x,y)\in \Sigma$.
The latter endpoint has coordinates $x=\eu^c$ and $y=y_1(x)$ corresponding to the branch of $\Sigma$ that goes to $1$ as $x$ tends to the puncture.

A toric $A$-brane in the quantum theory (at genus zero, or weak string coupling) is therefore labeled by a point on $\Sigma$.
A closed loop on $\Sigma$ based at $x$ defines a \emph{domain wall} for the $A$-brane, as each point along the path corresponds to a different choice of moduli that departs from $c$ and eventually comes back.
\footnote{If the projection of $\Sigma$ to $\BC^\ast_x$ has several branches, it is understood that we consider the lift of $x$ to the canonical branch where the superpotential $W_{\AKV}$ is always computed, namely the branch where $y_1(x) \to 1$ when $x$ tends to the puncture.
There are then two types of paths: those along which $\log y$ has trivial net branching and those along which the branching is non-trivial. The former type corresponds to closed BPS states, and are studied in \cite{Klemm:1996bj, Eager:2016yxd, Banerjee:2018syt}. Here we are concerned with paths of the second type.}
See \cite{Acharya:2001dz, Delmastro:2020dkz} for studies of superpotentials from domain walls from the viewpoint of the worldvolume QFT.

The second key piece of data in \eqref{eq:AKV-proposal} is the 1-form $\lambda_{\AKV}$.
In \cite{Aganagic:2001nx} it was argued that this differential is determined by the tension of certain domain walls for $L$, which play the role of appropriate flat coordinates for the open string. 
Consider a domain wall $D_{c\to c+2\pi\ii}$ obtained by taking a path along which $x=\eu^c$ shifts to $\eu^{c+2\pi\ii}$, this defines a flat coordinate 
\be\label{eq:hat-Y}
	\hat Y  = \frac{1}{2\pi\ii} \int_{D_{c\to c+2\pi\ii}} \log(- y)\, \dif\log x\,.
\ee
Similarly, there is a flat coordinate $\hat c$ arising as the tension of a domain wall that shifts $\log y$ by $2\pi\ii$.
The differential is then given by these as follows
\be
	\lambda_{\AKV} =  \hat Y \, \dif \hat c\,.
\ee
It is further argued in \cite{Aganagic:2001nx} that $\dif\hat c / \dif c=1$ where $c=
\log x$, so that only the quantum corrections present in $\hat Y$ contribute to $W_{\AKV}$.
In conclusion, the derivative of the superpotential corresponds to the tension of the domain wall $D_{c\to c+2\pi\ii}$
\be\label{eq:WAKV-hatY}
	\partial_c W_{\AKV} = \hat Y \,.
\ee

\subsection{4d \texorpdfstring{$\cN=1$}{N=1} superpotential and quantum cuts from D6 branes}

We now explain how the domain wall tension reviewed earlier enters in our construction.
Consider a semi-infinite domain wall $D_{c\to \star}$ corresponding to a path along which $x$ goes from $\eu^c$ to infinity along one of the punctures.
This path defines a family of special Lagrangian $A$-branes starting with the one supported at $L$ and ending with a purely classical brane, since all instanton corrections are suppressed near the puncture.
We can then consider another semi-infinite domain wall $D_{c+2\pi\ii \to \star}$, and take the concatenation with the previous one with inverted orientation
\be\label{eq:domain-wall-diff}
	D_{c\to  \star}\circ D_{\star\to c+2\pi\ii} \simeq D_{c\to c+2\pi\ii}\,.
\ee
By construction, the tension of the composite wall equals the flat coordinate \eqref{eq:hat-Y}.

The role of the semi-infinite domain wall $D_{c\to \star}$ is played in our story by the Calabi--Yau twofold that defines the symplectic cut.
Indeed as we explained in Section~\ref{sec:Abrane-cuts} there is a close relationship between Lagrangian $A$-branes and symplectic cuts.
Both definitions share a choice of hyperplane $h_1$ defined in \eqref{eq:Abrane-cut-planes}, which \emph{contains} the affine half-line $\ell$ that models the special Lagrangian $L$.
Both the CY2 and $L$ are $U(1)^2$ fibrations over the respective bases (the hyperplane $h_1$ or $\ell$).
In this regard, the CY2 that defines the symplectic cut of $X$ can be regarded as the support for a semi-infinite domain wall ending with the $A$-brane on $L$.

To make further progress and refine this picture, let us recall that the open string superpotential $W_{\AKV}$ has a target space interpretation as the 4d $\cN=1$ superpotential of the worldvolume QFT engineered by a D6 brane on $L\times \BR^4$, when embedding the topological string on $X$ into Type IIA string theory on $X\times \BR^4$ \cite{Ooguri:1999bv}.
The domain wall is then engineered by taking a family of D6 branes wrapping the 4-chain defined by the CY2-fold $X_0$ that defines the symplectic cut \eqref{eq:X0}.
The path $D_{c\to\star}$ that defines the domain wall is the image of the hyperplane $h_1$ under the Hori--Vafa mirror map projected to $\Sigma$, see \cite{Cassia:2024txc} for more details.
The quantum Lebesgue measure $\cH^D(\bt,c,\e)$ is the equivariant volume of the domain wall labeled by $D_{c\to\star}$.
Taking the difference of domain walls as in \eqref{eq:domain-wall-diff} therefore is expected to give the tension of the domain wall $D_{c\to c+2\pi\ii}$
\be\label{eq:match-with-AKV}
	\frac{1}{2\pi\ii}\left(\cH^D(\bt, c+2\pi\ii,\e) - \cH^D(\bt, c,\e)\right) 
	\sim
	\frac{1}{2\pi\ii} \int_{D_{c\to c+2\pi\ii}} \log (-y)\, \dif\log x\,.
\ee
Here $\sim$ denotes that this is not an equality in the strict sense, 
but that taking a suitable non-equivariant limit of the left hand side and selecting regular terms
should reproduce the right hand side.

Comparing with \eqref{eq:hat-Y}, it is natural to identify the l.h.s.\ of \eqref{eq:match-with-AKV} as $\hat Y(\e)$, the equivariant analogue of the open-string flat coordinate. 
Pushing this analogy, and recalling from \eqref{eq:WAKV-hatY} 
that $\hat Y =  \partial_cW_{\AKV}$,
it is therefore natural to expect that \eqref{eq:match-with-AKV} defines the derivative of an equivariant extension of $W_{\AKV}(c)$, namely $W(c,\e)$.
This is in perfect agreement with our formula \eqref{eq:dW-deltaHD}.

To conclude, there is a reasonable analogy between our definition of the equivariant superpotential and standard considerations based on open string mirror symmetry.
A key difference is that in both cases the relevant geometric objects are non-compact.
As a consequence of this, computations done in the non-equivariant setting usually neglect semiclassical contributions from constant maps, and the focus is entirely on the non-constant instanton contributions in each phase separately.
Instead in our setting we can keep track of semiclassical contributions as well, since they get regularized by equivariance. 
Thanks to this we obtain a function $\cH^D(\bt, c,\e)$ that is globally defined, and whose monodromies around different loops in the complexified $c$-plane encode all disk superpotentials $W_{\AKV}$ in the different phases.

\section{Summary and outlook}\label{sec:summary}

In this paper we introduced and studied a quantum uplift of the symplectic cut construction in the setting of toric Calabi--Yau threefolds.
Our approach is based on the equivariant gauged linear sigma model, whose target is the threefold after cutting.
One the main characters in our story is the quantum Lebesgue measure $\cH^D(\bt,c,\e)$ associated to the choice of cut. The measure corresponds to the equivariant quantum volume of the divisor that is invariant under the $U(1)$ action defining the cut.

The quantum measure encodes both closed and open topological string invariants.
Integrating $\cH^D(\bt, c,\e)$ over all values of the transverse modulus $c$ recovers the equivariant quantum volume $\cF^D(\bt,\e)$ of the ambient CY3
\be\label{eq:FD-conclusion}
 \cF^D(\bt,\e) = \int_{-\infty}^{+\infty} \cH^D(\bt, c,\e)\, \dif c\,,
\ee
which is further related to the closed Gromov--Witten potential.
The monodromy of $\cH^D(\bt,c,\e)$ along a circle in the complexified $c$-plane defines instead an equivariant open string superpotential $W(\bt, c,\e)$
\be\label{eq:W-conclusion}
	W(\bt, c, \e) = \frac{1}{2\pi\ii} \int^{c+2\pi\ii}_{c} \cH^D (\bt, s,\e) \dif s  \,.
\ee
The standard disk instanton potential for a toric Lagrangian associated to the cut can be recovered from $W(\bt,c,\e)$ by a suitable (phase dependent) Laurent expansion in the equivariant parameters $\e_i$.
Through equations \eqref{eq:FD-conclusion}-\eqref{eq:W-conclusion}, the quantum Lebesgue measure provides a direct relation between closed and open string invariants.

Equivariant partition functions $\cF^D, \cH^D$ are globally defined across the entire K\"ahler moduli space (extended by open string moduli for the latter).
It is interesting to consider how this works out from the viewpoint of series expansions near large radius points and near conifold points, such as the ones we obtained.
To this end, recall that equivariant partition functions include not only instantons, but also semiclassical contributions due to constant maps (which would diverge for non-compact targets without equivariance).
Neither the instanton nor the semiclassical contributions are globally defined on their own, since they both exhibit jumps across different phases of the geometry.
Remarkably, their sum turns out to be globally well-defined (see also Remark \ref{rmk:note-added} for a mathematical explanation of this fact). 
This perspective extends and unifies previous large-volume computations of GW invariants, in very concrete ways.
In particular, the partition function of the equivariant GLSM produces all-instanton resummed expressions for $\cF^D, \cH^D$ that can be written down in closed form and which can be analytically continued across different phases with no difficulty.
A prototypical example of this are the large-radius and the conifold-point expressions \eqref{eq:conifold-q-volume-1}-\eqref{eq:conifold-q-volume-2} and \eqref{eq:conifold-q-volume-3} in terms of hypergeometric functions.

\subsubsection*{Further developments}

Throughout this work we have come across several important questions that deserve to be studied in greater detail. 

\medskip

\emph{Normalization factors.}
On a purely technical level, there are certain issues with the normalization of the physical superpotentials obtained in the non-equivariant limit.
In this paper we have largely neglected these, since the normalization can be unambiguously fixed by hand by matching the first open-string instanton contribution with that of the toric brane in $\BC^3$.
Nevertheless, it would certainly be satisfying to find 
a coherent explanation for the overall factors.

\medskip

\emph{Mirror symmetry.}
The global nature of quantum volumes $\cF^D,\cH^D$ hinges in a crucial way on equivariance.
In our story this is naturally introduced by a deformation of the GLSM, where $\e_i$ correspond to masses of chiral fields.
A simple manipulation of the GLSM Coulomb branch integral leads to an equivariant generalization of the Hori--Vafa mirror geometry, as remarked around \eqref{disk-general-y}.
On the one hand, this would be key to place \eqref{eq:match-with-AKV} on a more firm footing, by providing the appropriate equivariant generalization of the right-hand side. 
On the other hand, studying the equivariant deformation of Hori--Vafa mirror symmetry would also open up the possibility to study non-compact $B$-branes as mirror non-compact $A$-branes. 
Due to multivaluedness of the holomorphic top-form, we expect calibration to produce non-compact mirror sLags very similar in nature to the non-compact $A$-branes studied in the context of wrapped Fukaya categories. 
For mirrors of toric threefolds these can be studied concretely by a straightforward extension of the analysis of compact $A$-branes initiated in \cite{Banerjee:2022oed}.

\medskip

\emph{Other phases, other cuts and framing.}
In this work we focused on a few examples, and analyzed the relation to open strings only in certain phases of the $c$-plane.
However a strength of our framework lies precisely in the global global nature of $\cH^D$, whose different monodromies tie together equivariant superpotentials associated to different phases.
Some of the additional phases are most naturally accessible by changing the type of cut under consideration. In addition to this, the open string superpotential is known to change under so-called framing of the brane. Although we have partially addressed some of these points in the examples considered in this paper, it would be very interesting to extend our discussion further to include all these global issues in a coherent and systematic fashion.

\medskip

\emph{Higher genus.}
Although we have restricted attention to genus-zero worldsheets, there are good reasons to expect an extension of our observations to all-genera.
From the viewpoint of GLSMs, the summation over genera incurs in notorious technical complications related to the holomorphic anomaly equations \cite{Bershadsky:1993cx}.
On the other hand, in recent years the study both closed and open instanton partition functions has advanced significantly through a variety of approaches, including the topological vertex \cite{Aganagic:2003db}, the TS/ST correspondence \cite{Grassi:2014zfa},
the study of resurgent structures of both closed and open strings \cite{Marino:2006hs, Marino:2008ya, Pasquetti:2010bps, Couso-Santamaria:2013kmu, Gu:2022sqc}, and geometric interpretations and extensions of the latter from in terms of Riemann-Hilbert problems \cite{Bridgeland:2016nqw, Bridgeland:2017vbr, Coman:2018uwk, Alim:2021mhp, Grassi:2022zuk, Alim:2022oll, Coman:2020qgf, Gavrylenko:2023ewx}. 
In many of these works, higher genus worldsheet instantons play an important role in the analytic continuation across the moduli space. In our work we observed that equivariance plays a similar role in the analytic continuation of genus zero string partition functions. It would therefore be interesting to explore the role of equivariance in the higher genus setting.

\medskip

\emph{Lift to M-theory.}
It would be certainly desirable to gain a better grasp on the physics behind quantum symplectic cuts in the context of string theory. 
A promising direction seems to be through an uplift to M-theory \cite{Aganagic:2001nx, Aganagic:2001ug} where the symplectic cut of $X$ appears naturally.
In presence of the D6 brane on $L\times \BR^4$, going to M-theory replaces $X$ with a $G_2$ manifold fibered by a circle over $X$ which shrinks at $L$. 
If we perform a 9/11 flip and compactify instead on a (choice of) $S^1$ from the $U(1)^3$ fibration of $X$ we obtain a new manifold whose topology resembles the symplectic cut \eqref{eq:CY3-cut-decomposition}.

\bigskip

We plan to continue our discussion of these and other questions in upcoming works \cite{Cassia:2024txc}.

\appendix

\section{Special function definitions and identities}

In this appendix we collect technical formulas about special functions which we use in the main text.

We define the Pochhammer symbol $(z)_n$ as the ratio of Gamma functions
\be
\label{eq:Pochhammer}
 (z)_n := \frac{\Gamma(z+n)}{\Gamma(z)}
\ee
which is well-defined for complex values of $z$ and $n$ provided $z$ and $z+n$ are not negative integers. For $n\in\mathbb{Z}$, the Pochhammer symbol can be written as the finite product
\be
 (z)_n = \prod_{k=0}^{n-1} (z+k)
\ee
if $n>0$, and as
\be
 (z)_n = \prod_{k=n}^{-1} \frac{1}{(z+k)}
\ee
if $n<0$.

The beta function has the following integral representations
\be\label{int-beta-function}
 B(a,b) = \frac{\Ga(a)\Ga(b)}{\Ga(a+b)} = \int^1_0 y^{a-1} (1-y)^{b-1} \dif y =
 \int^\infty_0 \frac{x^{a-1}}{(1+x)^{a+b}} \dif  x
\ee
where we have used the following change of variables $x=y/(1-y)$ under the integral.
The Appell $F_1$ function has the integral representation
\be\label{int-AppelF1-function}
 F_1(a;b_1,b_2;c;z_1,z_2) =
 \frac{\Ga(c)}{\Ga(a)\Ga(c-a)}
 \int_0^1 \frac{w^{a-1} (1-w)^{c-a-1}}{(1-w z_1)^{b_1}(1-w z_2)^{b_2}} \, \dif w
\ee
for $\Re(a)>0\land \Re(c-a)>0$.

We use the integral representation of the ${}_2F_1$ hypergeometric function
\be\label{HG-int-1}
 B(b,c-b) \, {}_2F_1 (a,b;c;z) = \int_0^1 x^{b-1} (1-x)^{c-b-1} (1-zx)^{-a}~\dif x
\ee
with $\Re(c)>\Re(b)>0$ and $\Re(z)\leq 1$.

Below we list some relevant identities for hypergeometric functions. First we have
\begin{multline}
\label{eq:rule2}
 {}_2F_1(a,b;c;z) = \frac{\Ga (c)\Ga (a+b-c)}{\Ga(a)\Ga(b)}
 \, {}_2F_1(c-a,c-b;c-a-b+1;1-z)\, (1-z)^{c-a-b} \\
 + \frac{\Ga (c) \Ga (c-a-b)}{\Ga (c-a) \Ga (c-b)}
 \, {}_2F_1(a,b;a+b-c+1;1-z)
\end{multline}
where $c-a-b$ is not an integer. Next we recall
\be\label{HG-z-zinverse}
 {}_2 F_1 \left (a,b;c;1-z \right ) = z^{-a} {}_2 F_1 \left (a, c-b; c; 1 -\frac{1}{z} \right )~,~~~~z \notin (0,\infty)
\ee
and
\begin{multline}
\label{eq:rule1}
 {}_2F_1(a,b;c;z) = \frac{\Ga (c) \Ga (b-a)}{(-z)^a \Ga (b) \Ga (c-a)}
 \, {}_2F_1\left(a,a-c+1;a-b+1;\frac{1}{z}\right) \\
 + \frac{\Ga (c) \Ga (a-b)}{(-z)^b \Ga (a) \Ga (c-b)}
 \, {}_2F_1\left(b,b-c+1;b-a+1;\frac{1}{z}\right)
\end{multline}
Finally, we have a different integral representation for the function ${}_2 F_1$ as
\be\label{int-HG-2}
 B(a, c+b-a) \, {}_2 F_1 (c,a; c+b ; 1-z)
 = \int_0^\infty \frac{x^{a-1}}{(1+x)^b (1+zx)^c}~\dif x
\ee
with $\Re(b+c) > \Re(a) > 0$ and $\Re(z) \geq 0$.

\section{Disk partition function for conifold}\label{a:conifold}

Using the conventions from Section~\ref{s:conifold} we define the disk function for $t>0$ as the contour integral
\be
 \cF^D (t, \e) = \oint_{(1),(2)} \frac{\dif\phi}{2\pi\ii} \,
 \eu^{t\phi} \Ga(\e_1+\phi) \Ga(\e_2+\phi) \Ga(\e_3-\phi) \Ga(\e_4-\phi) 
\ee
where the contour includes two towers of poles $-\e_1-n$ and $-\e_2-n$ for $n=0,1,2,\dots$.
After performing the integrals the answer can be rewritten as follows
\be\label{confiold-def1}
\begin{aligned}
 \cF^D (t, \e)  
 =&  \eu^{-\e_1 t} \Ga(\e_2-\e_1) \Ga(\e_1+\e_3) \Ga(\e_1+\e_4)
 \, {}_2F_1\left(\e_1+\e_3,\e_1+\e_4;\e_1-\e_2+1;\eu^{-t}\right) \\
 & + \eu^{-\e_2 t} \Ga(\e_1-\e_2) \Ga(\e_2+\e_3) \Ga(\e_2+\e_4)
 \, {}_2F_1\left(\e_2+\e_3,\e_2+\e_4;\e_2-\e_1+1;\eu^{-t}\right)\,.
\end{aligned}
\ee
The disk partition function satifies the following equivariant PF equation
\be
 \Big ( \cD_1 \cD_2 - \eu^{-t} \cD_3 \cD_4 \Big ) \cF^D (t, \e)
 \equiv \Big ( (\e_1 + \partial_t) (\e_2 + \partial_t)
 - \eu^{-t} (\e_3 - \partial_t) (\e_4 - \partial_t) \Big ) \cF^D (t, \e) =0
\ee
and semiclassically it goes to the equivariant volume (with some insertions of $\Ga$-class).  
It is natural to write it in the following way
\be
 \cF^D (t,\e) = \semicl +t \Li_2(\eu^{-t}) - 2 \Li_3 (\eu^{-t}) + O(\e)~,
\ee
where the semiclassical terms contain singularities in $\e$'s and instanton terms (the poles $n=1,2,\dots$)
are regular (for more details see \cite{Cassia:2022lfj}). 

Using the functional definition \eqref{confiold-def1} and the identity \eqref{eq:rule2} we can rewrite the disk function as
\begin{multline}
\label{conifold-PF-1}
 \cF^D (t, \e) =
 \frac{\Ga(\e_3+\e_2)\Ga(\e_4+\e_1)\Ga(\e_1+\e_3)\Ga(\e_2+\e_4)}
 {\Ga(\e_1+\e_2 + \e_3+\e_4)} \times \\
 \times \eu^{-\e_1 t} \, {}_2F_1 (\e_1+\e_4, \e_1+\e_3; \e_1+\e_2+\e_3+\e_4; 1-\eu^{-t})\,,
\end{multline}
or, equivalently, using \eqref{HG-z-zinverse} we can write 
\begin{multline}
\label{HP-one-1overz}
 \cF^D (t, \e) =
 \frac{\Ga(\e_1+\e_3)\Ga(\e_2+\e_3)\Ga(\e_1+\e_4)\Ga(\e_2+\e_4)}
 {\Ga(\e_1+\e_2+\e_3+\e_4)} \times \\
 \times \eu^{\e_3 t} \, {}_2F_1(\e_1+\e_3,\e_2+\e_3;\e_1+\e_2+\e_3+\e_4;1-\eu^t)~.
\end{multline}
This form of the disk partition function is naturally related to the chamber $t<0$ and can also be obtained via the following contour integral
\be\label{conifold-another-cham}
\begin{aligned}
 \cF^D (t, \e)
 =& -\oint_{(3)(4)} \frac{\dif\phi}{2\pi\ii} \,
 \eu^{t\phi} \Ga(\e_1+\phi) \Ga(\e_2+\phi) \Ga(\e_3-\phi) \Ga(\e_4-\phi) \\
 =&  \eu^{\e_3 t} \Ga(\e_1+\e_3) \Ga(\e_2+\e_3) \Ga(\e_4-\e_3)
 \, {}_2F_1\left(\e_1+\e_3,\e_2+\e_3;\e_3-\e_4+1;\eu^t\right) \\
 & + \eu^{\e_4 t} \Ga(\e_1+\e_4) \Ga(\e_2+\e_4) \Ga(\e_3-\e_4)
 \, {}_2F_1\left(\e_1+\e_4,\e_2+\e_4;\e_4-\e_3+1;\eu^t\right) \\
\end{aligned}
\ee
where one can use \eqref{eq:rule2} to related the above expression to \eqref{HP-one-1overz}. 
The extra minus in front of the contour integral \eqref{conifold-another-cham}
appears due to the fact that closing the contour along the imaginary
axis either to the left or to the right would carry opposite orientation. 

We remark that all these different formulas can obtained from the following integral expression  
\be
 \cF^D (t, \e) 
 = \int_{-\ii\infty}^{\ii\infty} \frac{\dif\phi}{2\pi\ii}
 \int_0^\infty\dif y_1\dif y_2\dif y_3\dif y_4 \,
 \eu^{t\phi-y_1-y_2-y_3-y_4} y_1^{\e_1+\phi-1} y_2^{\e_2+\phi-1} y_3^{\e_3-\phi-1} y_4^{\e_4-\phi-1}
\ee
which is the representation \eqref{disk-general-y} written for the conifold.

\section{Disk partition function for local \texorpdfstring{$\BP^2$}{P2}}\label{a:locP2}

Using the conventions from Section~\ref{s:P2} we define the disk function for local $\BP^2$ 
for $t>0$ by the contour integral
\be\label{P2-closed-contour}
 \cF^D (t, \e) = \oint_{(1),(2),(3)} \frac{\dif\phi}{2\pi\ii} \,
 \eu^{t\phi} \Ga(\e_1+\phi) \Ga(\e_2+\phi) \Ga(\e_3+\phi) \Ga(\e_4-3 \phi) ~,
\ee
where we take into account three towers of the poles: $(-\e_1-n)$, $(-\e_2-n)$ and $(-\e_3-n)$
for $n=0,1,\dots$. The disk partition function $\cF^D (t, \e)$ satisfies the following equivariant PF equation
\be
 \Big ( \cD_1 \cD_2 \cD_3 - \eu^{-t} (2+ \cD_4) (1+\cD_4) \cD_4 \Big )\cF^D (t, \e) =0~. 
\ee
The expansion in equivariant parameters $\e$ and the relation of $\cF^D (t, \e)$
to closed GW invariant is discussed in \cite{Cassia:2022lfj}.
Alternatively, we can write the disk partition function as
\be
\begin{aligned}
 \cF^D
 &= \int_{-\ii\infty}^{\ii\infty} \frac{\dif\phi}{2\pi\ii}
 \int_{0}^\infty \dif y_1 \dif y_2 \dif y_3 \dif y_4
 ~ \eu^{\phi t-\sum y_i} y_1^{\e_1+\phi-1} y_2^{\e_2+\phi-1} 
 y_3^{\e_3+\phi-1} y_4^{\e_4-3\phi-1} \\
 &= \int_{0}^\infty  \dif y_1 \dif y_2 \dif y_3 \dif y_4 
 ~ \eu^{-\sum y_i} \prod_i y_i^{\e_i-1}
 \delta\left(\log\left(\frac{\eu^{-t} y_4^3}{y_1 y_2 y_3}\right)\right) \\
 &=
 \eu^{-\e_1 t}
 \int_{0}^\infty \dif y_2 \dif y_3 \dif y_4~
 y_2^{\e_2-\e_1-1} y_3^{\e_3-\e_1-1} y_4^{\e_4+3\e_1-1}
 \exp\left(-\frac{\eu^{-t}y_4^3}{y_2y_3}-y_2-y_3-y_4\right) \\
 &=
 \eu^{-\e_1 t} \Ga(\e_2-\e_1) \Ga(\e_3-\e_1) \Ga(3\e_1+\e_4) \times \\
 & \quad \times \, {}_3F_2\left(\e_1+\frac{\e_4}{3},\e_1+\frac{\e_4}{3}+\frac13,\e_1+\frac{\e_4}{3}+\frac23;\e_1-\e_2+1,\e_1-\e_3+1;-27\eu^{-t}\right) \\
 & + \eu^{-\e_2 t} \Ga(\e_1-\e_2) \Ga(\e_3-\e_2) \Ga(3\e_2+\e_4) \times \\
 & \quad \times \, {}_3F_2\left(\e_2+\frac{\e_4}{3},\e_2+\frac{\e_4}{3}+\frac13,\e_2+\frac{\e_4}{3}+\frac23;\e_2-\e_1+1,\e_2-\e_3+1;-27\eu^{-t}\right) \\
 & + \eu^{-\e_3 t} \Ga(\e_1-\e_3) \Ga(\e_2-\e_3) \Ga(3\e_3+\e_4) \times \\
 & \quad \times\, {}_3F_2\left(\e_3+\frac{\e_4}{3},\e_3+\frac{\e_4}{3}+\frac13,\e_3+\frac{\e_4}{3}+\frac23;\e_3-\e_1+1,\e_3-\e_2+1;-27\eu^{-t}\right)~,
\end{aligned}
\ee
where we use the ${}_3 F_2$ hypergeometric function.
This explicit expression matches with the residue computation \eqref{P2-closed-contour} after resummation. 

\printbibliography

\end{document}